\documentclass[onecolumn,a4paper,accepted=2023-07-12]{quantumarticle}
\pdfoutput=1

\usepackage[backend=biber,style=alphabetic, maxbibnames=5, url=false]{biblatex}
\usepackage{amsmath}
\usepackage{amssymb}
\usepackage{amsthm}
\usepackage{amsfonts}
\usepackage[caption=false]{subfig}
\usepackage[colorlinks]{hyperref}
\usepackage[all]{hypcap}
\usepackage{tikz}
\usepackage{relsize}
\usepackage{color,soul}
\usepackage[utf8]{inputenc}
\usepackage{capt-of}
\usepackage{mathtools}
\usepackage{float}
\usepackage[section]{placeins}
\usepackage{listings}
\usepackage{svg}

\usepackage[T1]{fontenc}  
\usetikzlibrary{decorations.pathreplacing}
\usepackage{braket}

\theoremstyle{definition}
\newtheorem{definition}{Definition}[section]
\theoremstyle{definition}

\theoremstyle{definition}
\newtheorem{lemma}[definition]{Lemma}

\newcommand{\eq}[1]{\hyperref[eq:#1]{Equation~\ref*{eq:#1}}}
\renewcommand{\sec}[1]{\hyperref[sec:#1]{Section~\ref*{sec:#1}}}
\DeclareRobustCommand{\app}[1]{\hyperref[app:#1]{Appendix~\ref*{app:#1}}}
\newcommand{\fig}[2][]{\hyperref[fig:#2]{Figure~\ref*{fig:#2}#1}}
\newcommand{\tab}[1]{\hyperref[tab:#1]{Table~\ref*{tab:#1}}}
\newcommand{\theoremref}[1]{\hyperref[theorem:#1]{Theorem~\ref*{theorem:#1}}}
\newcommand{\definitionref}[1]{\hyperref[definition:#1]{Definition~\ref*{definition:#1}}}

\newcommand{\eqkak}{\overset{\mathrm{kak}}{=\joinrel=}}

\begin{document}

\title{Relaxing Hardware Requirements for Surface Code Circuits using Time-dynamics}

\date{\today}

\author{Matt McEwen}
\email{mmcewen@google.com}
\affiliation{Google Quantum AI, Santa Barbara, California 93117, USA}

\author{Dave Bacon}
\email{dabacon@google.com}
\affiliation{Google Quantum AI, Seattle, Washington 98103, USA}

\author{Craig Gidney}
\email{craig.gidney@gmail.com}
\affiliation{Google Quantum AI, Santa Barbara, California 93117, USA}

\maketitle

\begin{abstract}
The typical time-independent view of quantum error correction (QEC) codes hides significant freedom in the decomposition into circuits for execution on hardware.  Using the concept of detecting regions, we design time-dynamic QEC circuits directly instead of designing static QEC codes to decompose into circuits. In particular, we improve on the standard circuit constructions for the surface code, presenting new circuits 
that can embed on a hexagonal grid instead of a square grid,
that can use ISWAP gates instead of CNOT or CZ gates,
that can exchange qubit data and measure roles,
and that move logical patches around the physical qubit grid while executing.
All these constructions use no additional entangling gate layers and display essentially the same logical performance, having teraquop footprints within 25\% of the standard surface code circuit.
We expect these circuits to be of great interest to quantum hardware engineers,
because they achieve essentially the same logical performance as standard surface code circuits while relaxing demands on hardware.
\end{abstract}

\section{Introduction}\label{sec:introduction}

Traditionally, quantum error correcting (QEC) codes are defined by a static structure of stabilizers~\cite{shor_scheme_1995, calderbank_good_1996, steane_multiple-particle_1996, kitaev_fault-tolerant_1997}.
For example, the surface code \cite{bravyi_quantum_1998, dennis_topological_2002} is usually introduced without time-dynamics as an unchanging set of stabilizer terms to repeatedly measure. 
Although this time-independent approach to QEC is appealing in its simplicity, performing quantum computation on hardware requires dealing with time dynamics.
At the logical level, time dynamics are unavoidable because logical computation applies operations changing the structure of the QEC circuit as it is being executed \cite{bravyi_universal_2005, horsman_surface_2012}.
At the physical level, time dynamics are unavoidable because stabilizer measurements are not atomic operations available on hardware \cite{fowler_surface_2012}.
The required stabilizer measurements must be decomposed into layers of native hardware operations, which cause the state of the system to vary from moment to moment as the layers execute.
While not all approaches to fault-tolerance rely on repeated stabilizer measurements, other approaches like single-shot QEC~\cite{fujiwara_ability_2014, bombin_single-shot_2015} still involve the execution of a stabilizer circuit with non-trivial time dynamics.
Some recent works have begun to explore the time dynamics of codes, including in Floquet codes  \cite{hastings_dynamically_2021, haah_boundaries_2022, aasen_adiabatic_2022, paetznick_performance_2022, gidney_benchmarking_2022}. However, these approaches remain close to the current standard, focusing on the evolution of code stabilizers and the propagation of errors over time \cite{gottesman_opportunities_2022}.

Experimentally implementing QEC codes presents an impressive challenge.
The surface code is a popular candidate because it presents an excellent compromise of good logical performance and achievable demands on hardware \cite{fowler_surface_2012}. The surface code permits simple decomposition into a circuit via the addition of \emph{measure qubits}, with a resulting qubit grid requiring only four-fold local connectivity in 2D, and using a cycle depth of only four layers of CNOT or CZ gates.
At the same time, these circuits for the surface code display impressive logical performance under realistic error models \cite{krinner_realizing_2022, google_quantum_ai_suppressing_2022} and can be decoded efficiently by matching \cite{dennis_topological_2002, fowler_optimal_2013, higgott_pymatching_2021}.
Previous attempts to construct circuits with more relaxed hardware requirements have faced significant challenges. Lower connectivity either demands unreasonable overhead in the cycle depth or the use of alternative codes sacrificing logical performance \cite{bacon_operator_2005, chamberland_topological_2020, sundaresan_matching_2022}. Further challenges are presented by various non-ideal realities of hardware, such as the presence of leakage states. Modifications to improve the code's resilience to such effects also typically introduce additional overhead and harm logical performance \cite{fowler_coping_2013}. 

In this work we aim to provide an alternative foundation of concepts for reasoning about quantum error correcting \emph{circuits} as opposed to quantum error correcting \emph{codes}. 
In particular, we highlight the concepts of \emph{detectors} and \emph{detecting regions} as a generalization of code stabilizers to the time-dynamic circuit picture. As evidence that these concepts are useful, we present several circuits which are improvements over the standard circuit decompositions of the surface code. These improved circuits still implement the surface code and so enjoy essentially equivalent logical performance. At the same time, they relax demands made on the hardware that implement them. These circuits were all inspired by looking at and modifying the time dynamics of the detectors in the standard surface code circuit.

Due to the difficulty of implementing QEC experimentally, and the resulting breadth of hardware architectures aiming to implement QEC, progress is often focused on exploiting hypothetical hardware specifics for better error correction performance. This kind of work is essentially in the form ``If hardware can do X then we could do Y''. Recent examples of such progress include new codes exploiting strong noise bias \cite{tuckett_tailoring_2019, bonilla_ataides_xzzx_2021}, long-range connectivity \cite{breuckmann_balanced_2020, breuckmann_quantum_2021, baspin_connectivity_2022, panteleev_asymptotically_2021, roffe_bias-tailored_2022,}, or hardware parity measurements \cite{lalumiere_tunable_2010, divincenzo_multi-qubit_2012, royer_qubit_2018, reagor_hardware_2022, livingston_experimental_2022}.
Our work strives to be of the complementary form: ``Hardware doesn't need to do X because we can do Y'': We show that designing the time-dynamic QEC circuit directly can allow for relaxation of the hardware requirements.

In particular, we show that efficient surface code circuits can be constructed using three couplings per qubit rather than four, using ISWAP gates instead of CNOT or CZ gates, and with improved resilience to leakage by involving measurements on all physical qubits. All three of these exemplar circuits represent an improvement over the state of the art, performing the surface code with essentially the same logical performance, and maintaining the use of only four layers of entangling gates rather than adding overhead in circuit depth.
The improvements shown in each of these circuits are not mutually exclusive or exhaustive. We also discuss combining the improvements from these constructions, as well as benefits they provide beyond relaxing requirements on hardware such as enabling lower cost logical compilation.
That the existence of these circuits is surprising speaks to the usefulness of directly constructing QEC circuits rather than QEC codes.

\subsection{Organization}
This paper is organised as follows:
In \sec{concepts}, we provide some background on approaching fault-tolerant circuits using detecting regions, and the tools we used to explore those concepts and make the following constructions. We also discuss relevant hardware constraints that motivate our constructions.
In the next three sections, we present our major results in the form of three improved circuits for the surface code: 
In \sec{hex-grid}, we provide circuits for a hex-grid, requiring only three couplings per qubit to implement the surface code. 
In \sec{iswap}, we provide circuits that use ISWAP gates to perform the surface code.
In \sec{walking}, we provide circuits for the surface code that exchange the roles of data and measure qubits in each cycle. 
In all three cases, we explain the circuit construction in terms of detecting regions, and rigorously benchmark the performance of each circuit, demonstrating essentially the same logical performance as the standard surface code. 
Finally, in \sec{conclusion}, we discuss combining these constructions, present a table of all 24 included circuit constructions, and provide some outlook and commentary on future constructions using these concepts.

Following the main text, we provide several appendices discussing parts of our work in more detail. 
In \app{stabilizers} we provide more rigorous definitions regarding the propagation of detecting regions.
In \app{step_code}, we present the equivalent of two of our results in the repetition code for pedagogical purposes.
In \app{sliding}, we discuss an applications of the walking surface code circuits beyond relaxing hardware requirements, in improving logical circuit compilation. 
In \app{noise_model}, we detail the methodology we used for numerically benchmarking our constructions, including detailing the noise models used.
In \app{benchmarking}, we provide summary benchmarking of all our constructions.
Finally, in \app{crumble}, we provide some convenient links for opening circuits for our constructions in an online interactive tool.

Finally, we also provide a set of supplementary figures available as an \href{https://arxiv.org/src/2302.02192/anc/supplementary_figures.pdf}{ancillary file with this work}, including visualization of each circuit we benchmark along with more traditional plots of error correction performance. This file can also be found in our data repository found at \href{https://zenodo.org/record/7587578}{\texttt{zenodo.org/record/7587578}} \cite{mcewen_data_2023}, along with a full description of each circuit we benchmarked, the results of our benchmarking, and all major assets used in this paper. The code used to produce the circuits, perform the benchmarking, and make figures is available in our code repository at \href{https://github.com/Strilanc/midout}{\texttt{https://github.com/Strilanc/midout}}.

\section{Concepts and Tools}\label{sec:concepts}

Fault-tolerance is the property that a circuit overall can be more reliable than the faulty gates that make it up. 
This is a very desirable property for circuits we want to run on hardware, where noisy gates might otherwise limit our chances of success to unacceptably low levels.
When a QEC code is described as fault-tolerant, we usually mean that it provides us some strategy to make fault-tolerant circuits.
The task of constructing fault-tolerant circuits is often approached in this way: First, choosing code stabilizers; second, producing a \emph{cycle circuit} that measures those stabilizers; third, repeating that cycle circuit to build a locally fault-tolerant circuit chunk implementing the code; and fourth, applying some strategy to execute logical computation without disturbing that local fault-tolerance.
In this section, we aim to introduce an alternative paradigm for approaching fault-tolerant circuits more directly than via choosing stabilizers, without disturbing the desirable properties of codes or requiring a new strategy for decoding or logical computation. 

We start by recasting the typical approach to static stabilizer codes in terms of propagating stabilizers in the circuit, highlighting how the state of the system changes as the stabilizers are measured by the cycle circuit. 
We then introduce the concepts of detectors and detecting regions, using the same propagation rules to understand the sensitivity of the circuit to errors. 
Finally, we use these concepts to address the fault-tolerance of the circuit directly, rather than via the measurements of stabilizers. We argue that detecting regions are a useful primitive to not only the stabilizers but other key concepts in QEC as well.
This explanation aims to be pedagogical and rooted in the circuit picture, but we provide more rigorous definitions in \app{stabilizers}.

Following this, we discuss the software tools we used in exploring these constructions and some of the relevant hardware constraints that motivated us in constructing new circuits.

\subsection{Mid-cycle states}

\begin{figure}[p]
    \centering
    \resizebox{\linewidth}{!}{
        \includegraphics{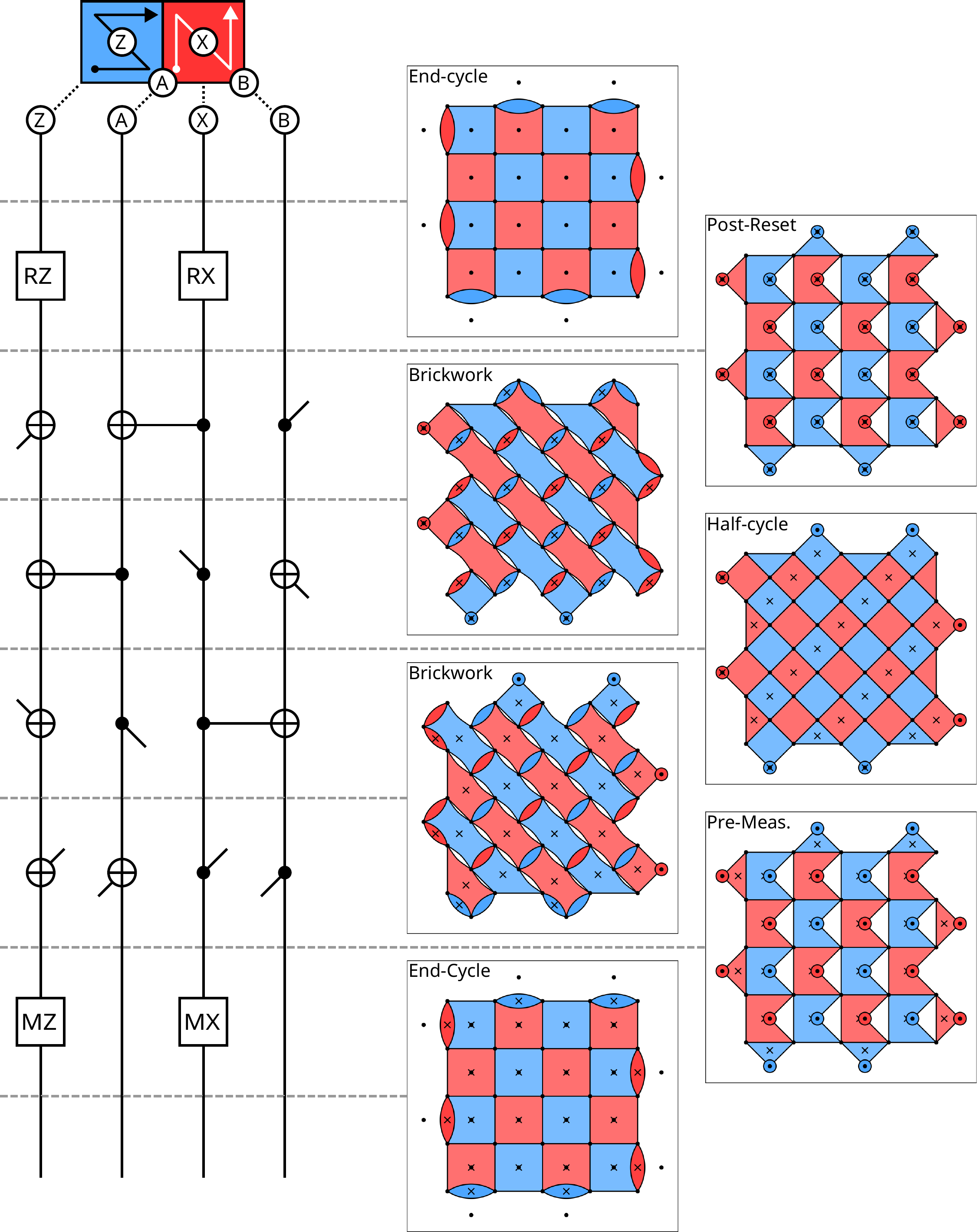}
    }
    \caption{
    \textbf{Mid-cycle States in the Standard Surface Code Circuit.}
    Left: The standard circuit for the surface code, shown using CNOT gates on two measure qubits (X and Z) and two data qubits (A and B). Time proceeds down the page. Diagram at the top indicates the order in which the data qubits are interacted with.
    Right: A visual representation of the state at each point during the circuit. Each shape represents a single stabilizer term consisting of either Z (blue) or X (red) Pauli elements. The corners of the shape indicate the qubits included in the term, with circles indicating single qubit terms.
    Stabilizer terms introduced by the reset gates are marked with an X.
    }
    \label{fig:surface_code_cycle}
\end{figure}

The stabilizer formalism \cite{gottesman_stabilizer_1997} permits analysis of Clifford circuits by simply following Pauli terms around the circuit. 
We track the state of the system using Pauli terms that stabilize the current state, and each applied operation transforms these stabilizers \cite{gottesman_heisenberg_1998}. We define these transformations via their \emph{stabilizer flows}, which describe how Pauli terms before and after a gate operation are related, as detailed in \app{stabilizers}.
The simplicity of this approach underlies the popularity of stabilizer codes for quantum error correction.

Consider the CSS surface code:
It is typically defined and depicted as a checkerboard of interlocking weight-4 X and Z stabilizers. 
We can define a simple circuit using an ancillary \emph{measure qubit} and four layers of CNOT gates to measure a stabilizer, which we refer to as the \emph{standard surface code circuit} throughout and show in \fig{surface_code_cycle}. 
When this cycle circuit is considered as a whole, we can see that it preserves the code stabilizers, leaving us at the end of the cycle in the same checkerboard state we started in. In this picture, the code stabilizers are persistent and unchanging. 

However, during the execution of the cycle circuit, the code stabilizers are transformed by each layer of operations, only returning to the state described by the initial checkerboard at the end of the cycle. Further, the code stabilizers represent only around half of the stabilizers present at any given time. At the start of the cycle, each measure qubit provides an additional single qubit stabilizer enforced by the measurement or reset\footnote{We have $d^2$ data qubits and $d^2-1$ measure qubits in a distance $d$ patch. We have $1$ logical observable, and $d^2-1$ code stabilizers touching the data qubits, and so obviously need another $d^2-1$ stabilizers. We have one such stabilizer on each measure qubit, as noted.}. \fig[b]{surface_code_cycle} shows how both the code stabilizers and ancillary stabilizers evolve as the circuit is executed. For reasons we will explain later in \sec{detecting_regions}, we have made a motivated choice not to show the measure qubit stabilizers before the reset layer and to include the measure qubit stabilizers into the code stabilizers after the layer of reset operations; for now, this is simply a change of generators which doesn't affect the described state.

For convenience, we name these \emph{mid-cycle} states as follows:
Between the measurements and resets, the \emph{end-cycle} state is shown as the familiar checkerboard pattern. 
Immediately following the reset gates, we see the flag-like pattern of the \emph{post-reset} state. Here, we show a full generating set of stabilizers, including the measure qubit stabilizers. This state is equivalent to the end-cycle state when the single measure qubit stabilizers are included, as the weight-4 square code stabilizers and the weight-5 flag-like terms differ only by those single qubit stabilizers. The full set of stabilizers in both cases generate the same overall stabilizer group.
This pattern is transformed by each subsequent layer of CNOT gates, first into a brick-wall-like pattern we call the \emph{brickwork} state, then into a checkerboard pattern rotated by 45 degrees that we call the \emph{half-cycle} state, then into a modified version of the brickwork state, and finally back into a flag-like pattern at the \emph{pre-measure} state, returning us to the checkerboard pattern for the end-cycle state after measurements.
The half-cycle state is remarkable for being an unrotated surface code state, as originally proposed for the surface code \cite{kitaev_fault-tolerant_1997, bravyi_quantum_1998}. The half-cycle state is a surface code state with the same distance as the end-cycle state, but twice the number of qubits involved and twice the number of stabilizers - the usually ignored measure qubits stabilizers have become included into the code state.

Recognising these mid-cycle states is helpful for three reasons.
Firstly, they provide helpful checkpoints for understanding circuits. The mid-cycle states break the problem into smaller pieces which can be analysed and understood separately, rather than thinking of the entire cycle as a single operation that preserves the code stabilizers. 
Secondly, it shifts the focus from the qubits and their roles in the circuit toward the detecting regions. How errors are detected becomes more clear, and the traditional assignment of specific roles to specific qubits becomes de-emphasised; all qubits are equally covered by detecting regions.
Finally, these states provides some insight into what a surface code circuit should achieve, and inspiration on how it might be done differently. For instance, the half-cycle state does not ``remember where it came from'' and can be mapped back to an end-cycle state in more ways than the familiar one embodied in the standard circuit. 

To explore these freedoms, and to better approach the necessity of fault-tolerance provided by repeating the cycle circuit, we need to reach for a more general concept rooted in circuit approach.

\subsection{Detectors and Detecting Regions}\label{sec:detecting_regions}

In fault-tolerant stabilizer circuits, errors are identified by noticing violations of ideal circuit behaviour. 
In particular, circuits often feature small sets of measurements that should display a deterministic parity under noiseless execution. 
We refer to these sets of measurements as \emph{detectors}. 
In an experiment, recording a detector's measurements with an overall parity different to the expected parity is clear evidence of an error occurring, which we call a \emph{detection event}. This approach is most natural for stabilizers codes achieving fault-tolerance via repeated measurement of the stabilizers, but also works for related stabilizer QEC strategies, including single-shot QEC~\cite{fujiwara_ability_2014, bombin_single-shot_2015}, cluster-state based QEC~\cite{raussendorf_fault-tolerant_2006, bombin_logical_2021}, subsystem codes~\cite{bacon_operator_2005,  bravyi_subsystem_2012} and flag-qubits~\cite{chao_quantum_2018, chamberland_fault-tolerant_2019, chao_flag_2020}. 

For any given detector, we can ask where errors could be inserted into the circuit to affect that detector; essentially the region of the circuit where the detector is sensitive to errors. We follow Gottesman~\cite{gottesman_opportunities_2022} in considering a circuit as made up of \emph{locations}, essentially qubits between operations.
Each detector will be sensitive to specific errors in a finite set of locations in the circuit only, which we call the \emph{detecting region}.

\subsubsection{Examples of detecting regions}

\begin{figure}[p]
    \centering
    \resizebox{\linewidth}{!}{
        \includegraphics{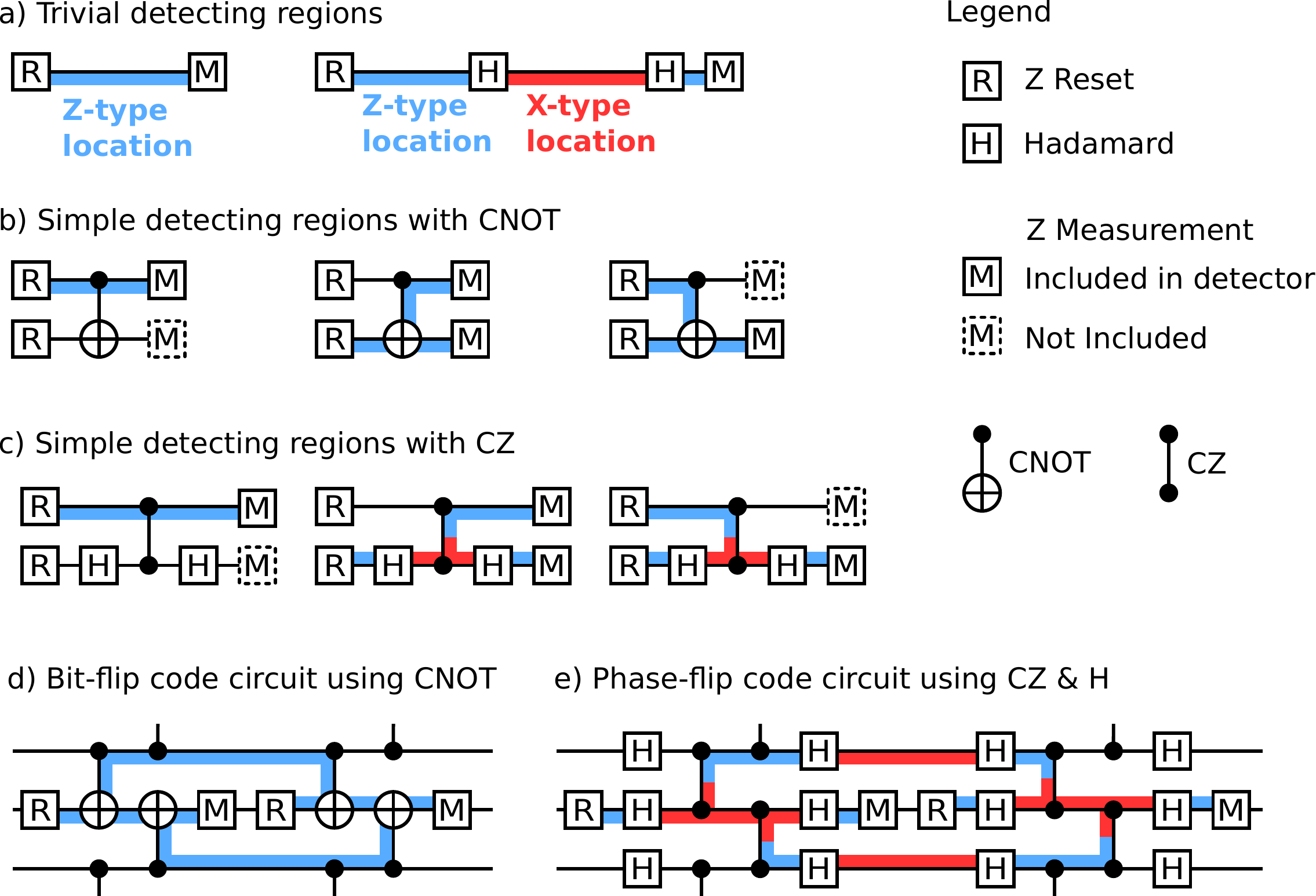}
    }
    \caption{
    \textbf{Examples of detecting regions.}
    a) Detecting regions for 1-qubit circuits. 
    Here, the detectors are just the single measurement in the circuit, which should have a deterministic outcome in the absence of noise. 
    Z Reset gates produce states that are stabilized by Z, indicated by a Z-type section of the detecting region (blue).
    Z Measurements terminate Z-type sections of the detecting region. 
    Hadamard gates change the type of the detecting region as it moves through the circuit, producing X-type locations (red).
    b) Three copies of the same simple circuit with a CNOT gate, indicating the three possible detectors and their detecting regions. Any two of these detectors form a generating set for the full set of detectors.
    c) The same circuit and detectors compiled into CZ and Hadamard gates. 
    Note that CZ gates connect Z-type (blue) locations to X-type locations.
    d) A part of a circuit for a bit-flip repetition code, using CNOT and Reset gates.
    The two consecutive measurements shown form a detector. 
    The associated detecting region is highlighted (blue), and is Z-type throughout.
    The region covers neighbouring data qubits during the measurement, reflecting the bit-flip code's ZZ code stabilizers.
    e) A part of a phase-flip repetition code, using CZ and Reset gates.
    Similarly, the two consecutive measurements shown form a detector. 
    The associated detecting region is highlighted, containing locations of both Z-type and X-type. 
    The regions include the data qubits as X-type locations during measurement, reflecting the phase-flip code's XX code stabilizers. 
    }
    \label{fig:simple_det_regions}
\end{figure}

\fig{simple_det_regions} shows some simple detecting regions and associated detectors.
A Z-basis measurement included in a detector is sensitive to errors immediately before it that anti-commute with Z. 
We can represent this by marking those locations in the circuit with Z-type.
An isolated reset and measurement in the same basis forms the smallest non-empty detector, where the detecting region covers the one location in the circuit, as illustrated in \fig[a]{simple_det_regions}.
All operations relate parts of the detecting region in the same way they relate Pauli terms being commuted through them, which we refer to as their \emph{stabilizer flows}. We define and discuss this concept in detail in \app{stabilizers}. 
Operations with stabilizer flows that relate different Pauli types will also change the type of the detecting region as it is propagated; for example, Hadamard gates with a Z-type on one side have X-type on the other.

A detecting region then consists of a set of locations in the circuit, along with a Pauli type at each location.
A detection event will be caused by any inserted error that anti-commutes with the type of the detecting region at its location. 
We say that the detecting region is \emph{sensitive} to such an error.

Detectors and detecting regions underlie the fault-tolerance of repetitively measuring code stabilizers. 
Consecutive measurements of the same stabilizer will agree under noiseless execution, and so pairs of consecutive measurements of the same code stabilizer form a detector in the bulk of the circuit.

Detecting regions are easier to visualise in 1-dimensional classical codes such as repetition codes.
\fig[d]{simple_det_regions} and \fig[e]{simple_det_regions} show the detecting regions for bit- and phase-flip repetition codes respectively, defined by neighbouring ZZ and XX code stabilizers respectively. The detecting regions in the bulk of these codes consist of a small, closed and local part of the circuit, visually indicating how much of the circuit each detector is responsible for.

We can transform easily between detectors and detecting regions. Given a detector, we can find the detecting region by simply propagating Pauli types backward from the included measurements, following stabilizer flows in reverse and backward-terminating on appropriate resets. This process is deterministic (given that the measurement parities are deterministic), and will produce the detecting region. Given a detecting region, the corresponding detector is simply the measurements the region terminates on. We note that the task of choosing appropriate detectors and detecting regions given an un-annotated circuit is a non-trivial task, and one that can have a large impact on logical performance.

\subsubsection{Overlapping structure of detecting regions}

\begin{figure}[p]
    \centering
    \resizebox{\linewidth}{!}{
        \includegraphics{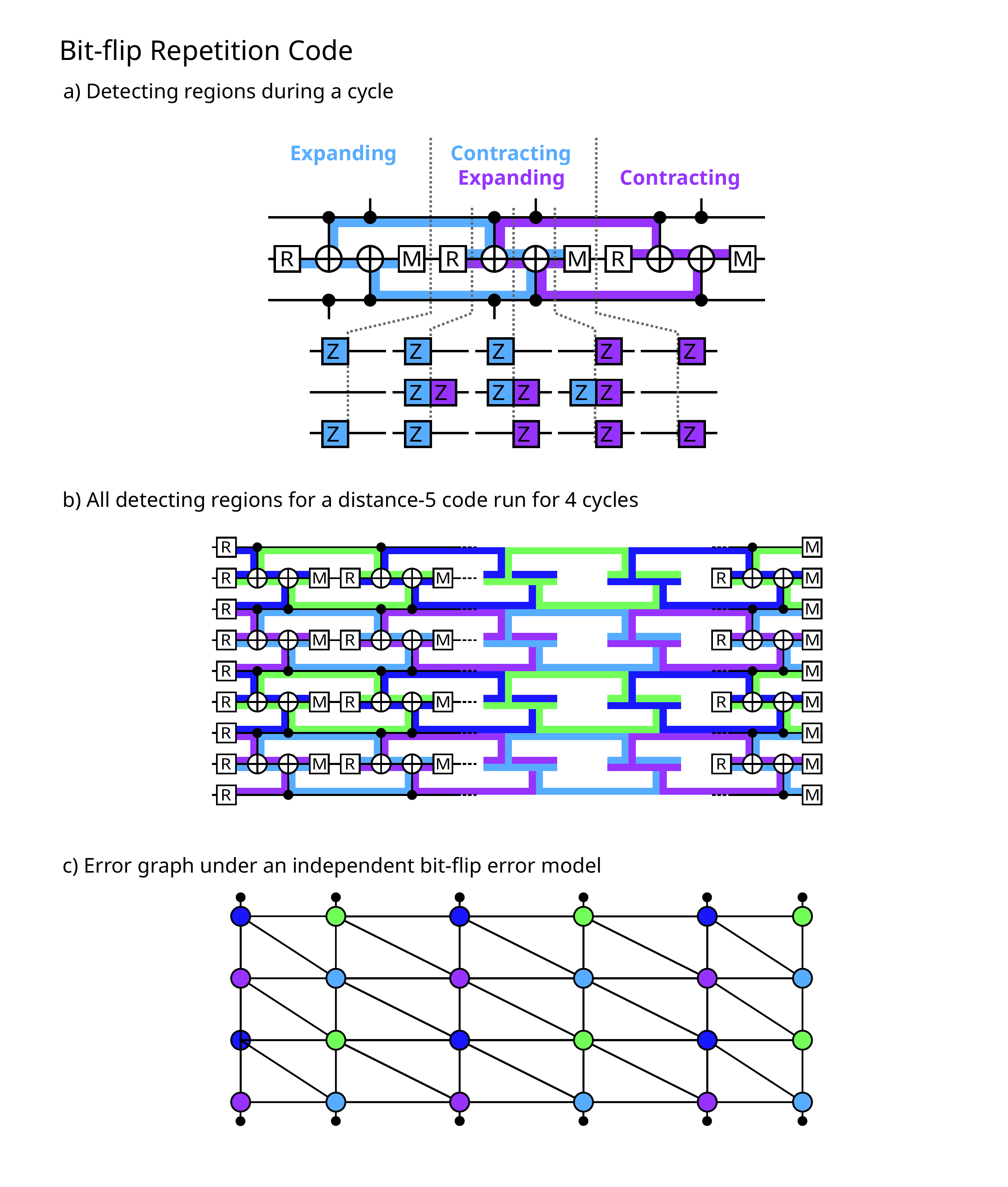}
    }
    \caption{
    \textbf{Overlapping detecting regions in the repetition code.}
    All locations of all regions shown here are Z-type.
    a) Two neighbouring detecting regions of the bit-flip repetition code (blue, purple), shown over three cycles. Each detecting region covers two cycles: In the first cycle, it emerges from a reset on the measure qubit and expands to cover the code stabilizer. In the second cycle, it contracts from the code stabilizer down to terminate on a measurement. During the middle cycle, where the two shown regions co-exist, time-slices of the detecting regions are shown.
    b) All detecting regions for a distance-5 bit-flip repetition code. The detection regions overlap such that all locations in the circuit are covered by two detecting regions, except the boundaries which are covered by one. Note the smaller detecting regions at the start of the code that emerge from resets on both measure and data qubits. On cycles three and four, we omit the circuit elements to emphasise that the detecting region shapes alone are sufficient to understand the implementation. Note the final detecting regions terminate on measurements on both the measure and data qubits.
    c) The associated error graph for the distance-5 bit-flip code, illustrating the correspondence between error graph edges and overlaps in the detecting regions.
    Nodes on the boundary feature single-ended edges, corresponding to sections of the circuit not overlapping with any other detecting region.
    }
    \label{fig:overlapping_det_regions}
\end{figure}

Given that each detecting region is responsible for some part of the circuit, it stands to reason that a good set of detectors should have detecting regions that cover all relevant parts of the circuit. These detecting regions should also overlap, such that any individual error is detected by multiple regions and provides a usable syndrome. 

In the repetition and surface codes, the bulk detecting regions cover locations in two neighbouring cycles, rather than one cycle as one might naively expect. They also cover the code stabilizer at the time-slice between those cycles. \fig[a]{overlapping_det_regions} illustrates this for the bit-flip repetition code, where the time-slice between measure and reset gate layers shows the existing region covering the ZZ stabilizer. During a single cycle, two detecting regions coexist; one emerging from that cycle's reset gate and \emph{expanding} to cover the stabilizer, and one \emph{contracting} from the code stabilizer to terminate on that cycle's measurement gate. Halfway through the cycle, we can see the regions collectively produce the same pattern of stabilizers as the typical code state (ZZ on neighbouring qubits), but involving all measure and data qubits, rather than only the data qubits. This is the equivalent of the half-cycle state previously discussed for the surface code. 

In the bulk of the repetition code, each location is covered by exactly two detecting regions, meaning an inserted bit-flip error will be noticed by two detectors, as illustrated in \fig[b]{overlapping_det_regions}. This is typically represented in terms of the resulting error graph, shown in \fig[c]{overlapping_det_regions}. Here, each detector is represented by a node. Edges between nodes indicate a possible bit-flip error that will be noticed by these two detectors. Under an independent bit-flip error model, these edges correspond directly to overlaps between the relevant detecting regions. 
The structure of these overlaps, and the resulting structure of the error graph, makes the repetition and surface codes amenable to decoding by matching~\cite{dennis_topological_2002, fowler_optimal_2013, higgott_pymatching_2021}. 

Detecting regions are a useful primitive concept, as they single-handedly define relevant QEC concepts, as follows:
\begin{itemize}
    \item The measurements that the region terminates on defines the corresponding detector.
    \item The overlaps between regions define the edges of the error graph under an independent Pauli error model, just as the detectors define the nodes of the error graph.
    \item The shapes of the regions define the circuit operations via the implied stabilizer flows.
\end{itemize}
As such, provided with only an overlapping structure of detecting regions, we have defined both the code and the circuit in a natural way.

\subsubsection{Detecting Region Slices}

\begin{figure}[htb]
    \centering
    \resizebox{\linewidth}{!}{
        \includegraphics{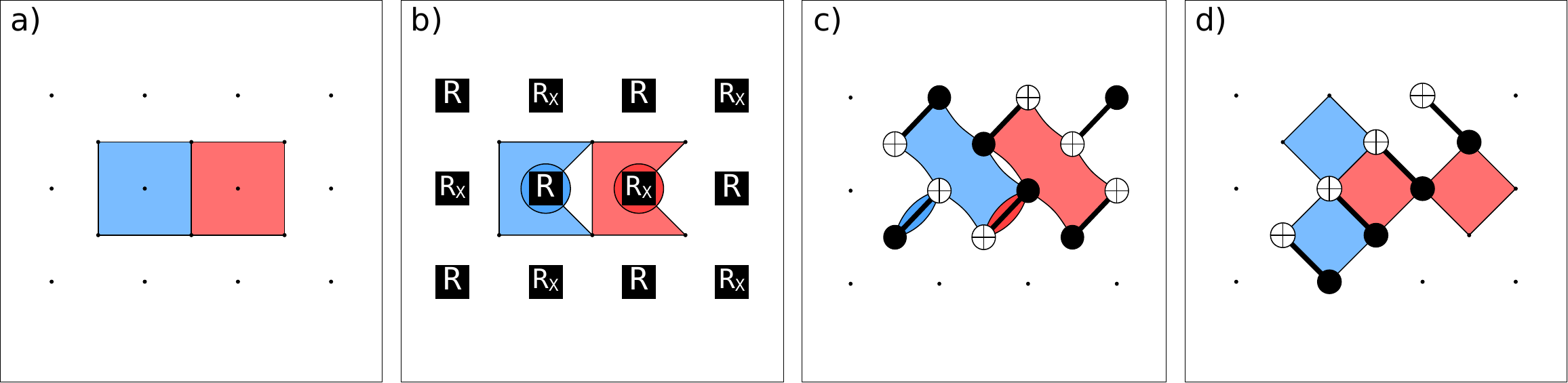}
    }
    \caption{
    \textbf{Detecting region slices in a 2D circuit. }
    Dots indicate qubits locations in a 2D grid. Each shape corresponds to a slice of a detecting region including the qubits at each vertex. Colors indicate the Pauli type, with X-type regions being red and Z-type regions being blue. 
    a) Two detecting region slices representing detectors as they are covering the code stabilizers of the surface code.
    b,c,d) Four detecting region slices in each subsequent time-slice of the standard circuit, with intervening gate layers shown as an overlay. 
    Two new detecting regions introduced by the reset gates in (b) are shown, in addition to the two from (a).
    }
    \label{fig:det_slices}
\end{figure}

Time-slices of the detecting regions are also helpfully related to the code stabilizer picture. A time-slice of a detecting region is a stabilizer of the state at that point in the circuit. \fig{det_slices} shows four detecting region slices after subsequent layers of the standard surface code circuit.

In the language introduced by Aaronson and Gottesman \cite{aaronson_improved_2004} for simulating stabilizer circuits, a time-slice of a detecting region is a possible term in the stabilizer tableau, one that additionally corresponds to a specific detector. 
In the language introduced by Hastings and Haah \cite{hastings_dynamically_2021} for Floquet codes, a time-slice of a detecting region is a generating term of the instantaneous stabilizer group (ISG) that additionally corresponds to a specific detector.
The mid-cycle patterns discussed earlier were represented as time-slices of detecting regions, rather than as any other choice of stabilizer generators. This makes them both more aesthetically pleasing and more meaningful - again, each shape corresponds directly to a specific detector. 

However, it is not the case that the time-slices of the detecting regions always form a complete set of terms for the stabilizer tableau or ISG. 
Some circuit locations are not covered by any detecting region, despite still being in the support of a stabilizer term. Examples include immediately after measurement but before reset, as in \fig[a]{det_slices}, and on gauge qubits in subsystem codes. These are locations where an inserted error cannot contribute to a logical error and does not need to be detected, and so does not need to be included in any detecting region. 
This motivated our choice to ignore the measure qubit stabilizers at the end-cycle state in \fig{surface_code_cycle}; they are not relevant to error correction until after the reset gate, where they are again included in detecting regions. 

The development of new code circuits, and especially understanding their behaviour at boundaries, can be greatly aided by constructing them in terms of detecting regions. 
This is especially true when considering how different detecting regions fit together to cover the space-time of the circuit with an appropriate overlapping structure, as we will discuss in our main results.

\subsection{Detecting Regions in the Surface Code}\label{sec:standard_det_regions}

\begin{figure}[p]
    \centering
    \resizebox{\linewidth}{!}{
        \includegraphics{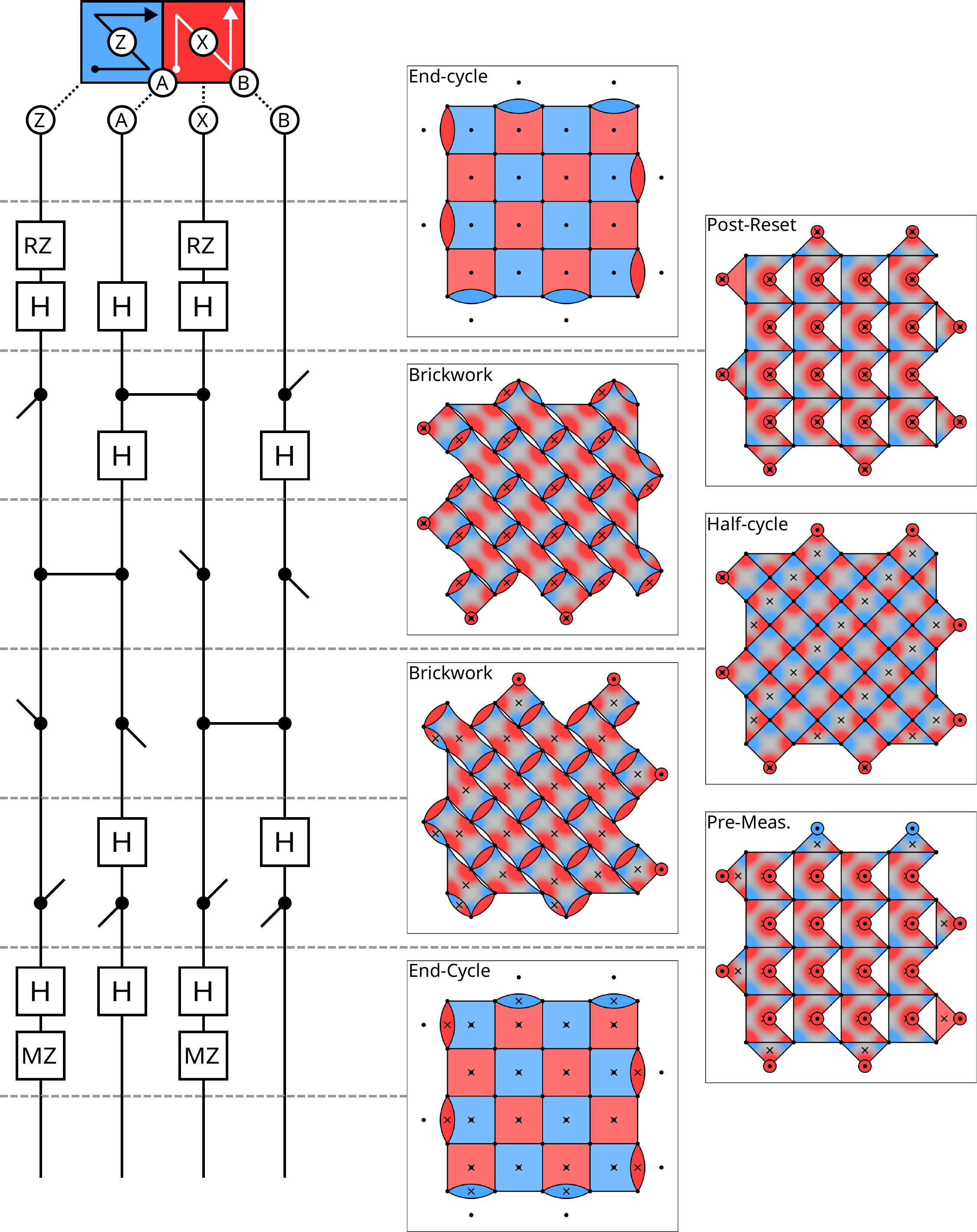}
    }
    \caption{
    \textbf{Mid-cycle States in the CZ Surface Code Circuit.}
    Left: The circuit for the surface code compiled for hardware using H and CZ gates, shown on two measure qubits (X and Z) and two data qubits (A and B). Time proceeds down the page. Diagram at the top indicates the order in which the data qubits are interacted with.
    Right: A visual representation of the state at each point during the circuit. Each shape represents a slice of a detecting region. The corners of the shape indicate the qubits included in the term, with circles indicating single qubit terms. Colors indicate the Pauli elements Z (blue) or X (red), either for the whole shape or for each qubit in the slice.
    Expanding detecting regions are indicated by an X marker.
    Note the inclusion of unnecessary Hadamard gates on qubit A at the start and end of each round; these make CSS stabilizers at measurement, and canceling them yields XZZX/ZXXZ stabilizers at measurement.
    }
    \label{fig:surface_code_cycle_cz}
\end{figure}

We now revisit a standard surface code circuit. 
To avoid retreading ground and to introduce some new possibilities, we now consider the circuit as compiled for hardware using only Z-type reset and measurements, CZ and Hadamard gates as shown in \fig{surface_code_cycle_cz}.
We now interpret the states directly as time-slices of the detecting regions.

Compared to the standard circuit using CNOTs, this circuit displays regions with different Pauli types at locations at the same time slice. 
The regions at each time-slice display a nice overlapping structure by construction. Each location in the circuit is covered by four detecting regions, two with Z-type and two with X-type. This produces the two connected components of the surface code error graph under X and Z errors respectively.
It also makes clear the correlated nature of Y errors; wherever they are inserted, they anti-commute with four detecting regions rather than two.

We can also see that the mid-cycle states display more than just two kinds of detecting region slices. We further distinguish detecting regions by whether they are expanding or contracting; whether they emerged from a reset at the start of this round, or emerged in the previous round and began this round covering the code stabilizers. These two regions evolve differently, with the contracting regions getting smaller as the cycle circuit continues and the expanding regions growing to cover the code stabilizers. 
In our constructions, tracking which regions are expanding and contracting will prove important. 
Failing to contract a detecting region that has already expanded will make a larger detecting region, one responsible for more of the circuit, overlapping with more detecting regions and worsening logical performance. 

The detecting regions picture allows us to emphasise that there are several equivalent ways of looking at the surface code cycle. The familiar end-cycle picture involved starting with the data qubits involved in a surface code state and the measure qubits un-entangled and ready to be used. 
Starting at the half-cycle state, we have a picture where all qubits are involved in a surface code state. During the cycle circuit, half of the stabilizers (those corresponding to the contracting regions) are transformed to occupy only single measure qubits and are measured (while the stabilizers corresponding to expanding regions transform to cover only data qubits), before we return to the half-cycle state. Naively, a surface code without measure qubits where only half the stabilizers are measured in each cycle sounds worse than a code with measure qubits where all stabilizers are measured in each cycle; in fact these are two descriptions of the same circuit.

Alternatively, we can also start with the pair of brickwork states: The two-body stabilizers can be measured via a common decomposition of a pair-measurement into a CNOT followed by single-qubit measurement, with another CNOT allowing us to reconstruct a (slightly different) brickwork pattern. Two layers of CNOTs allow us to move between the two brickwork patterns, essentially exchanging which stabilizers are two- and six-body. Again, this is simply an unusual description of the standard surface code circuit.
The advantage of these alternative approaches to the cycle circuit will become more clear as we use them to explain new constructions. 

\subsection{Boundaries}\label{sec:boundaries}

So far, we have focused on detecting regions in the bulk of the circuit, but now turn to explicitly address what happens at boundaries, which can have a significant influence of the final circuit performance. First, we address the general strategy we use for temporal boundaries in all surface code circuits, and then provide some notes on the more complex spatial boundaries.

First, given a bulk circuit for the surface code, temporal boundaries (namely logical initialization and measurement) can be introduced very simply. Due to the CSS nature of the standard surface code circuit, we can implement logical measurement in the Z basis as follows: At any point in the circuit (but traditionally at the end of the cycle where measurements are already occurring) measure all the qubits in the Z basis. Delete any detecting region that anti-commutes with these new measurements, and any detecting region that commutes with them now terminates on them; these are the correct final round detectors to ensure the logical measurement is fault tolerant. Logical measurement is the X basis is very similar, but the qubits should be measured in the X basis. Logical initialisation is essentially the same process but time-reversed, with resets taking the place of measurements: At any point (but again traditionally at the start of the cycle where resets are already occurring) reset all the qubits in the Z or X basis for Z or X logical initialization. Again, delete or terminate the bulk detecting regions appropriately. We use this strategy for the temporal boundaries of all our constructions. Even when the circuit is compiled to gates that mix X and Z stabilizer terms, this strategy remains essentially the same. The circuit state is always only single qubit rotations away from a CSS-like state\footnote{The states reached by standard circuit for the surface code in CZs therefore violate the letter but not the spirit of the CSS property of the surface code. This suggest we should more often consider the class of CSS-up-to-single-qubit-rotations codes}.

Spatial boundaries can also be found using a similar strategy, but this is far from optimal. In the original unrotated surface code~\cite{kitaev_fault-tolerant_1997, bravyi_quantum_1998}, drawing a square box containing $(2d-1)^2$ total qubits and simply removing all gates that cross this boundary is sufficient to make the circuit for a distance-$d$ patch from a tiled bulk circuit\footnote{To draw out the equivalence to the temporal boundary strategy, this is equivalent to performing single qubit measurements on all qubits outside the box between each layer of gates.}. However, unlike the temporal case, these spatial boundaries are wasteful. The rotated surface code~\cite{bombin_optimal_2007} significantly improves the number of qubits used by exploiting a more complex boundary; appropriately including half the measure qubits around a square patch for $2(d^2)-1$ total qubits for a distance-$d$ logical patch. This exemplifies the major differences in performance that changes in boundaries can make to a surface code construction.

For all of our constructions, finding efficient spatial boundaries presented a more interesting and difficult challenge than finding the bulk circuit. 
For each construction, we present a choice of boundaries that preserves the graph-like code distance~\cite{gidney_benchmarking_2022} (i.e. the code distance considering only errors that produce pairs of detection events). This is not to say that other boundaries are not possible, or even preferable. Various possible constraints or considerations, including from hardware (for example, which qubits are measurable in parallel, such as all qubits or only non-nearest-neighbouring qubits) or from logical considerations (e.g. densely packing logical qubit patches in a distillation factory) can place different demands on the circuit, which can be variously accommodated by different boundaries. Our strategy for finding boundaries involved simply iterating on circuit details, especially which gates to omit or include at the boundaries, until the detecting structure preserved the code distance and benchmarked well. As such, we consider developing spatial boundaries still more art than science, and we relied heavily on tools that sped up iteration over any deep insights into how to pick good spatial boundaries.

\subsection{Software Tools}\label{sec:software}

A major challenge to exploring QEC circuits is the intricate and interlocking nature of the fault-tolerant constructions.
Getting a seemingly small detail wrong, from choosing the wrong measurements for a detector to choosing the wrong order for operations, tends to radically alter the logical performance of the circuit. While simple error correction circuits can be designed by hand, the constructions we present here necessitated the use of software tools to visualise and analyse the circuits. 
Software tools also proved vital in verifying the correctness of the circuit (such as in terms of code distance) and in benchmarking the true performance of the circuit. 

In particular, we made extensive use of Stim~\cite{gidney_stim_2021} and the tools it provides for representing and manipulating QEC circuits, for verifying them and for benchmarking them. 
Over the course of this work, Stim has been improved to include new tools for visualizing circuits and stabilizer states during circuits, which we made extensive use of in both exploring and explaining these constructions. Stim now also includes a \href{https://github.com/quantumlib/Stim/blob/main/glue/crumble/README.md}{prototype of an interactive circuit editor \emph{Crumble}}, which is also accessible at \href{https://algassert.com/crumble}{algassert.com/crumble}. This tool was especially useful for finding appropriate boundary configurations for these code circuits, as it enabled rapid rearranging, inclusion and omission of entangling gates at the boundaries, showing the resulting change in the detecting structure. In \app{crumble}, we provide some convenient links for opening the circuits for our constructions in Crumble, which we hope will serve as a convenient jumping off point. 

These tools were crucial in several ways. 
First, the tools automated critical tasks like verifying that a construction preserves the code distance, permitting cheap exploration because mistakes would be caught early, and helping build intuition for what changes to the circuit constitute mistakes. 
Second, the tools revealed new insights by making visualization and interaction easier. Visualizing the interlocking nature of many detection regions was the initial impetus for all the presented constructions. Crumble proved crucial both in finding effective circuits, especially in finding appropriate boundary detecting regions, and in understanding and explaining our circuits to each other. 
Finally, it proved important that our tools were flexible enough to handle these constructions.
Despite ostensibly being designed to simulate traditional QEC code circuits, Stim operates on general annotated circuits and was able to analyse and benchmark our constructions without any redesign.
A tool built specifically for the current paradigm of measuring stabilizer operators would be much less likely to achieve this without modification. 

The importance of software tools used in this kind of work is often under appreciated, so we have chosen to emphasise the tools we used here with the aim of encouraging wider use and further development of such tools. 

\subsection{Hardware Requirements}

The constraints under which we design error correction circuits come from difficulties and challenges we face in designing hardware. These constraints can depend sensitively on the hardware architecture being targeted. We choose to focus our examples on the paradigm of superconducting
qubit arrays. As previously mentioned, the popularity of the surface code rests on the relatively acceptable demands it makes on hardware: square lattice connectivity, and a cycle containing only
four layers of entangling CZ or CNOT gates.
Further relaxing these hardware constraints generally comes at a significant cost, typically in compromising the error correction performance by introducing additional overhead in the cycle circuit.

Lower connectivities are generally easier to achieve in hardware, usually permitting higher couplings, lower crosstalk, fewer on-chip structures and control lines, and relaxing other constraints such as the frequency arrangement of qubits. Connectivity in a square lattice presents an achievable but difficult design modality~\cite{google_quantum_ai_suppressing_2022, krinner_realizing_2022}, but cutting edge architectures featuring lower connectivity have also been produced~\cite{sundaresan_matching_2022}. However, lower connectivity has historically required an unreasonable swap overhead to implement the surface code, usually frustrating performance below threshold. Use of alternative codes better suited for lower connectivity generally feature comprimises in the logical performance, such as the heavy-hex code displaying a threshold in only one logical observable~\cite{chamberland_topological_2020}. We present a circuit for the surface code that embeds on the lower connectivity hex grid without swap overhead in \sec{hex-grid}.

Circuit decompositions are usually expressed in terms of CNOT and CZ gates. However, other entangling operations are often naturally achievable in hardware, such as the ISWAP gate. For example, in superconducting architectures with tunable couplers, the ISWAP gate is less demanding in terms of frequency arrangement and may be performed at a higher fidelity than the CZ gate~\cite{foxen_demonstrating_2020}.
However, an error correcting code must be compiled to ISWAP gates without significant overhead before this can be taken advantage of. We present a circuit for the surface code using ISWAP gates in \sec{iswap}.

Finally, QEC circuit design generally neglects important non-idealities of the hardware, such as significant sources of correlated errors. In many architectures, including transmon qubit arrays, the ability of the qubit to be promoted out of the computational states into a higher energy leakage
state presents such a source of correlated errors~\cite{fowler_coping_2013, ghosh_understanding_2013}. Such errors are not considered at the level of designing the QEC circuit itself, with attempts to address leakage generally involving adding minimal operations to the standard circuit attempting not to disturb the logical structure while achieving some suppression of the correlated error effect~\cite{ghosh_leakage-resilient_2015, brown_leakage_2019, battistel_hardware-efficient_2021, mcewen_removing_2021, miao_overcoming_2022}. As such, these strategies add overhead to the standard circuit, producing a trade-off
between leakage error suppression and the errors induced by the included operations themselves.
We present a circuit that achieves such a benefit without any additional gate layers in \sec{walking}.

In all three cases, the standard circuit places demands on the hardware that are considered achievable, but the circuit decomposition itself proceeds without feedback from what is preferable for the hardware. 
We regard this work as opening up new possibilities for tailoring the circuit decomposition of a code to the strengths of hardware designs, with the hope that this frees up further improvements in performance.
\section{Hex-grid Surface Code Circuits}\label{sec:hex-grid}

Surface codes are typically compiled to circuits on a square grid (the grid with Schl\"afli symbol \{4,4\}).
In this layout, each qubit has four neighbors.
Without using more layers of operations, it seems difficult to use a sparser connectivity because every qubit interacts with all of its neighbors in each cycle of the circuit.
Measurement qubits interact with their four adjacent data qubits, in order to acquire all the parts of the four-body code stabilizer. The data qubits interact with their four adjacent measurement qubits, in order to make their contribution to the four stabilizers involving that data qubit.

Here, we show a circuit for the surface code with the same code distance and the same number of entangling layers as the usual circuit, but that executes on a hex grid (the grid with Schl\"afli symbol \{6,3\}). When instantiated in a 2D architecture, this means that each qubit needs only three couplers to neighbouring qubits, rather than the typical four. 

We first address the circuit in the bulk, for which we provide two complementary explanations: a bottom-up circuit-focused approach centered around the half-cycle state and a top-down stabilizer-focused approach centered around the brickwork states.
We then talk about the challenge of finding appropriate boundaries for surface code patches using this bulk circuit. Finally, we perform numerical benchmarking of the presented circuit.

\subsection{Bottom-up Circuit-Focused Construction}

Following our previous discussion of the half-cycle state in \sec{standard_det_regions}, we can approach this circuit as a new strategy for measuring half of the stabilizers of the half-cycle state without using any additional qubits.

As in the standard surface code, a four body contracting stabilizer can be measured by using two layers of CNOT gates; the first layer folds the stabilizer in half, from four body to two body; the second folds it again into a one-body stabilizer, which can be directly measured. This half-cycle circuit is pictured in \fig[a]{hex_half_cycle_picture}.
Distinct from the standard surface code, we then apply the same gates in reverse order, returning to the half-cycle state by the same route we left it. 
This basic action has required only three of the four nearest neighbour couplings on the square grid to be used.
The same action can be used on a neighbouring X stabilizer at the same time; the first layer of CNOT gates folds the four-body X stabilizer in the opposite direction. The subsequent layer also folds the X stabilizer into a single qubit stabilizer which can be measured. 

\begin{figure}[ht]
    \centering
    \resizebox{\linewidth}{!}{
        \includegraphics{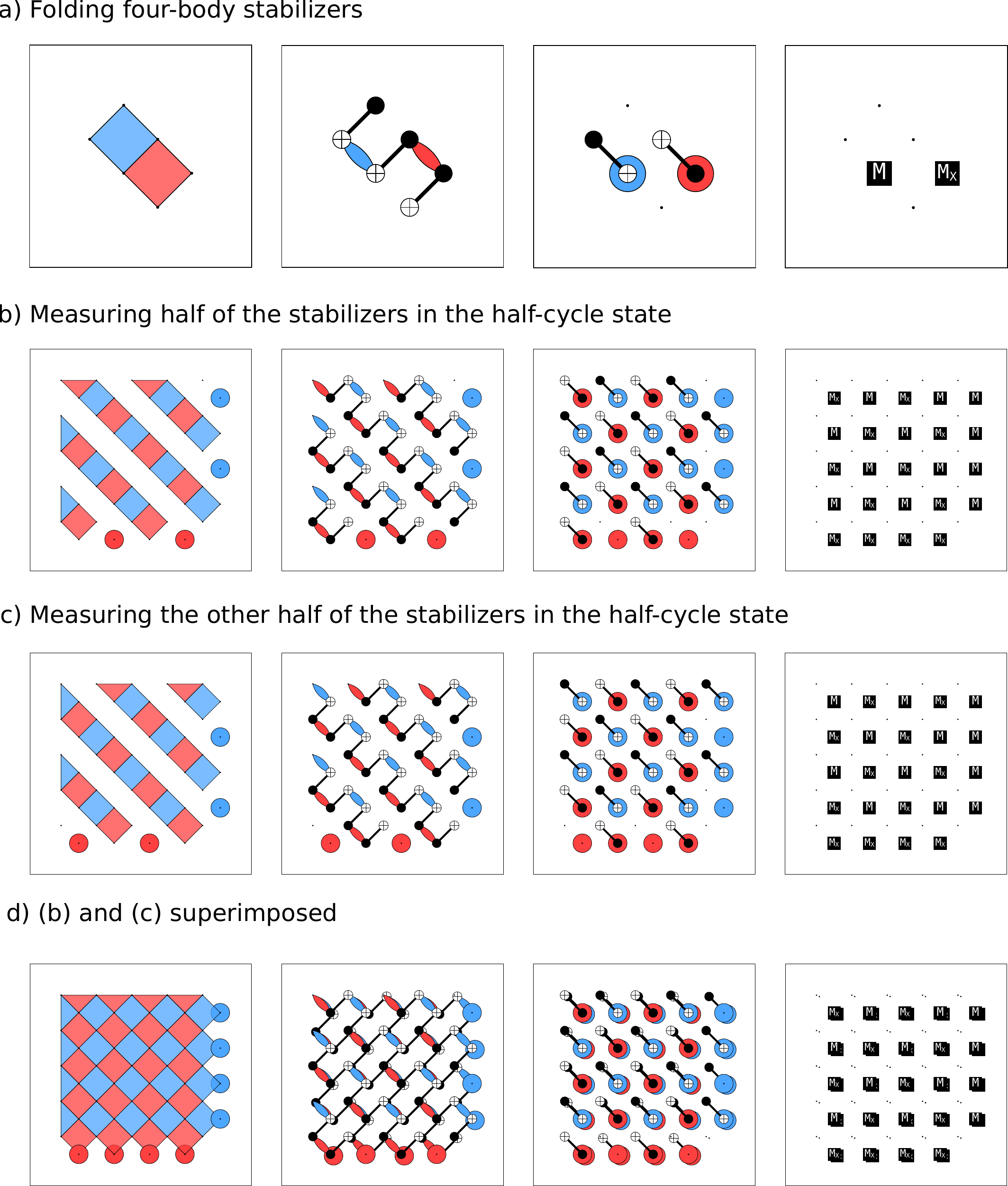}
    }
    \caption{
    \textbf{Half-cycle Picture for the Hex-Grid Circuit.}
    a) A schematic indication of how to measure neighbouring square four-body stabilizers of different types using gates on only three out of four edges. 
    b) An illustration of how folding can measure half of the stabilizers in the half-cycle state. These operations can be reversed to reconstruct the half-cycle state, but with expanding and contracting regions exchanged.
    c) The same strategy applied to the other half of the stabilizers in the half-cycle state. Again, these operations are reversed to reconstruct the half cycle state.
    d) Visually superimposes the two cycles (b) and (c), illustrating on the left that all stabilizers in a half-cycle state are measured by the two cycles, and that the resulting pattern of gates requires only a hex grid to be implemented.
    }
    \label{fig:hex_half_cycle_picture}
\end{figure}

These fold-based half-cycle circuits tile nicely.
It's possible to simultaneously measure entire diagonal columns of square stabilizers, alternating between X type and Z type, as shown in \fig[b]{hex_half_cycle_picture}. This lets us measure of half of the stabilizers in the half-cycle state, which is the goal of a surface code cycle circuit. 

Using this half-cycle and then its reverse differs from the standard circuit in two important ways: First, as mentioned already, it only uses three of the four available couplings in each square stabilizer to measure a column of stabilizers; and second, the cycle being reversed expands the stabilizer back to where it was contracted from, essentially exchanging which detecting regions in the half-cycle state are expanding and contracting. In the subsequent cycle, rather than executing the same cycle circuit, we must use a different cycle circuit aiming to measure the stabilizers corresponding to the contracting detecting regions, which are now in the columns adjacent to the ones we just measured. We can use the same strategy we have just presented to measure those columns as well; use two layers of CNOT gates to fold those stabilizers down to one-body stabilizers to be measured, and then reverse those layers to return top the half-cycle state with expanding and contracting regions exchanged. This is illustrated in \fig[c]{hex_half_cycle_picture}.

Taking stock, we have constructed two cycle circuits starting and ending at the half-cycle state. One measures the even numbered diagonal columns of alternating Z and X stabilizers, and the other measures the odd numbered diagonal columns. If we alternate these two cycles, we measure all the half-cycle stabilizers every two rounds. This means we return to the original half-cycle state with the same pattern of expanding and contracting regions every two rounds. 
This distinguishes the cycles here from the usual surface code circuit, which returns to the same half-cycle state in every round, including the locations of the expanding and contracting regions. 

These cycles must alternate: after one cycle, we return to a half-cycle pattern but with the expanding and contracting detecting regions switched. In the absence of noise, the state described by these two patterns is identical; they have the same stabilizers. However, returning to the same noise-free state obviously does not imply that we can simply repeat one of the two cycle circuits, never learning half the stabilizers. In the presence of noise, these two states are not the same; different stabilizers will have been measured more recently and so have less accumulated error. In the detecting region picture, this is made even more obvious: at the half-cycle state, half of the detecting regions are expanding and have so far existed for only around 1/4 of their overall lifetime, whereas the contracting regions have already lived for around 3/4 of their lifetime. The change in where the expanding and contracting regions are located at the half-cycle distinguishes the two half-cycle patterns and enforces an overall 2-round-periodicity in the circuit. 

The final important insight is that we can lay out the two cycle circuits in a compatible way, having them agree on which nearest neighbouring coupling to avoid using. This is shown schematically in \fig[d]{hex_half_cycle_picture}. Each of the two cycles only uses half of the couplings in the direction of the second folding, with the other half being used in both cycle circuits. The resulting connectivity actually used is therefore only a hex grid, rather than the usual square grid. 

\subsection{Top-down Stabilizer-Focused Construction}

\begin{figure}[ht]
    \centering
    \resizebox{\linewidth}{!}{
        \includegraphics{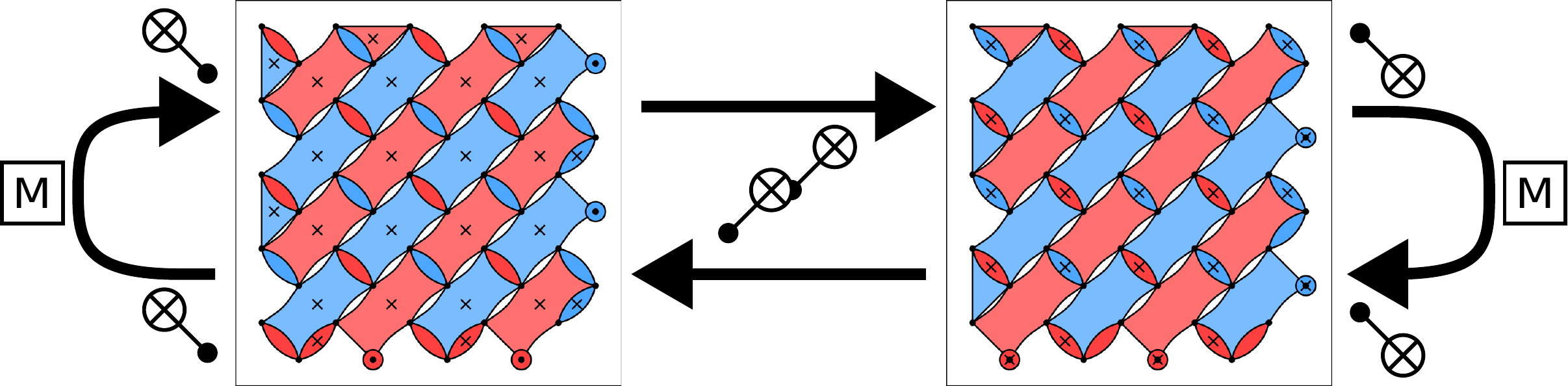}
    }
    \caption{
    \textbf{Brickwork-state Picture for the Hex-grid Circuit.}
    A schematic indicating how the hex-grid circuit measures the appropriate stabilizers.
    Starting on the left, the two body stabilizers can be measured simply using a common decomposition; applying a CNOT between the two qubits, measuring one of them, and applying the same CNOT again, as indicated by the u-turn-like arrow.
    Changing between the two brickwork patterns can be easily achieved by applying two layers of CNOTs along the direction perpendicular to the two body stabilizers, as indicated by the straight arrows (center). X markers indicate six-body stabilizers that are contracted into two-body stabilizers when moving to the right.
    These two-body stabilizers can then be measured, as indicated by the u-turn-like arrow on the right.
    }
    \label{fig:hex_brickwork_picture}
\end{figure}

Considering the brickwork patterns of the surface code (as shown in \fig{hex_brickwork_picture}), we see a pattern of two- and six-body stabilizers. 

We can non-destructively measure a two body X stabilizer or a two body Z stabilizer via a common circuit decomposition: performing a CNOT, then a single qubit X or Z basis measurement, then undoing the CNOT.
Since the two-body stabilizers in the brickwork states don't overlap, we can measure all these stabilizers in parallel. Both layers of CNOT gates operate along the two body stabilizers, and so do not connect the two middle qubits in any six-body stabilizer.

To construct a complete cycle, we also need to transition between the two brickwork states. As occurs in the standard surface code circuit, we can exchange the two- and six-body stabilizers using two layers of CNOTs, as indicated in \fig{hex_brickwork_picture}.
This switching is visually clear in the circuit using CNOT gates, where we can see visually that the locations of the two body stabilizers in the two brickwork patterns are the same, but their Pauli type has switched. Both layers of CNOT gates operate perpendicular to the two body stabilizers, and again do not connect the two middle qubits in any six-body stabilizer.

Putting this together, we can execute the surface code cycle on a hex grid. 
Starting in a brickwork state, we:
\begin{enumerate}
    \item Measure the two-body stabilizers using a CNOT layer, a measurement layer, and a CNOT layer
    \item Exchange the two-body and six-body stabilizer using two CNOT layers
    \item Measure the two-body stabilizers, again using a CNOT layer, measurement layer, and a CNOT layer
    \item Exchange stabilizers again, returning to the original brickwork state, using two CNOT layers
\end{enumerate}
This strategy uses two distinct cycles, and naturally embeds on a hex grid.

\subsection{Boundaries}

Having now addressed the bulk circuit, we turn to defining appropriate boundaries. We use the strategy described in \sec{boundaries} for temporal boundaries. The spatial boundaries can be most easily understood by looking at the half-cycle states. By comparison, the standard surface code half-cycle state (\fig{surface_code_cycle}) features alternating weight-3 and weight-4 stabilizers around all four boundaries. 

\begin{figure}[p]
    \centering
    \resizebox{\linewidth}{!}{
        \includegraphics{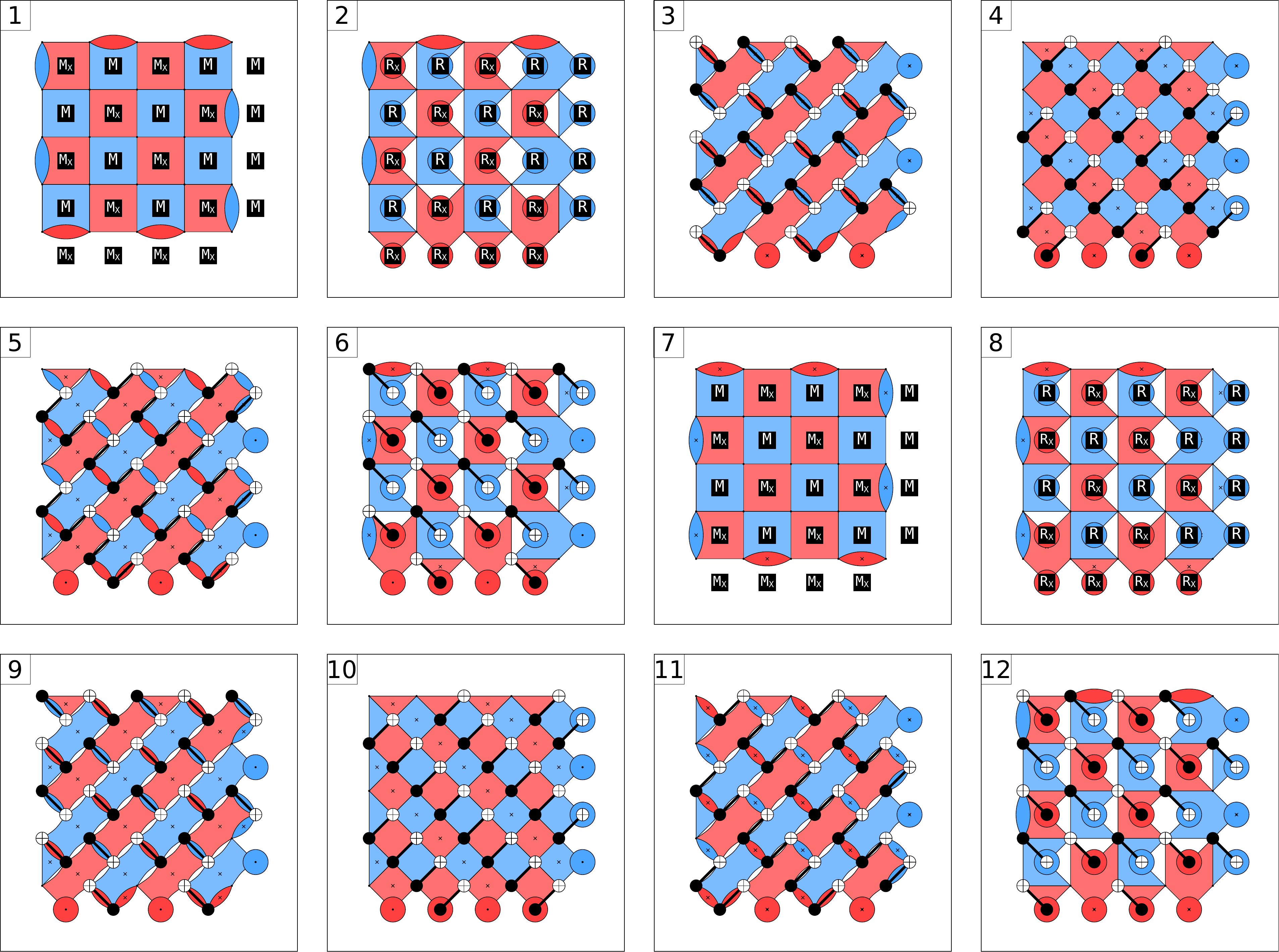}
    }
    \caption{
    \textbf{The Mid-cycle States in the Hex-grid Circuit.}
    All 12 layers of the hex-grid surface code circuit over both distinct cycles. Pictured behind each layer of gates are the detecting regions slices immediately after that layer. Detecting regions introduced in layer 2 are marked with an X. They are expanding from layer 2 until layer 7, where they cover the usual surface code stabilizers, and are subsequently contracting.
    }
    \label{fig:hex_boundaries}
\end{figure}

\fig{hex_boundaries} shows slices of the full detecting regions for a logical qubit patch executing the three-coupler surface code circuit, illustrating the shape of detecting regions around the boundaries.
The boundary construction can be most easily understood at the half-cycle state, for instance \fig[.4]{hex_boundaries}. Here, we see two \emph{flat} boundaries on the top and left where detecting regions are truncated to weight-3, and two \emph{spiky} boundaries featuring weight-1 detecting regions on the right and bottom. This contrasts with the standard surface code boundaries at the half-cycle shown in \fig{surface_code_cycle}, which has all four boundaries featuring alternating flat and spiky detecting regions. 
This gives us complete diagonal columns of detecting regions (also visualized in \fig{hex_half_cycle_picture}) to measure out; specifically, the alternating pattern on gates in each diagonal column terminate nicely with a weight-3 stabilizer at one end (the flat boundary) but use a weight-1 region at the other end (the spiky boundary). Which boundaries are spiky is therefore determined by the spatial direction for the gate layers adjacent to measurement. The full circuit is provided in our data repository~\cite{mcewen_data_2023}. The supplementary figures include a visualization of the full cycle circuit including only 4 detecting regions in the bulk (Supp. Fig. 5), which may aid the reader in tracking the evolution of one detecting region through the circuit. For this purpose, we also recommend opening the circuit in Crumble using the links provided in \app{crumble}.

We found appropriate boundary constructions using an iterative (or  brute-force) approach. Starting with the detecting region slices at the half-cycle state, we iteratively removed gates and manipulated the detecting region termination points while attempting to preserve the graph-like code distance~\cite{gidney_benchmarking_2022}. To restrict the search space, we constrained ourselves to exploring circuits without changing the number of gate layers, without breaking the regular patterns of the gate layers, and where possible preserving that measurement layers apply only to qubits on one of the two neighboring sub-grids (often called data and measure qubits in the standard circuit). To enable faster exploration of the impact of adding and removing gates on the detecting structure, we recommend the use of interactive tools such as Crumble (as detailed in \sec{software}).

\subsection{Benchmarking}\label{sec:hex_benchmarking}

Armed with the full description of the circuit, including boundaries, we now benchmark the circuit by numerically simulating a quantum memory experiment. The full details of our benchmarking strategy, including the noise model, are discussed in \app{noise_model}. In particular, we benchmark the circuit as compiled for superconducting hardware, using relative error rates representative of current experimental errors~\cite{gidney_benchmarking_2022, google_quantum_ai_suppressing_2022}. We compile to single qubit rotations, CZ gates and Z-type measurements and resets only, with the primary error parameter $p$ being equal to the CZ gate error rate. 

For brevity, we skip over intermediate benchmarking results and discuss only the \emph{teraquop footprint}. This is the number of physical qubits required to produce a logical qubit displaying a one-in-a-trillion logical error rate over a $d\times d\times d$ space-time block. 
This metric encapsulates the most important information regarding the performance of the code: First, the vertical asymptote where the footprint diverges vs. error rate indicates the threshold of the code, which is important for estimating near term performance while hardware error rates remains near threshold. Second, the relative value of the footprint at aspirational physical error rates (such as around $p=1\times10^{-3}$) indicates the sub-threshold scaling of the code. 
Given the very low logical error rate involved, the procedure for estimating the teraquop footprint involves linearly extrapolating the log logical error to higher code distances. The procedure used is included in our code repository. Further benchmarking plots for this code circuit are included in \app{benchmarking}, including the more traditional plot of logical error rate versus physical error rate from which teraquop footprint is extrapolated.

\begin{figure}[p]
    \centering
    \resizebox{\linewidth}{!}{
        \includegraphics{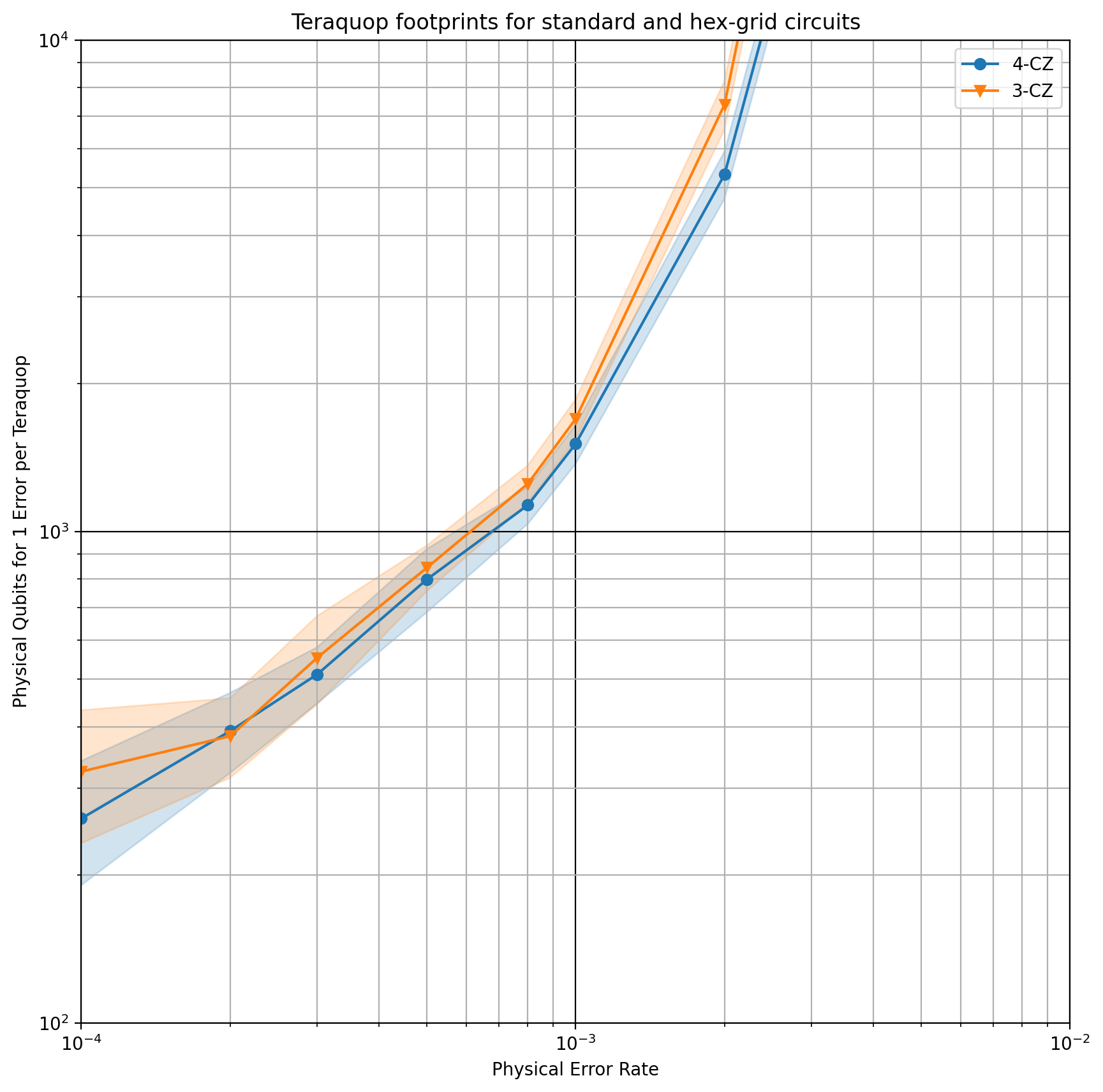}
    }
    \caption{
    \textbf{Teraquop footprint estimates for the hex-grid surface code circuit. }
    The number of physical qubits required for a single code patch to achieve a logical error rate of $1\times10^{-12}$ over a $d \times d \times d$ space-time block.
    Each curve combines memory experiments in both Z and X basis, and uses the \texttt{SI1000} noise model as described in \app{noise_model}. 
    \texttt{4-CZ} (blue circles) is the standard surface code circuit compiled to CZ gates, single qubit rotations, and Z-type measurements and resets.
    \texttt{3-CZ} (orange triangles) is the hex-grid circuit, using the same gateset.
    }
    \label{fig:hex_benchmarking}
\end{figure}

\subsection{Summary}

The hex-grid surface code circuit presents an appealing alternative to the traditional surface code circuit by reducing the connectivity required to implement the code at a minimal cost of final logical performance. 
Lower connectivity in hardware generally reduces costs in implementing an architecture, from relaxing frequency constraints to freeing up additional footprint on chip. Given the minimal change to the threshold, we expect near-term superconducting hardware tailored for the surface code can take advantage of the hex-grid circuit to improve physical error rates over those achievable using a square grid. However, the use of a hex grid also brings additional challenges; qubit and coupler yield is likely to be more costly in such an architecture, as we expect compatible subsystem codes to reduce the effective code distance more than in the square grid surface code case. In a similar vein, using the hex grid circuit in a square grid architecture with broken couplings provides additional freedom to avoid a reduction in code distance that imperfect yield would typically entail. 

The approach taken by this circuit makes clear that the surface code does embed naturally on a hex grid, which opens the way to other similar constructions.
In \app{benchmarking}, we additionally show benchmarking results for compiling the surface code to a heavy-hex grid~\cite{chamberland_topological_2020, sundaresan_matching_2022} and semi-heavy-hex grid, both of which further relaxes the connectivity requirements relative to a hex grid. The circuits themselves are included in the supplementary figures and in our data repository. We also provide a compilation to an architecture providing heterogeneous entangling operations: The hex-grid circuit naturally decomposes into two-qubit entangling measurements along the connections for measuring two-body stabilizers and two-qubit entangling gates in the orthogonal direction for exchanging brickwork states. Given progress in conducting hardware two-qubit parity measurements~\cite{lalumiere_tunable_2010, divincenzo_multi-qubit_2012, royer_qubit_2018, reagor_hardware_2022, livingston_experimental_2022}, this kinds of architecture may also provide a compelling alternative to the traditional square grid.

\section{ISWAP Surface Code Circuits}\label{sec:iswap}

Surface code are traditionally constructed using CNOT-like gates. 
CNOT gates provide the theoretically simplest circuit implementation, corresponding most closely with the original introduction of the surface code~\cite{kitaev_fault-tolerant_1997, bravyi_quantum_1998}, and usually referred to as the CSS surface code. 
These operations naturally assemble the stabilizer value to be measured onto the measure qubit, producing circuit implementations that are easy to understand and manipulate. 

Here, we show a circuit for the surface code using ISWAP gates. This circuit has the same code distance and the same number of entangling layers as the usual circuit. This permits the surface code to be operated without overhead on architectures that provide a native ISWAP-like gate, rather than demanding a CNOT-like native interaction. 

First, we address the background of the CNOT-like and ISWAP-like gate classes in more detail. We then introduce the bulk circuit from the perspective of the half-cycle state, describe a choice of patch boundaries that preserves the graph-like code distance, and numerically benchmark the resulting code as a memory experiment. 

\subsection{Entangling Gates in Error Correction Circuits}

Up to local single qubit Clifford gates, there are only four classes of Clifford two-qubit gates: Identity-like, CNOT-like, SWAP-like, ISWAP-like.
This equivalence is made clear by the KAK decomposition of these gates, as discussed in detail in \app{kak}. 
The Identity-like and SWAP-like gates are not entangling and cannot be used to construct a QEC circuit alone.
CNOT-like gates are the traditional gates used to construct QEC circuits. 
Using the remaining gate class, the ISWAP-like gates, is addressed in this construction.

ISWAP-like gates can be understood as the product of a SWAP-like and a CNOT-like gate. 
Gates with ISWAP-like behaviour move the quantum information around in addition to producing the necessary entanglement, making it non-obvious how to assemble the requisite stabilizer information for measurement.

Different hardware implementations generally provide different native entangling gates, but generally target a CNOT-like gate when aiming to perform error correction. In particular, superconducting qubit arrays traditionally target either a native CNOT such as via a cross-resonance gate~\cite{paraoanu_microwave-induced_2006, rigetti_fully_2010}, or a native CZ via a fixed or tunable capacitive coupling~\cite{yan_tunable_2018, foxen_demonstrating_2020}. As shown in \fig{surface_code_cycle_cz}, the surface code can be trivially adapted to use CZ gates, producing ZXXZ/XZZX stabilizers at measurement after canceling extraneous single qubit gates~\cite{wen_quantum_2003, bonilla_ataides_xzzx_2021, google_quantum_ai_suppressing_2022}. Such superconducting hardware can also naturally produce ISWAP-like gate~\cite{arute_quantum_2019}, but these have not been used in demonstrations of error correction due to a lack of appropriate circuit decomposition.
Here we present a circuit for the surface code using ISWAP gates. 


\subsection{Circuit Construction}

For simplicity, we first consider the construction using the CXSWAP gate rather than ISWAP. The CXSWAP gate is equal in action to a CX gate followed by a SWAP gate. It has the same KAK decomposition coefficients as the ISWAP gate, but has the conceptual advantage that its stabilizer flows each involve only Z or X terms and do not exchange them. Like circuits expressed in CNOT gates, the resulting detecting regions have only one Pauli type throughout, simplifying visualization. The circuit expressed in ISWAP gates differs only by single qubit gates, as described in \app{kak}. 

The complexity in this circuit construction arises because of the swapping behaviour of the entangling interaction. In the standard picture, assembling the stabilizer information to be measured is complicated by the movement of the relevant qubit states around the grid as the entangling layers are executed. By looking at detecting regions as a whole rather than focusing on moving qubit states, we can identify patterns of detectors that tile nicely and reproduce the relevant properties of the surface code. 

\begin{figure}[htb]
    \centering
    \resizebox{\linewidth}{!}{
        \includegraphics{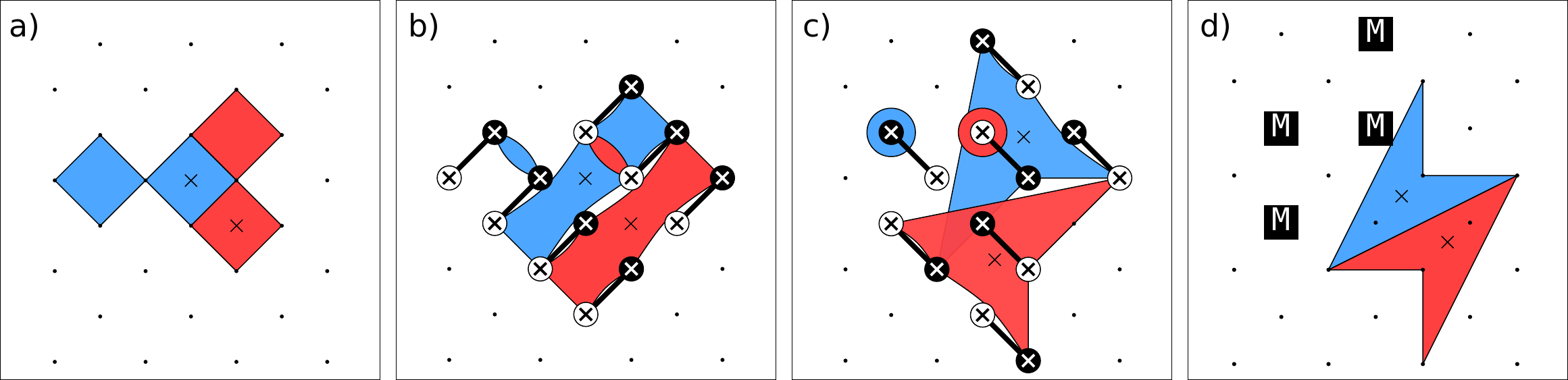}
    }
    \caption{
    \textbf{Half-cycle picture for the CXSWAP circuit. }
    CXSWAP preserves detecting region type and is equivalent to ISWAP up to single qubit gates. The layer of CXSWAP gates immediately before each detecting region slice is shown. If the CXSWAP is decomposed into a CX followed by a SWAP, the black circle indicates the control qubit of the CX.
    a) Four detecting regions in half cycle state are shown, two of X and Z type, and two expanding (marked by an X) and two contracting (unmarked). 
    b) The modified brickwork state using CXSWAP as the entangling layer.
    c) The modified state immediately prior to measurement, having contracted the two body stabilizers using CXSWAPs.
    d) The end-cycle state immediately after measurement, showing dart-shaped four-body stabilizers.
    }
    \label{fig:cxswap_states}
\end{figure}

Again, we start by considering the half-cycle state, as shown in \fig[a]{cxswap_states}. 
Rather than the next layer of entangling gates constructing a standard brickwork state, the swapping behaviours elongates the six body stabilizers. \fig[b]{cxswap_states} shows four of the detecting regions in this modified brickwork state; showing all the detecting region slices is visually complicated because they now overlap when drawn as detecting region slices. The next layer of entangling gates aims to contract the two body stabilizers to one body for subsequent measurement, as shown in \fig[c]{cxswap_states}, and produces an even more complex five-body stabilizer out of the previously extended six-body stabilizers. The subsequent measurement of the one-body stabilizer terms produces a relatively nice pattern of dart shaped four-body stabilizers. While this looks quite different to the normal surface code state at measurement, it displays the same overlapping structure of the stabilizers in the bulk, illustrating that errors on the unmeasured qubits will produce the same pattern of detection events as the standard code. We can then reverse these circuit steps (exchanging measurements for reset gates) to return to the half-cycle state. This cycle has the effect of contracting half of the detecting regions in that state and replacing them with new regions that expanded from the reset gates. The other half of the detecting regions expanded to cover the dart-shaped four-body stabilizers after measurement, and then returned to the half-cycle state.
Similar to the hex-grid circuit, we use an analogous cycle to then measure those detecting regions, with the newly expanded detecting regions now expanding to cover the dart-shaped stabilizers at measurement. By alternating these two cycles, we implement the surface code. 

\begin{figure}[ht]
    \centering
    \resizebox{\linewidth}{!}{
        \includegraphics{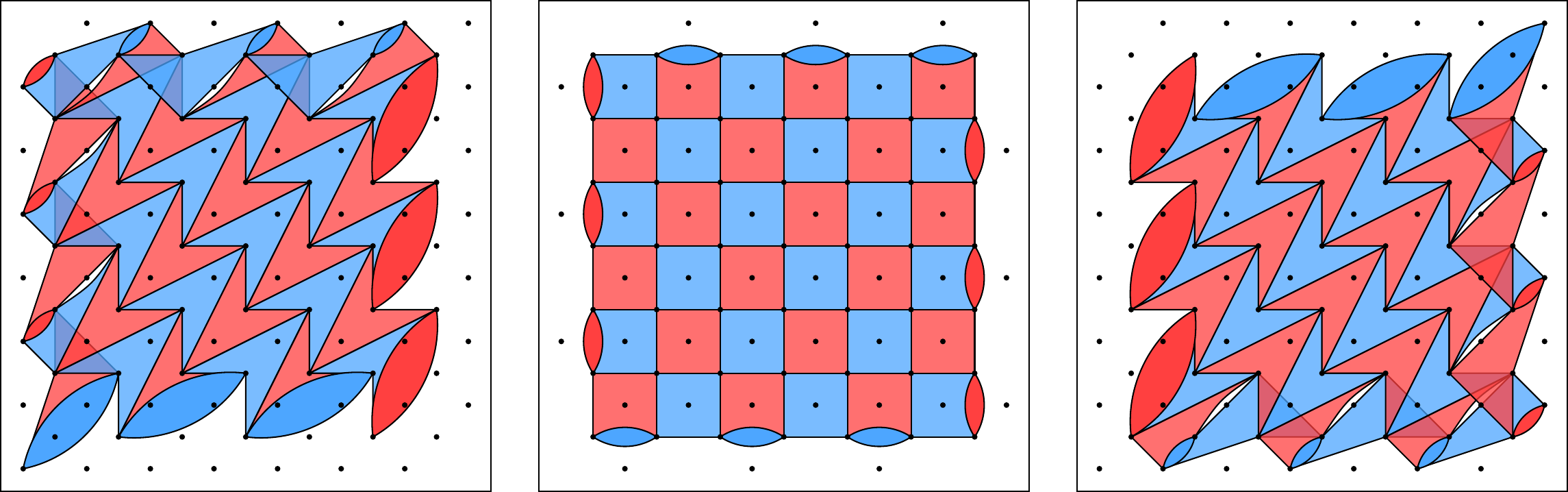}
    }
    \caption{
    \textbf{Patch distortion in the CXSWAP/ISWAP circuit. }
        Center: The end-cycle state for the standard surface code circuit. Left, Right: The end-cycle states for the CXSWAP/ISWAP circuit, showing data qubits shifted in alternating directions along diagonal lines in the bulk. Corresponding measure qubits are shifted in the opposite direction to data qubits. Boundaries feature more complex distortions.
        The CXSWAP/ISWAP circuit alternates between two cycles, reaching both end-cycle states shown.
    }
    \label{fig:cxswap_distortion}
\end{figure}

The pattern of stabilizers in the ISWAP circuit can also be understood as a distortion of the standard surface code patch. \fig{cxswap_distortion} shows this distortion visually; each diagonal line of data qubits is shifted in alternating directions. Along the same lines, measure qubits are distorted in the opposite direction, as they are exchanged with data qubits by the action of the ISWAP gates. 

\subsection{Boundaries}

\begin{figure}[ht]
    \centering
    \resizebox{\linewidth}{!}{
        \includegraphics{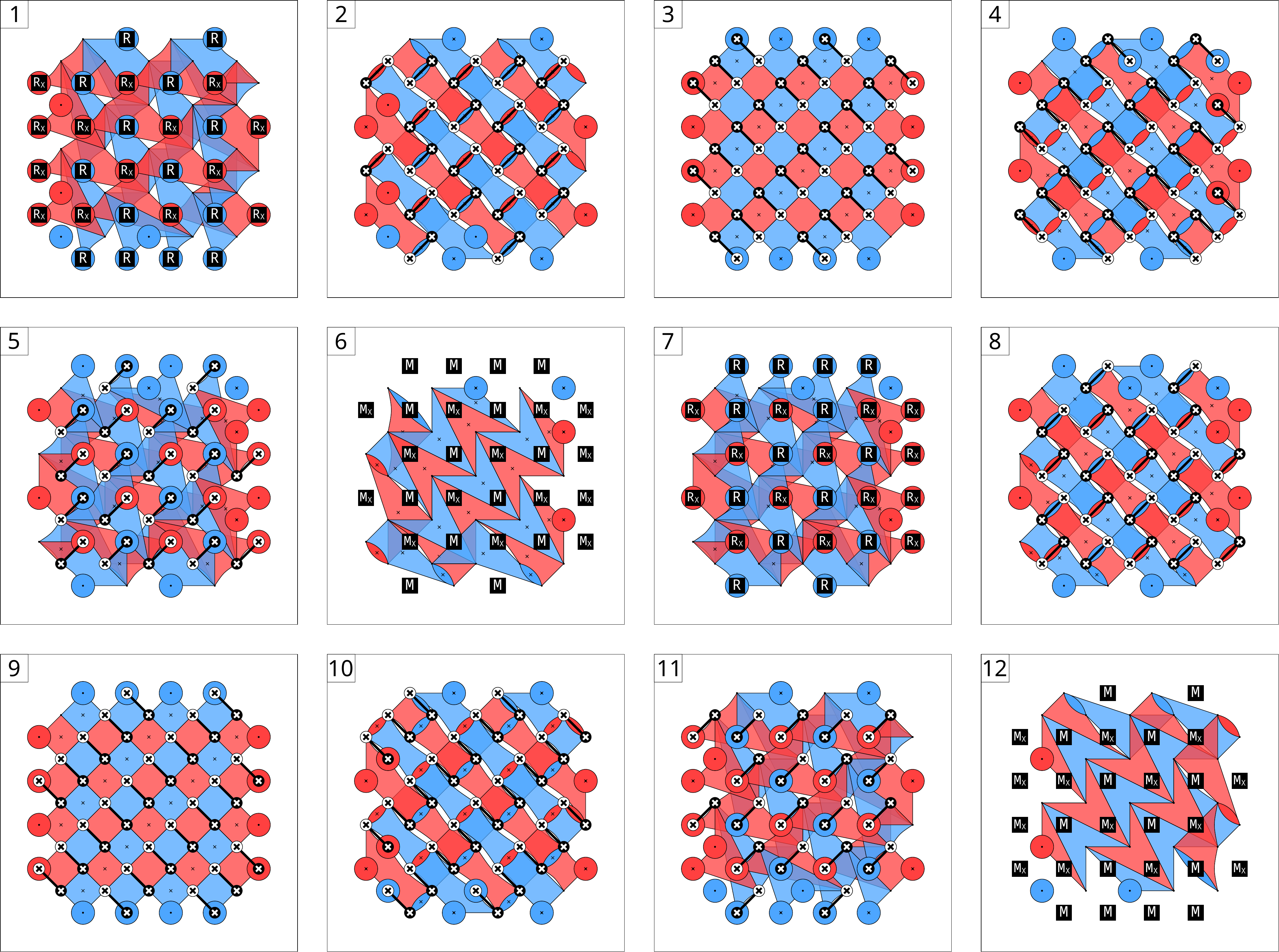}
    }
    \caption{
    \textbf{Both cycles for the CXSWAP circuit. }
    Detecting region slices are shown after each layer of gates in the cycle. Regions introduced by the resets in panel 1 are marked with a black X.
    Detecting regions overlap in the bulk, as shown more simply in \fig{cxswap_states}, but tile nicely at the mid-cycle (3 and 9) and end-cycle  (6 and 12) states. 
    Red regions are X stabilizers, blue regions are Z stabilizers.
    }
    \label{fig:cxswap_boundaries}
\end{figure}

As with other constructions, the boundaries prove more complex than the bulk circuit. At the measurement layer, they correspond to the standard rotated surface code boundaries modified by the same distortion applied to stabilizers shown in \fig{cxswap_distortion}. The full circuit schedule for the CXSWAP circuit is shown in \fig{cxswap_boundaries}. While all of the overlapping detecting regions in the bulk are visually complex, focusing on the detecting regions around the outer edges of the patch help to understand the boundaries. In particular, the mid-cycle states display helpful symmetry and a simple boundary shape consisting of terminating weight-1 detecting regions. Our process for finding the boundary construction was similar to that of the hex-grid circuit; starting with a guess at good detecting slices at the mid-cycle, we iteratively removed gates and manipulated the detecting region terminations while preserving the graph-like code distance, eventually arriving at the provided construct.

Slightly different boundaries are used in the ISWAP circuit we benchmarked, as included in the supplementary figures; two additional qubits were left in at the corners and measurements gates were not preferentially arranged to occur on the same qubit sub-grid (i.e. not on any neighbouring qubits simultaneously). Given both the CXSWAP and ISWAP circuits benchmark with essentially the same performance as their standard circuit counterpart, we believe the differences in boundaries does not significantly impact performance. That said, the boundary shown for the CXSWAP circuit is slightly simpler to understand visually and would likely perform better where hardware imposes costs on simultaneously measuring neighbouring qubits. 

\subsection{Benchmarking}

We now numerically benchmark the circuit by simulating a quantum memory experiment using both the standard and ISWAP circuits.
The full details of our benchmarking strategy, including the noise model, are discussed in \app{noise_model}. In particular, we set the primary error component $p$ in both cases to be the CZ or ISWAP gate for the standard and ISWAP circuits respectively. 

\begin{figure}[p]
    \centering
    \resizebox{\linewidth}{!}{
        \includegraphics{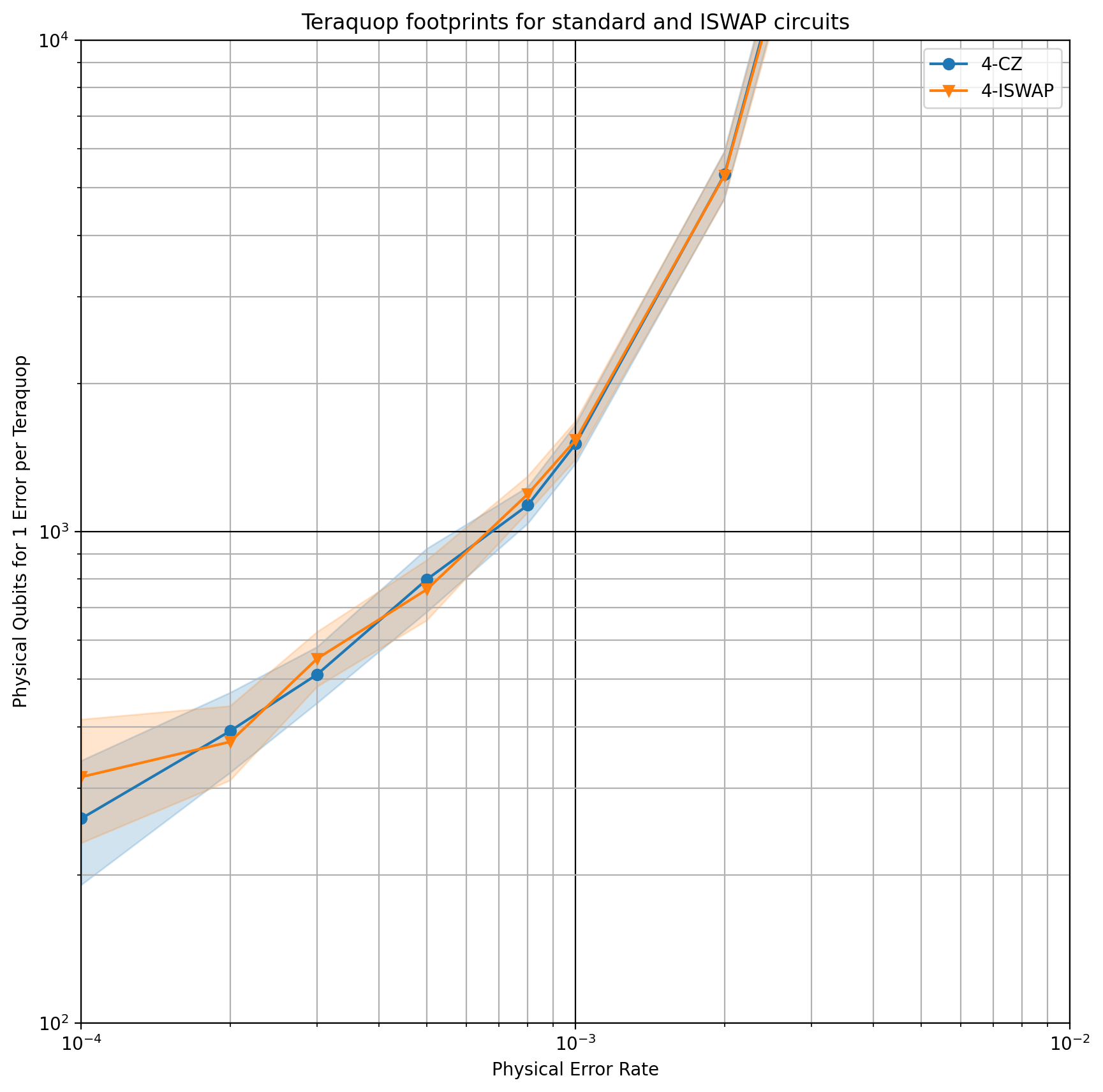}
    }
    \caption{
    \textbf{Teraquop footprints for the ISWAP surface code circuit. }
    The number of physical qubits required for a single code patch to achieve a logical error rate of $1\times10^{-12}$ over a $d \times d \times d$ space-time block. 
    Each curve combines memory experiments in both Z and X basis, and uses the \texttt{SI1000} noise model as described in \app{noise_model}. 
    \texttt{4-CZ} (blue circles) is the standard surface code circuit compiled to CZ gates, single qubit rotations, and Z-type measurements and resets.
    \texttt{4-ISWAP} (orange triangles) is the ISWAP circuit, using ISWAP gates, single qubit rotations, and Z-type measurements and resets.
    }
    \label{fig:iswap_benchmarking}
\end{figure}

We find that the teraquop footprint for the ISWAP circuit is essentially identical to the standard circuit. As before, such a qualitative statement is sensitive to the details of the error model. However, we expect that this circuit will be of interest for experimental implementations under the assumption that the ISWAP gate may provide a path to lower error rates than a CNOT or CZ gate, or provide other benefits not reflected in our simple error model. 

\subsection{Summary}

The ISWAP circuit presents a compelling alternative circuit to target for hardware architecture with a natural ISWAP gate, alleviating the need to target a CNOT-like entangling gate for implementing the surface code. This construction completes the possible implementations of the surface code in terms of the 2Q CLifford KAK gate classes. 

\section{Walking Surface Code Circuits}\label{sec:walking}

Surface code circuits typically aim to provide a method of measuring static stabilizers, leaving any logical time dynamics to a higher level of code manipulation. 
Compiling logical algorithms in the surface code, such as by lattice surgery~\cite{horsman_surface_2012}, typically manipulates the underlying circuit only subtractively, that is by deciding which stabilizers to simply not measure. 
In this view, the roles of the physical qubits in the circuit are fixed, and the only variable is whether the operations dictated by the template circuit are performed or not at any given location in space and time. 
With the freedom provided by additional detecting region shapes, it is possible to break the fixed allocation of qubit roles to physical qubits, and thereby break the fixed location of code stabilizers on the underlying qubit grid. 

Here, we present various circuits for walking a surface code patch, moving it on the underlying qubit grid without increasing overhead in circuit layers or reducing the code distance. This has the effect of exchanging the roles of data and measure qubits, which is a desirable primitive operation in a code circuit for the purposes of mitigating leakage~\cite{fowler_coping_2013, ghosh_leakage-resilient_2015, brown_leakage_2019}.

First, we discuss the relevance of qubit roles to the problem of correlated errors induced by leakage.
We then introduce the walking circuit and its detecting structure, describe the moving patch boundaries, and numerically benchmark the code circuit. Finally, we discuss the implications for leakage mitigation in hardware experiments. 

\subsection{Leakage Mitigation in QEC Experiments}

Most hardware implementations for quantum computing feature higher energy states nearby the chosen computational subspace, which can be erroneously populated in what is generally referred to as a \emph{leakage error}. Leakage errors are especially problematic for quantum error correction, as the states they produce are typically long lived and not accounted for in the physical interactions used to implement gates~\cite{fowler_coping_2013}. 
Leakage errors generally induce a large number of equivalent uncorrelated Pauli errors, providing an out-sized contribution for the error budget in a quantum error correction experiment~\cite{google_quantum_ai_suppressing_2022}. 

Various strategies are typically used to reduce the prevalence of leakage errors in gate implementations~\cite{motzoi_simple_2009, chen_measuring_2016, foxen_demonstrating_2020}, and in removing or suppressing the correlated effect of leakage errors when they do occur in the typical cycle circuit.
Leakage is generally easier to remove from measure qubits, as immediately after measurement they are not holding any important quantum information and can be unconditionally reset~\cite{magnard_fast_2018, battistel_hardware-efficient_2021, mcewen_removing_2021, zhou_rapid_2021, }. 

Removing leakage from data qubits presents a more difficult challenge. 
Data qubit leakage can be removed directly by the use of gate operations that specifically affect the leakage state without disturbing the computational states~\cite{miao_overcoming_2022}, but such additional operations introduce new error sources to the error budget. 

Alternatively, the roles of measure and data qubits can be regularly exchanged by the addition of SWAP or CNOT-like operations~\cite{fowler_coping_2013, ghosh_leakage-resilient_2015, brown_leakage_2019}. Assuming that entangling interactions do not move leakage between qubits, this permits leakage removal on each physical qubit after it is measured in every second cycle, curtailing the possible spread of leakage in time.
However, this strategy also introduces additional operations necessary to exchange the qubit roles, impacting the final error budget.
The circuit discussed here provides the ability to exchange qubit roles without adding any additional gates. 
While this strategy does not resolve leakage errors entirely, it provides a compelling alternative to known strategies for exchanging roles and could compliment direct removal strategies by alleviating some of the burden of leakage errors. 

\subsection{Circuit Construction}

\begin{figure}[p]
    \centering
    \resizebox{0.8\linewidth}{!}{
        \includegraphics{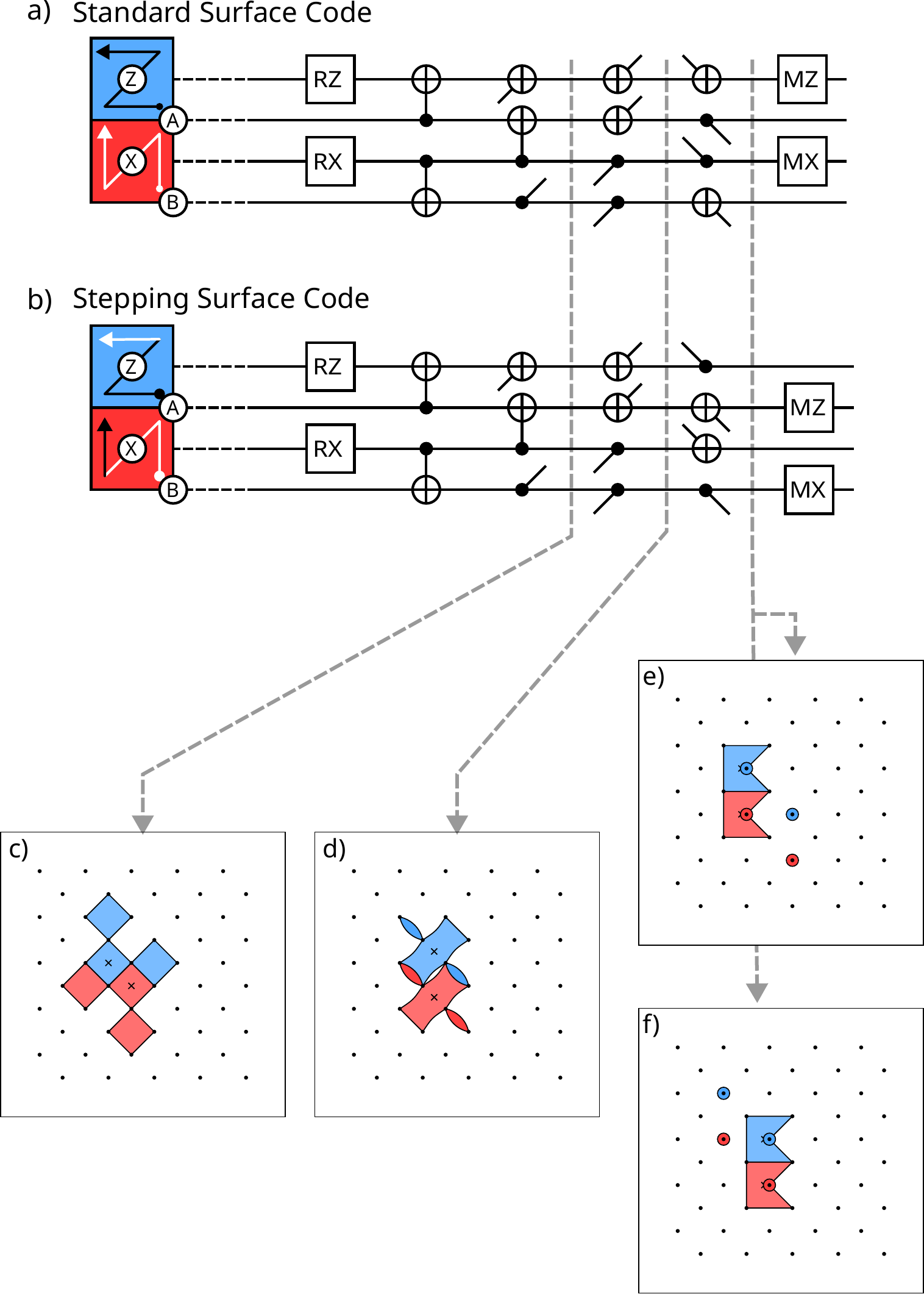}
    }
    \caption{
    \textbf{Comparing the circuit and mid-cycle states for standard and stepping circuits.}
    a) The standard circuit for the surface code on two measure qubits (Z, X) and two data qubits (A,B). The measure qubits are reset, assemble stabilizer information from their four neighbouring measure qubits in the order indicated by the Z-shaped arrows, and are measured.
    b) The stepping circuit, which is identical to the standard circuit until the last layer of CNOT gates.
    This layer is instead the same as the first layer of CNOT gates with target and control reversed. At the measurement layer, the roles of qubits have been exchanged, with qubits that began the round as data qubits (A,B) now being measured.
    c) The half-cycle state for both circuits, showing six detecting regions. Two expanding regions are marked with an X. The expanding regions have a choice of whether to pair up with the upper-left or lower-right contracting regions of the same Pauli type.
    d) The second brickwork state for both circuits. The choice of pairing for the expanding region is now aligned along the diagonal axis. 
    e) The flag-like state immediately before measurement for the standard circuit, corresponding to pairing the expanding region with the upper-left contracting region.
    f) The flag-like state for the stepping circuit, corresponding to pairing the expanding region with the lower-right contracting region.
    }
    \label{fig:walking_comparison}
\end{figure}

\fig{walking_comparison} shows the standard circuit alongside the \emph{stepping} circuit that exchanges measure and data qubits.
The initial insight for constructing this circuit can be found in the half cycle state, illustrated in \fig[c]{walking_comparison}. In the standard surfaces code circuit, pairs of detecting regions begin centered around the same measure qubit, one expanding one covering just the measure qubit and one contracting one covering four data qubits and the measure qubit, as illustrated previously in \fig{det_slices}. By the half-cycle, these have both been mapped to four-body stabilizers involving their shared measure qubit. 

The half-cycle state is highly symmetric, and this makes clear that maintaining this pairing and measuring out the contracting region on that measure qubit is only one possible option. \fig[c]{walking_comparison} shows a second possible pairing of expanding and contracting regions. Choosing to pair expanding and contracting regions around a different qubit choice has the effect that the contracting regions terminate on what was previously a data qubit, and the expanding regions cover stabilizers involving what were previously measure qubits. This exchanges the underlying qubit roles on the physical grid while preserving the logical structure of the code. 

In fact, this different pairing requires only a very simple modification; changing the directions of the CNOT gates in the last layer, which we refer to as the \emph{stepping trick}.
That this trick is sufficient can be seen from the symmetry of the weight-2 detecting region slices immediately before the last layer of entangling gates. 
The direction of the last layer of CNOTs determines which of the two qubits the detecting region will contract down to. 
This trick is relatively generic, and can be easily applied to other circuits. 
In this section, we focus on applying the stepping trick to the standard circuit and emphasise the resulting ability to move a logical patch in arbitrary directions on the underlying grid. 
The same trick can be easily applied to the hex circuit to exchange qubit roles, as we discuss further in \sec{conclusion}. 
The trick also applies to repetition codes, as we illustrate in \app{step_code}.

\subsection{Boundaries}

\begin{figure}[p]
    \centering
    \resizebox{\linewidth}{!}{
        \includegraphics{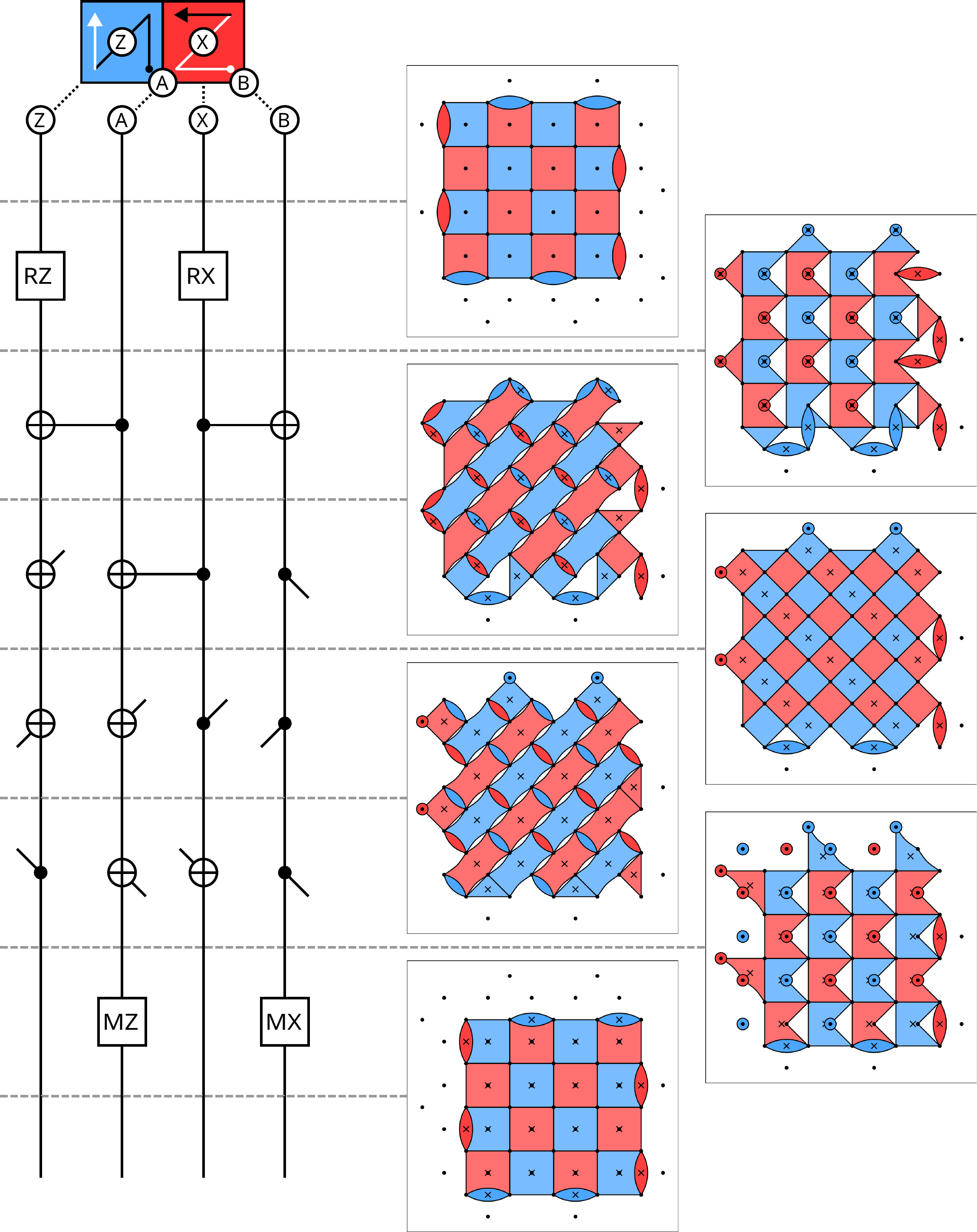}
    }
    \caption{
    \textbf{Mid-Cycle States in the Walking Circuit.}
    Left: The circuit for the walking surface code, shown on two measure qubits (X and Z) and two data qubits (A and B). Time proceeds down the page.
    Right: The detecting region slices at each layer of the circuit. Each shape represents a single stabilizer term consisting of either Z (blue) or X (red) Pauli elements. The corners of the shape indicate the qubits included in the term, with circles indicating single qubit terms.
    The expanding detecting regions introduced in this cycle are marked with an X.
    }
    \label{fig:walking_boundaries}
\end{figure}

The boundaries prove more complex to construct than the bulk circuit. The walking circuit boundary must be chosen carefully to appropriately create new detecting regions on the \emph{leading edges}, the two spatial boundaries in the direction of movement. We must also have detecting regions appropriately measured out on the \emph{trailing edges}, the two spatial boundaries facing away from the direction of movement. In \fig{walking_boundaries}, we provide a choice of boundaries that preserves the graph-like code distance,  prioritizing not introducing any problematic hook errors. 

One important detail we do not address visually here is the behaviour of the logical observable. As the code patch moves, the logical observable can be moved along with it by the inclusion of measurements at the trailing boundary, as expressed in the benchmarked circuits included in our data and code repository~\cite{mcewen_data_2023}. 

We refer to the behaviour exhibited by one application of this circuit as taking a \emph{step} and a single cycle as a \emph{step cycle} to distinguish it from the standard cycle where the logical patch remains in place. The step cycle circuit can be trivially shifted in space to follow the patch, allowing further steps to be taken, and rotated in space to permit steps to be taken in the four cardinal directions relative to the physical qubit square grid. 

Interestingly, compared to the standard circuit, the walking cycle circuit and its time inverse are more distinct. In the standard circuit, running the circuit in reverse is equivalent to a 180-degree spatial rotation of the same circuit; in terms of only the represented stabilizers, the two flag-like patterns are identical and the two brickwork patterns are equivalent up to rotation. This is not the case in the walking cycle, where all the patterns are noticeably different. The time-reversed cycle circuit, with measurements and resets exchanged, provides another pattern of boundaries which results in the patch taking a step (in the opposite direction to the time-forward circuit), and can similarly be shifted and rotated to permit steps to be taken in any direction. We only use the forward-going cycle shown in \fig{walking_boundaries} in the benchmarked circuits.

\subsection{Behaviours for the Logical Patch}

\begin{figure}[ht]
    \centering
    \resizebox{\linewidth}{!}{
        \includegraphics{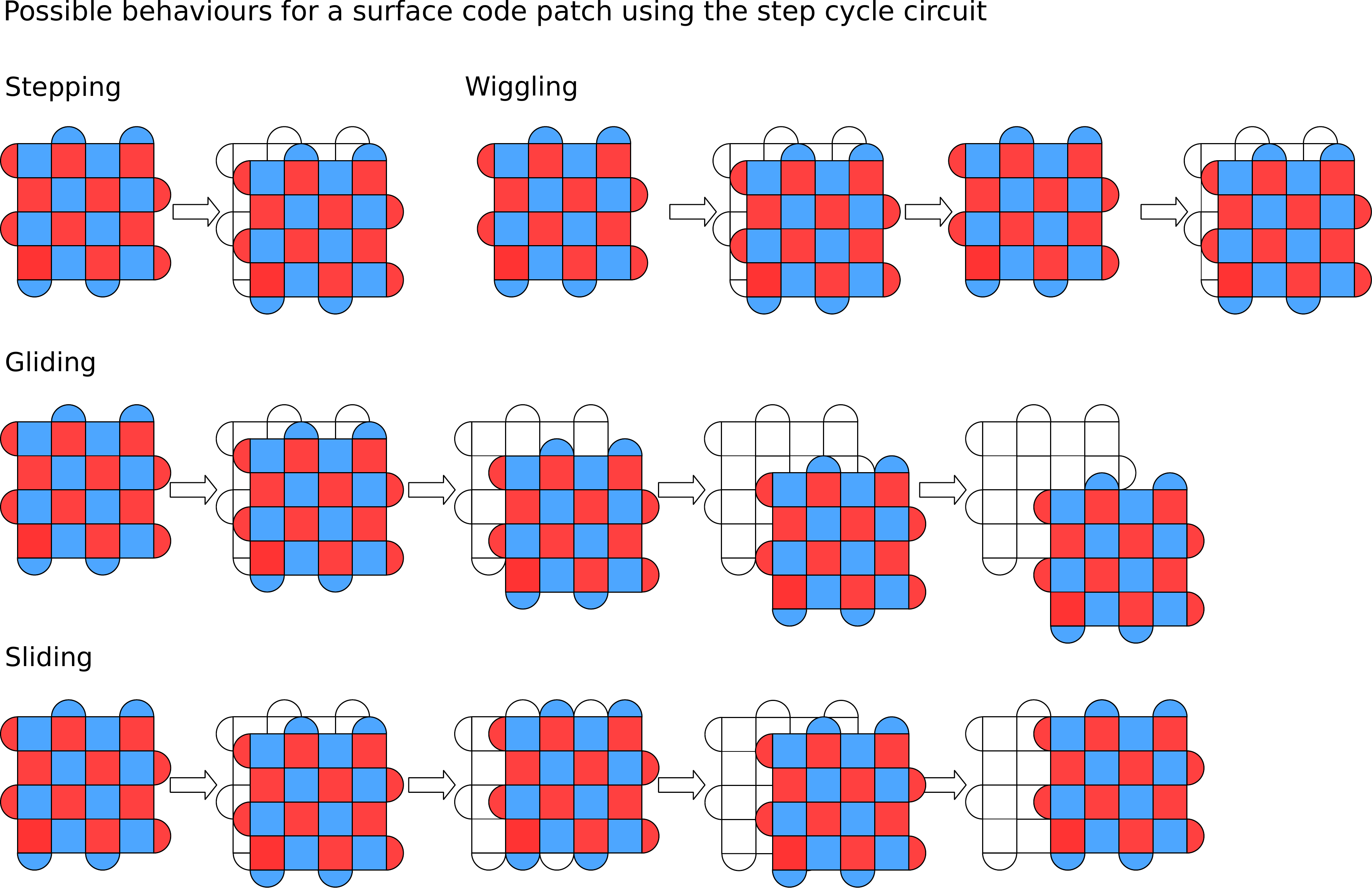}
    }
    \caption{
    \textbf{Behaviours for the logical patch using step cycle circuits.}
    Given the capability of taking a single step (Stepping), we can chain adjacent cycles of steps to produce additional behaviours. These include the patch stepping back and forth in place, returning to its starting position every second cycle (Wiggling), continuing to step in the same direction (Gliding) or taking alternating perpendicular steps to move the patch laterally (Sliding).
    }
    \label{fig:walking_behaviours}
\end{figure}

Given the capability of taking a step in any direction in each cycle, we can define additional behaviours for the logical qubit patch which can be helpful. \fig{walking_behaviours} shows four possible macroscopic behaviours which can be achieved using step cycle circuits, which we benchmark in the following section. The most basic behaviour, \emph{wiggling}, involves simply taking steps back and forth on the spot, returning to the same position every second cycle. This presents a minimal overhead in terms of number of qubits to achieve the swapping of qubit roles in the bulk. Continuing to take additional steps in the same direction continues to move the patch through the underlying qubit grid, which we call \emph{gliding}. Another specific behaviour more compatible with current layouts for logical algorithms is \emph{sliding}, where the patch takes alternating steps in perpendicular directions, allowing patches to move laterally. This could be especially convenient when considering routing problems, especially when all other qubits are engaging in wiggling behaviour. An application of this behaviour to aiding logical compilation is discussed in \app{sliding}.

\subsection{Benchmarking}

We now numerically benchmark the circuit by simulating a quantum memory experiment using stepping circuits in each of the three behaviours presented in \fig{walking_behaviours}.
As before, the full details of our benchmarking strategy, including the noise model, are discussed in \app{noise_model}. 

\begin{figure}[p]
    \centering
    \resizebox{\linewidth}{!}{
        \includegraphics{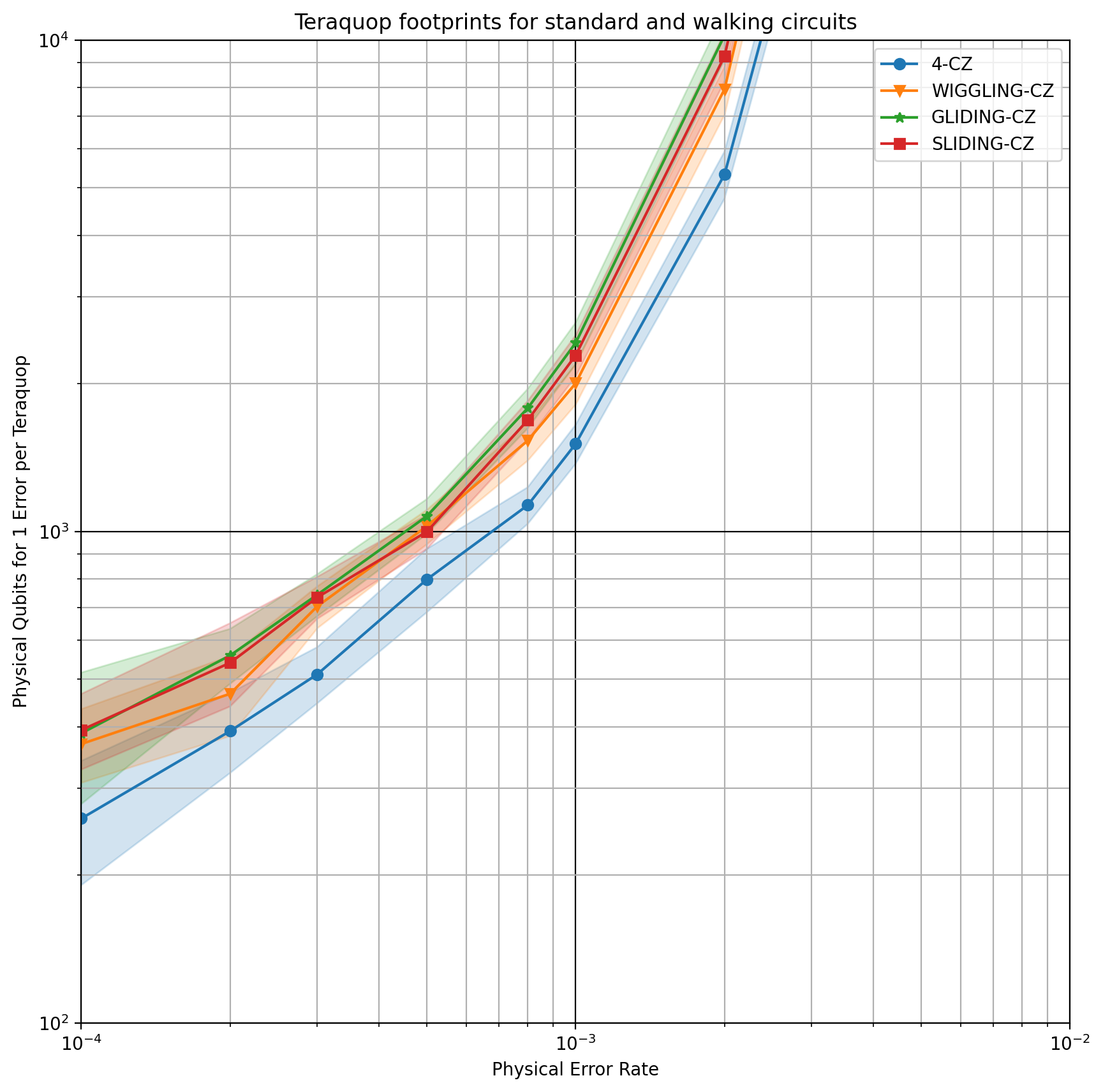}
    }
    \caption{
    \textbf{Teraquop footprints for the walking surface code circuits. }
    The number of physical qubits required for a single code patch to achieve a logical error rate of $1\times10^{-12}$ over a $d \times d \times d$ space-time block. 
    Each curve combines memory experiments in both Z and X basis, and uses the \texttt{SI1000} noise model as described in \app{noise_model}. 
    \texttt{4-CZ} (blue circles) is the standard surface code circuit compiled to CZ gates, single qubit rotations, and Z-type measurements and resets.
    \texttt{WIGGLING-CZ} (orange triangles), \texttt{GLIDING-CZ} (green stars), and \texttt{SLIDING-CZ} (red squares) are walking circuits for wiggling, gliding and sliding behaviours respectively.
    }
    \label{fig:walking_benchmarking}
\end{figure}

The teraquop footprints for all four behaviours are qualitatively very similar, with footprints between 1000 and 3000 qubits at an aspirational limiting error rate of $1\times10^{-3}$. The standard surface code displays slightly better performance than the walking codes, which we attribute to the additional qubits that the step cycle moves into as it steps, resulting in a larger number of qubits used for the same distance and error suppression. We also notice that the wiggling behaviour displays slightly better performance than the gliding and sliding behaviours, which we attribute to changes in the number of long error chains in the gliding and sliding behaviours through time. Overall though, the relative performance of these codes is subject to our assumed error model, and we would expect an error model that included leakage to affect the relative performance of these codes substantially. 

\subsection{Summary}

In summary, we have presented a circuit that achieves the exchange of roles of measure and data qubits without additional gate overhead. Additionally, this circuit moves the logical patch on the underlying physical qubit grid, opening new possible behaviours for the logical patch. 
In the absence of leakage errors, the circuit performs comparably to the traditional surface code circuit. We expect when considering leakage errors that this circuit will display additional advantages, given the exchange of roles coupled with existing progress on removing leakage from measure qubits. Simulating this circuit in the presence of leakage and quantifying the advantage is important future work. We expect that this technique will compliment current work on direct removal of leakage on data qubits, rather than solve the presence of leakage outright. The low additional overhead to implement this technique in terms of error rates makes it a compelling addition to practical error correction experiments. In addition to the relevance of walking to leakage removal, it also has implications for higher-level logical compilation, which we discuss in \app{sliding}.

\section{Conclusion and Outlook}\label{sec:conclusion}

The three circuit constructions we have presented in detail represent three applications of this approach to QEC circuit construction.
In this section, we address combining these connectivity, gate and behaviour benefits, address some additional related constructions, and provide some outlook for these techniques to be applied more broadly.

\subsection{Further Constructions}

The presented circuits provide distinct benefits from the view of simplifying the required hardware, but these  are not mutually exclusive. We can combine the constructions above to generate circuits for each combination of benefits, including a circuit that embeds on a hex grid, uses ISWAP gates, and exchanges data and measure qubits between neighbouring cycles. While we do not address these combined circuits in the same level of detail, we do provide the exact circuit and benchmarking results.

\begin{table}[ht]
    \centering
    \resizebox{\linewidth}{!}{
    \begin{tabular}{|r|l|l|l|l|}
    \hline
    & &&  & \textbf{Qubits} \\
    \textbf{Circuit Type} & \textbf{Circuit Label} & \textbf{Grid} & \textbf{Entangling Gate} & \textbf{Exchange} \\
     & & & & \textbf{Roles} \\
    \hline
    
\hline
    \textbf{Standard} & \texttt{4-CX} & Square & CX & No \\
    \hline
    & \texttt{4-CZ} & Square & CZ & No \\
    \hline
    \hline
    \textbf{Hex-grid} (\sec{hex-grid}) & \texttt{3-CX} & Hex & CX & No \\
    \hline
    & \texttt{3-CZ} & Hex & CZ & No \\
    \hline
    & \texttt{3-CX-wiggle} & Hex & CX & Yes \\
    \hline
    & \texttt{3-CZ-wiggle} & Hex & CZ & Yes \\
    \hline
    \hline
    \textbf{ISWAP} (\sec{iswap}) & \texttt{4-CXSWAP} & Square & CXSWAP & No \\
    \hline
    & \texttt{4-ISWAP} & Square & ISWAP & No \\
    \hline
    & \texttt{3-CXSWAP} & Hex & CXSWAP & No \\
    \hline
    & \texttt{3-ISWAP} & Hex & ISWAP & No \\
    \hline
    & \texttt{3-CXSWAP-wiggle} & Hex & CXSWAP & Yes \\
    \hline
    & \texttt{3-ISWAP-wiggle} & Hex & ISWAP & Yes \\
    \hline
    \hline
    \textbf{Walking} (\sec{walking}) & \multicolumn{4}{c|}{Capable of arbitrary movement direction} \\
    \hline
    Wiggling & \texttt{WIGGLING-CX} & Square & CX & Yes \\
    \hline
    & \texttt{WIGGLING-CZ} & Square & CZ & Yes \\
    \hline
    Gliding & \texttt{GLIDING-CX} & Square & CX & Yes \\
    \hline
    & \texttt{GLIDING-CZ} & Square & CZ & Yes \\
    \hline
    Sliding & \texttt{SLIDING-CX} & Square & CX & Yes \\
    \hline
    & \texttt{SLIDING-CZ} & Square & CZ & Yes \\
    \hline
    \hline
    \textbf{Toric} & \multicolumn{4}{c|}{Presented without boundary constructions} \\
    \hline
    CSS Toric Code & \texttt{TORIC-4-CX} & Square & CX & No \\
    \hline
    Heavy-hex & \texttt{TORIC-3\_HEAVY-CX} & Heavy-hex & CX & No \\
    \hline
    Semi-heavy-hex & \texttt{TORIC-3\_SEMI\_HEAVY-CX} & Semi-heavy-hex& CX & No \\
    \hline
    \hline
    \textbf{Parity Assisted} & \multicolumn{4}{c|}{$1/3$ of edges do parity measurements instead of unitary interactions} \\
    \hline
    & \texttt{TORIC-3-CX\_MXX\_MZZ} & Modified Hex & CX, MXX, MZZ & N/A \\
    \hline
    & \texttt{3-CX\_MXX\_MZZ} & Modified Hex & CX, MXX, MZZ & N/A \\
    \hline
    & \texttt{3-CZ\_MZZ} & Modified Hex & CZ, MZZ & N/A \\
    \hline
\end{tabular}

    }
    \caption{\textbf{All circuit constructions.}
        An exhaustive list of circuit constructions included in this work.
        Benchmarking for all circuits is included in \app{benchmarking}.
    }
    \label{tab:big_table}
\end{table}

\tab{big_table} presents each of the constructions that we have benchmarked. We have included summary benchmarking for each circuit in \app{benchmarking}, and detailed benchmarking as supplementary figures available as an ancillary file and in our data repository~\cite{mcewen_data_2023}. 

In addition to the circuits that we presented in detail and their combinations, we include some additional related circuits which we briefly discuss here. We present these additional circuits in the toric case without boundary constructions, but hope that finding appropriate boundaries will not prove especially difficult in light of the examples and methods discussed for the main constructions.

While we explicitly addressed embedding on a hex grid, cutting edge hardware architectures have targeted the related \emph{heavy-hex} grid, where an additional qubit is placed on each edge of the hex grid, reducing the average connectivity below three.
Using similar techniques to the hex grid circuits, we built a circuit for the surface code on the heavy-hex grid.
This circuit uses six layers of entangling gates per measurement cycle.
(The number of layers of entangling gates can be reduced from six to four by repurposing flag qubits for teleportation, but we assumed this would make performance worse and so did not benchmark that variant of the circuit for this paper.)
Our heavy-hex circuit outperforms the heavy-hex code~\cite{chamberland_topological_2020, sundaresan_matching_2022}, and also the heavy-square code~\cite{chamberland_topological_2020}.

We also include a circuit for the \emph{semi-heavy-hex} grid, where additional qubits are included on only a third of the hex-grid edges, which we use as measure qubits. This grid represents a middle ground between hex and heavy-hex grids, where the average connectivity is still lower than three, and qubits requiring measurement during the cycle have lower connectivity than the remaining qubits. This circuit requires only three layers of entangling gates per measurement cycle, taking advantage of the additional qubits to increase parallelization when compared to the hex-grid circuit.

Finally, we present additional circuits considering more unusual underlying qubit hardware, specifically hybrid circuits using a hardware native parity measurement on a third of the edges in a hex grid. This represents a middle ground between an architecture targeting the hex-grid circuit and one targeting only pair measurements~\cite{hastings_dynamically_2021, gidney_pair_2022}.  Each qubit in the hybrid circuit must participate in only one pair measurement and two standard entangling gates rather than three pair measurements, possibly simplifying the implementation of the architecture in hardware. Hardware implementing pair measurements has continued to improve~\cite{lalumiere_tunable_2010, divincenzo_multi-qubit_2012, royer_qubit_2018, reagor_hardware_2022, livingston_experimental_2022}, and this circuit provides an opportunity to use such hardware for the surface code.
Under our assumed error model, which gives the same two-qubit depolarizing strength to two qubits gates and pair measurements, these circuits display better performance than equivalent circuits using only two qubit gates. However, we expect details of the true error model in hardware to determine their relative performance in practice. We include them as inspiration for alternative hardware architectures to target the surface code without additional overhead. 

\subsection{Open Problems}

These constructions collectively represent only the first steps toward tailoring QEC circuit decompositions to relax or improve hardware requirements. Here, we discuss some directions for future research we hope are encouraged by the results presented here.

First, the surface code is only one error correction code. We chose to focus on it for its relevance to current experimental implementations of error correction. Applying these ideas to other codes is important open work, particularly to the color code, dynamical codes such as the honeycomb code, and to codes designed with biased noise in mind.

Second, the spatial compatibility of these code circuits with each other and with other standard circuits warrants further exploration. In particular, finding circuits for implementing sub-system codes that are compatible with these code circuits will be necessary for future implementation of these circuits on large grids of imperfect qubits.

Third, further refining the simple boundaries we have presented may be necessary to optimize performance and obey additional hardware constraints. In the context of logical algorithms, some boundary constructions may pack better than others, and behaviours like sliding may be useful in reducing costs. Other boundary constructions may be useful to respect constraints on which qubits may be measured simultaneously, even with an additional overhead in qubit number or packing ability.

Finally, the benchmarking we presented here is deficient in that it fails to account for the primary ways we expect our circuit constructions to do better than the standard circuit. Benchmarking them with more realistic noise models, especially those including hardware accurate gate errors, leakage, crosstalk and other effects of relaxing requirements on the underlying device architecture is an important next step for qualifying the advantages made possible by these circuits. 

\subsection{Outlook}

In this work, we have introduced an approach to constructing quantum error correction circuits more directly than via considering static stabilizer codes and only then forming their circuit decomposition. This work complements a wider thrust toward considering the full space-time dynamics of quantum error correction circuits~\cite{gottesman_opportunities_2022}.  

We emphasise the concept of detecting regions as a helpful basis for considering stabilizers at all points during QEC circuits. We used the insights produced by slicing detecting regions to generate new circuits for implementing QEC codes. These provide new freedoms to hardware engineers designing devices to implement QEC in experiments. We hope that these initial constructions serve as certificates that this approach is worth pursuing further.

\subsubsection{A note on recent progress in the space-time approach}
Following this work first appearing online, two additional works were published presenting progress in the space-time approach to fault-tolerance. We would like to draw attention to these works specifically, as we feel they compliments the concepts discussed in this work, and hope to encourage future developments along these lines.

First, \cite{bombin_unifying_2023} presents a lower-level approach to topological stabilizer fault tolerance, demonstrating that fault-tolerance in circuit-based, measurement-based, fusion-based and Floquet-based quantum computing can be understood as implementations of the same underlying fault-tolerant structure. In contrast to our work, where we restricted ourselves to the language of circuits, this work uses the language of the ZX calculus. In particular, the \emph{Pauli web} of a check in a ZX-graph is the direct analogue of a detecting region in a circuit. We chose to focus on the circuit picture for its applicability to hardware implementation, in particular being able to directly benchmark the circuit performance under hardware noise models. For those free of such considerations, we highly recommend the language of the ZX calculus and this work.

Second, \cite{delfosse_spacetime_2023} presents a generalised scheme for correcting faults in Clifford circuits by constructing an appropriate \emph{spacetime code} over that circuit. In contrast to our work, where we consider only optimising the surface code circuit, this work considers the more general problem in arbitrary Clifford circuits and of decoding the resulting code. As the authors note, the stabilizer terms of the spacetime code correspond to detectors in our work. We appreciate and endorse the authors suggestion that their approach could push the circuit-centric approach further, especially in searching for promising new quantum codes and circuit constructions. 

We feel these works serve as further evidence that considering the spacetime structure of circuits is a powerful new paradigm for approaching fault-tolerance. Moving beyond stabilizer codes reveals new connections between different fault-tolerance strategies, opens new avenues for exploration, and gives us hope for future improvements in the best strategies for implementing useful quantum computation in practice.

\section*{Contributions}
Matt McEwen and Craig Gidney conceived of using the detecting region picture to produce new circuits for the surface code.
Craig Gidney developed software tools, including Stim, that were vital in exploring and developing these concepts and constructions.
Matt McEwen conceived and built the initial wiggling and walking surface code circuits.
Dave Bacon and Craig Gidney conceived and built the initial ISWAP surface code circuits.
Craig Gidney conceived and built the initial hex-grid surface code circuits.
All authors participated in synthesizing these ideas into a coherent narrative centered around overlapping structures of detecting regions.
Matt McEwen compiled the manuscript.

\section*{Acknowledgements}

We thank Michael Newman and Cody Jones for valuable discussions around these concepts as they were being developed.
We thank Austin Fowler for writing the internal software tool we used for syndrome decoding, and for helpful review of the manuscript. 
We thank Alexis Morvan and Oscar Higgot for helpful review of the manuscript. 
We thank Alex Townsend-Teague for pointing out an error in the walking circuit figures. 
Finally, we would like to thank the entire Google Quantum AI team for making an environment where this work is possible.

\section*{Availability of data and code}
The circuits that were benchmarked and the resulting raw samples are available on Zenodo at \href{https://zenodo.org/record/7587578}{\texttt{https://zenodo.org/record/7587578}} \cite{mcewen_data_2023}.
The supplementary figures, along with the latex source and assets for this manuscript, are also available in the data repository and with the ArXiv submission. 
Code used to produce the circuits and to run the benchmarking is available on GitHub at \href{https://github.com/Strilanc/midout}{\texttt{https://github.com/Strilanc/midout}}.
Stim, including new tools for visualizing circuits and stabilizers, is available on GitHub at \href{https://github.com/quantumlib/Stim}{\texttt{https://github.com/quantumlib/Stim}} \cite{gidney_stim_2021}.

\printbibliography

@article{fowler_surface_2012,
	title = {Surface codes: Towards practical large-scale quantum computation},
	volume = {86},
	issn = {1050-2947, 1094-1622},
	url = {https://link.aps.org/doi/10.1103/PhysRevA.86.032324},
	doi = {10.1103/physreva.86.032324},
	pages = {032324},
	number = {3},
	journaltitle = {Physical Review A},
	shortjournal = {Phys. Rev. A},
	author = {Fowler, Austin G. and Mariantoni, Matteo and Martinis, John M. and Cleland, Andrew N.},
	date = {2012-09-18},
	note = {Publisher: American Physical Society},
}

@article{motzoi_simple_2009,
	title = {Simple Pulses for Elimination of Leakage in Weakly Nonlinear Qubits},
	volume = {103},
	issn = {0031-9007, 1079-7114},
	url = {https://link.aps.org/doi/10.1103/PhysRevLett.103.110501},
	doi = {10.1103/PhysRevLett.103.110501},
	pages = {110501},
	number = {11},
	journaltitle = {Physical Review Letters},
	shortjournal = {Phys. Rev. Lett.},
	author = {Motzoi, F. and Gambetta, J. M. and Rebentrost, P. and Wilhelm, F. K.},
	date = {2009-09-08},
	note = {Publisher: American Physical Society},
}

@article{chen_measuring_2016,
	title = {Measuring and Suppressing Quantum State Leakage in a Superconducting Qubit},
	volume = {116},
	issn = {0031-9007, 1079-7114},
	url = {https://link.aps.org/doi/10.1103/PhysRevLett.116.020501},
	doi = {10.1103/PhysRevLett.116.020501},
	pages = {020501},
	number = {2},
	journaltitle = {Physical Review Letters},
	shortjournal = {Phys. Rev. Lett.},
	author = {Chen, Zijun and Kelly, Julian and Quintana, Chris and Barends, R. and Campbell, B. and Chen, Yu and Chiaro, B. and Dunsworth, A. and Fowler, A. G. and Lucero, E. and Jeffrey, E. and Megrant, A. and Mutus, J. and Neeley, M. and Neill, C. and O’Malley, P. J. J. and Roushan, P. and Sank, D. and Vainsencher, A. and Wenner, J. and White, T. C. and Korotkov, A. N. and Martinis, John M.},
	date = {2016-01-13},
	note = {Publisher: American Physical Society},
}

@article{fowler_coping_2013,
	title = {Coping with qubit leakage in topological codes},
	volume = {88},
	issn = {1050-2947, 1094-1622},
	url = {https://link.aps.org/doi/10.1103/PhysRevA.88.042308},
	doi = {10.1103/PhysRevA.88.042308},
	pages = {042308},
	number = {4},
	journaltitle = {Physical Review A},
	shortjournal = {Phys. Rev. A},
	author = {Fowler, Austin G.},
	date = {2013-10-08},
	note = {Publisher: American Physical Society},
}

@article{arute_quantum_2019,
	title = {Quantum supremacy using a programmable superconducting processor},
	volume = {574},
	issn = {0028-0836, 1476-4687},
	url = {http://www.nature.com/articles/s41586-019-1666-5},
	doi = {10.1038/s41586-019-1666-5},
	pages = {505--510},
	number = {7779},
	journaltitle = {Nature},
	shortjournal = {Nature},
	author = {Arute, Frank and Arya, Kunal and Babbush, Ryan and Bacon, Dave and Bardin, Joseph C. and Barends, Rami and Biswas, Rupak and Boixo, Sergio and Brandao, Fernando G. S. L. and Buell, David A. and Burkett, Brian and Chen, Yu and Chen, Zijun and Chiaro, Ben and Collins, Roberto and Courtney, William and Dunsworth, Andrew and Farhi, Edward and Foxen, Brooks and Fowler, Austin and Gidney, Craig and Giustina, Marissa and Graff, Rob and Guerin, Keith and Habegger, Steve and Harrigan, Matthew P. and Hartmann, Michael J. and Ho, Alan and Hoffmann, Markus and Huang, Trent and Humble, Travis S. and Isakov, Sergei V. and Jeffrey, Evan and Jiang, Zhang and Kafri, Dvir and Kechedzhi, Kostyantyn and Kelly, Julian and Klimov, Paul V. and Knysh, Sergey and Korotkov, Alexander and Kostritsa, Fedor and Landhuis, David and Lindmark, Mike and Lucero, Erik and Lyakh, Dmitry and Mandrà, Salvatore and {McClean}, Jarrod R. and {McEwen}, Matthew and Megrant, Anthony and Mi, Xiao and Michielsen, Kristel and Mohseni, Masoud and Mutus, Josh and Naaman, Ofer and Neeley, Matthew and Neill, Charles and Niu, Murphy Yuezhen and Ostby, Eric and Petukhov, Andre and Platt, John C. and Quintana, Chris and Rieffel, Eleanor G. and Roushan, Pedram and Rubin, Nicholas C. and Sank, Daniel and Satzinger, Kevin J. and Smelyanskiy, Vadim and Sung, Kevin J. and Trevithick, Matthew D. and Vainsencher, Amit and Villalonga, Benjamin and White, Theodore and Yao, Z. Jamie and Yeh, Ping and Zalcman, Adam and Neven, Hartmut and Martinis, John M.},
	date = {2019-10-24},
	note = {Publisher: Nature Publishing Group},
}

@article{foxen_demonstrating_2020,
	title = {Demonstrating a Continuous Set of Two-qubit Gates for Near-term Quantum Algorithms},
	volume = {125},
	issn = {0031-9007, 1079-7114},
	url = {https://link.aps.org/doi/10.1103/PhysRevLett.125.120504},
	doi = {https://doi.org/10.1103/PhysRevLett.125.120504},
	pages = {120504},
	number = {12},
	journaltitle = {Physical Review Letters},
	shortjournal = {Phys. Rev. Lett.},
	author = {Foxen, B. and Neill, C. and Dunsworth, A. and Roushan, P. and Chiaro, B. and Megrant, A. and Kelly, J. and Chen, Zijun and Satzinger, K. and Barends, R. and Arute, F. and Arya, K. and Babbush, R. and Bacon, D. and Bardin, J. C. and Boixo, S. and Buell, D. and Burkett, B. and Chen, Yu and Collins, R. and Farhi, E. and Fowler, A. and Gidney, C. and Giustina, M. and Graff, R. and Harrigan, M. and Huang, T. and Isakov, S. V. and Jeffrey, E. and Jiang, Z. and Kafri, D. and Kechedzhi, K. and Klimov, P. and Korotkov, A. and Kostritsa, F. and Landhuis, D. and Lucero, E. and McClean, J. and McEwen, M. and Mi, X. and Mohseni, M. and Mutus, J. Y. and Naaman, O. and Neeley, M. and Niu, M. and Petukhov, A. and Quintana, C. and Rubin, N. and Sank, D. and Smelyanskiy, V. and Vainsencher, A. and White, T. C. and Yao, Z. and Yeh, P. and Zalcman, A. and Neven, H. and Martinis, J. M. and {Google AI Quantum}},
	date = {2020-09-15},
	note = {\_eprint: 2001.08343},
}

@article{magnard_fast_2018,
	title = {Fast and Unconditional All-Microwave Reset of a Superconducting Qubit},
	volume = {121},
	issn = {0031-9007, 1079-7114},
	url = {https://link.aps.org/doi/10.1103/PhysRevLett.121.060502},
	doi = {10.1103/PhysRevLett.121.060502},
	pages = {060502},
	number = {6},
	journaltitle = {Physical Review Letters},
	shortjournal = {Phys. Rev. Lett.},
	author = {Magnard, P. and Kurpiers, P. and Royer, B. and Walter, T. and Besse, J.-C. and Gasparinetti, S. and Pechal, M. and Heinsoo, J. and Storz, S. and Blais, A. and Wallraff, A.},
	date = {2018-08-07},
	note = {Publisher: American Physical Society},
}

@article{yan_tunable_2018,
	title = {Tunable Coupling Scheme for Implementing High-Fidelity Two-Qubit Gates},
	volume = {10},
	issn = {2331-7019},
	url = {https://link.aps.org/doi/10.1103/PhysRevApplied.10.054062},
	doi = {10.1103/PhysRevApplied.10.054062},
	pages = {054062},
	number = {5},
	journaltitle = {Physical Review Applied},
	shortjournal = {Phys. Rev. Applied},
	author = {Yan, Fei and Krantz, Philip and Sung, Youngkyu and Kjaergaard, Morten and Campbell, Daniel L. and Orlando, Terry P. and Gustavsson, Simon and Oliver, William D.},
	date = {2018-11-28},
	note = {Publisher: American Physical Society},
}

@article{google_quantum_ai_exponential_2021,
	title = {Exponential suppression of bit or phase errors with cyclic error correction},
	volume = {595},
	issn = {0028-0836, 1476-4687},
	url = {http://www.nature.com/articles/s41586-021-03588-y},
	doi = {https://doi.org/10.1038/s41586-021-03588-y},
	pages = {383--387},
	number = {7867},
	journaltitle = {Nature},
	shortjournal = {Nature},
	author = {{Google Quantum AI} and Chen, Zijun and Satzinger, Kevin J. and Atalaya, Juan and Korotkov, Alexander N. and Dunsworth, Andrew and Sank, Daniel and Quintana, Chris and McEwen, Matt and Barends, Rami and Klimov, Paul V. and Hong, Sabrina and Jones, Cody and Petukhov, Andre and Kafri, Dvir and Demura, Sean and Burkett, Brian and Gidney, Craig and Fowler, Austin G. and Paler, Alexandru and Putterman, Harald and Aleiner, Igor and Arute, Frank and Arya, Kunal and Babbush, Ryan and Bardin, Joseph C. and Bengtsson, Andreas and Bourassa, Alexandre and Broughton, Michael and Buckley, Bob B. and Buell, David A. and Bushnell, Nicholas and Chiaro, Benjamin and Collins, Roberto and Courtney, William and Derk, Alan R. and Eppens, Daniel and Erickson, Catherine and Farhi, Edward and Foxen, Brooks and Giustina, Marissa and Greene, Ami and Gross, Jonathan A. and Harrigan, Matthew P. and Harrington, Sean D. and Hilton, Jeremy and Ho, Alan and Huang, Trent and Huggins, William J. and Ioffe, L. B. and Isakov, Sergei V. and Jeffrey, Evan and Jiang, Zhang and Kechedzhi, Kostyantyn and Kim, Seon and Kitaev, Alexei and Kostritsa, Fedor and Landhuis, David and Laptev, Pavel and Lucero, Erik and Martin, Orion and McClean, Jarrod R. and McCourt, Trevor and Mi, Xiao and Miao, Kevin C. and Mohseni, Masoud and Montazeri, Shirin and Mruczkiewicz, Wojciech and Mutus, Josh and Naaman, Ofer and Neeley, Matthew and Neill, Charles and Newman, Michael and Niu, Murphy Yuezhen and O’Brien, Thomas E. and Opremcak, Alex and Ostby, Eric and Pató, Bálint and Redd, Nicholas and Roushan, Pedram and Rubin, Nicholas C. and Shvarts, Vladimir and Strain, Doug and Szalay, Marco and Trevithick, Matthew D. and Villalonga, Benjamin and White, Theodore and Yao, Z. Jamie and Yeh, Ping and Yoo, Juhwan and Zalcman, Adam and Neven, Hartmut and Boixo, Sergio and Smelyanskiy, Vadim and Chen, Yu and Megrant, Anthony and Kelly, Julian},
	date = {2021-07-15},
	note = {\_eprint: 2102.06132},
}

@article{brown_leakage_2019,
	title = {Leakage mitigation for quantum error correction using a mixed qubit scheme},
	volume = {100},
	issn = {2469-9926, 2469-9934},
	url = {https://link.aps.org/doi/10.1103/PhysRevA.100.032325},
	doi = {10.1103/PhysRevA.100.032325},
	pages = {032325},
	number = {3},
	journaltitle = {Physical Review A},
	shortjournal = {Phys. Rev. A},
	author = {Brown, Natalie C. and Brown, Kenneth R.},
	date = {2019-09-18},
	note = {Publisher: American Physical Society},
}

@article{krinner_realizing_2022,
	title = {Realizing repeated quantum error correction in a distance-three surface code},
	volume = {605},
	issn = {0028-0836, 1476-4687},
	url = {https://www.nature.com/articles/s41586-022-04566-8},
	doi = {10.1038/s41586-022-04566-8},
	pages = {669--674},
	number = {7911},
	journaltitle = {Nature},
	shortjournal = {Nature},
	author = {Krinner, Sebastian and Lacroix, Nathan and Remm, Ants and Di Paolo, Agustin and Genois, Elie and Leroux, Catherine and Hellings, Christoph and Lazar, Stefania and Swiadek, Francois and Herrmann, Johannes and Norris, Graham J. and Andersen, Christian Kraglund and Müller, Markus and Blais, Alexandre and Eichler, Christopher and Wallraff, Andreas},
	date = {2022-05-26},
	note = {Publisher: Springer Science and Business Media {LLC}},
}

@article{sundaresan_matching_2022,
	title = {Matching and maximum likelihood decoding of a multi-round subsystem quantum error correction experiment},
	rights = {Creative Commons Attribution 4.0 International},
	url = {https://arxiv.org/abs/2203.07205},
	doi = {https://doi.org/10.48550/arXiv.2203.07205},
	author = {Sundaresan, Neereja and Yoder, Theodore J. and Kim, Youngseok and Li, Muyuan and Chen, Edward H. and Harper, Grace and Thorbeck, Ted and Cross, Andrew W. and Córcoles, Antonio D. and Takita, Maika},
	date = {2022},
	doi = {10.48550/ARXIV.2203.07205},
}

@article{google_quantum_ai_suppressing_2022,
	title = {Suppressing quantum errors by scaling a surface code logical qubit},
	rights = {Creative Commons Attribution 4.0 International},
	url = {https://arxiv.org/abs/2207.06431},
	doi = {https://doi.org/10.48550/arXiv.2207.06431},
	author = {{Google Quantum AI} and Acharya, Rajeev and Aleiner, Igor and Allen, Richard and Andersen, Trond I. and Ansmann, Markus and Arute, Frank and Arya, Kunal and Asfaw, Abraham and Atalaya, Juan and Babbush, Ryan and Bacon, Dave and Bardin, Joseph C. and Basso, Joao and Bengtsson, Andreas and Boixo, Sergio and Bortoli, Gina and Bourassa, Alexandre and Bovaird, Jenna and Brill, Leon and Broughton, Michael and Buckley, Bob B. and Buell, David A. and Burger, Tim and Burkett, Brian and Bushnell, Nicholas and Chen, Yu and Chen, Zijun and Chiaro, Ben and Cogan, Josh and Collins, Roberto and Conner, Paul and Courtney, William and Crook, Alexander L. and Curtin, Ben and Debroy, Dripto M. and Barba, Alexander Del Toro and Demura, Sean and Dunsworth, Andrew and Eppens, Daniel and Erickson, Catherine and Faoro, Lara and Farhi, Edward and Fatemi, Reza and Burgos, Leslie Flores and Forati, Ebrahim and Fowler, Austin G. and Foxen, Brooks and Giang, William and Gidney, Craig and Gilboa, Dar and Giustina, Marissa and Dau, Alejandro Grajales and Gross, Jonathan A. and Habegger, Steve and Hamilton, Michael C. and Harrigan, Matthew P. and Harrington, Sean D. and Higgott, Oscar and Hilton, Jeremy and Hoffmann, Markus and Hong, Sabrina and Huang, Trent and Huff, Ashley and Huggins, William J. and Ioffe, Lev B. and Isakov, Sergei V. and Iveland, Justin and Jeffrey, Evan and Jiang, Zhang and Jones, Cody and Juhas, Pavol and Kafri, Dvir and Kechedzhi, Kostyantyn and Kelly, Julian and Khattar, Tanuj and Khezri, Mostafa and Kieferová, Mária and Kim, Seon and Kitaev, Alexei and Klimov, Paul V. and Klots, Andrey R. and Korotkov, Alexander N. and Kostritsa, Fedor and Kreikebaum, John Mark and Landhuis, David and Laptev, Pavel and Lau, Kim-Ming and Laws, Lily and Lee, Joonho and Lee, Kenny and Lester, Brian J. and Lill, Alexander and Liu, Wayne and Locharla, Aditya and Lucero, Erik and Malone, Fionn D. and Marshall, Jeffrey and Martin, Orion and McClean, Jarrod R. and Mccourt, Trevor and McEwen, Matt and Megrant, Anthony and Costa, Bernardo Meurer and Mi, Xiao and Miao, Kevin C. and Mohseni, Masoud and Montazeri, Shirin and Morvan, Alexis and Mount, Emily and Mruczkiewicz, Wojciech and Naaman, Ofer and Neeley, Matthew and Neill, Charles and Nersisyan, Ani and Neven, Hartmut and Newman, Michael and Ng, Jiun How and Nguyen, Anthony and Nguyen, Murray and Niu, Murphy Yuezhen and O'Brien, Thomas E. and Opremcak, Alex and Platt, John and Petukhov, Andre and Potter, Rebecca and Pryadko, Leonid P. and Quintana, Chris and Roushan, Pedram and Rubin, Nicholas C. and Saei, Negar and Sank, Daniel and Sankaragomathi, Kannan and Satzinger, Kevin J. and Schurkus, Henry F. and Schuster, Christopher and Shearn, Michael J. and Shorter, Aaron and Shvarts, Vladimir and Skruzny, Jindra and Smelyanskiy, Vadim and Smith, W. Clarke and Sterling, George and Strain, Doug and Szalay, Marco and Torres, Alfredo and Vidal, Guifre and Villalonga, Benjamin and Heidweiller, Catherine Vollgraff and White, Theodore and Xing, Cheng and Yao, Z. Jamie and Yeh, Ping and Yoo, Juhwan and Young, Grayson and Zalcman, Adam and Zhang, Yaxing and Zhu, Ningfeng},
	date = {2022},
	doi = {10.48550/ARXIV.2207.06431},
}

@article{zhou_rapid_2021,
	title = {Rapid and unconditional parametric reset protocol for tunable superconducting qubits},
	volume = {12},
	issn = {2041-1723},
	url = {https://www.nature.com/articles/s41467-021-26205-y},
	doi = {10.1038/s41467-021-26205-y},
	pages = {5924},
	number = {1},
	journaltitle = {Nature Communications},
	shortjournal = {Nat Commun},
	author = {Zhou, Yu and Zhang, Zhenxing and Yin, Zelong and Huai, Sainan and Gu, Xiu and Xu, Xiong and Allcock, Jonathan and Liu, Fuming and Xi, Guanglei and Yu, Qiaonian and Zhang, Hualiang and Zhang, Mengyu and Li, Hekang and Song, Xiaohui and Wang, Zhan and Zheng, Dongning and An, Shuoming and Zheng, Yarui and Zhang, Shengyu},
	date = {2021-12},
	note = {Publisher: Springer Science and Business Media {LLC}},
}

@article{gidney_how_2021,
	title = {How to factor 2048 bit {RSA} integers in 8 hours using 20 million noisy qubits},
	volume = {5},
	issn = {2521-327X},
	url = {https://quantum-journal.org/papers/q-2021-04-15-433/},
	doi = {10.22331/q-2021-04-15-433},
	pages = {433},
	journaltitle = {Quantum},
	shortjournal = {Quantum},
	author = {Gidney, Craig and Ekerå, Martin},
	date = {2021-04-15},
	note = {Publisher: Verein zur Förderung des Open Access Publizierens in den Quantenwissenschaften},
}

@article{mcewen_removing_2021,
	title = {Removing leakage-induced correlated errors in superconducting quantum error correction},
	volume = {12},
	rights = {All rights reserved},
	issn = {2041-1723},
	url = {http://www.nature.com/articles/s41467-021-21982-y},
	doi = {10.1038/s41467-021-21982-y},
	pages = {1761},
	number = {1},
	journaltitle = {Nature Communications},
	shortjournal = {Nat Commun},
	author = {{McEwen}, Matt and Kafri, D. and Chen, Z. and Atalaya, J. and Satzinger, K. J. and Quintana, C. and Klimov, P. V. and Sank, D. and Gidney, C. and Fowler, A. G. and Arute, F. and Arya, K. and Buckley, B. and Burkett, B. and Bushnell, N. and Chiaro, B. and Collins, R. and Demura, S. and Dunsworth, A. and Erickson, C. and Foxen, B. and Giustina, M. and Huang, T. and Hong, S. and Jeffrey, E. and Kim, S. and Kechedzhi, K. and Kostritsa, F. and Laptev, P. and Megrant, A. and Mi, X. and Mutus, J. and Naaman, O. and Neeley, M. and Neill, C. and Niu, M. and Paler, A. and Redd, N. and Roushan, P. and White, T. C. and Yao, J. and Yeh, P. and Zalcman, A. and Chen, Yu and Smelyanskiy, V. N. and Martinis, John M. and Neven, H. and Kelly, J. and Korotkov, A. N. and Petukhov, A. G. and Barends, R.},
	urldate = {2022-10-05},
	date = {2021-12},
	langid = {english},
}

@article{gidney_benchmarking_2022,
	title = {Benchmarking the Planar Honeycomb Code},
	volume = {6},
	issn = {2521-327X},
	url = {https://quantum-journal.org/papers/q-2022-09-21-813/},
	doi = {10.22331/q-2022-09-21-813},
	pages = {813},
	journaltitle = {Quantum},
	shortjournal = {Quantum},
	author = {Gidney, Craig and Newman, Michael and {McEwen}, Matt},
	urldate = {2022-10-06},
	date = {2022-09-21},
	langid = {english},
}

@misc{gidney_pair_2022,
	title = {A Pair Measurement Surface Code on Pentagons},
	rights = {Creative Commons Attribution 4.0 International},
	url = {https://arxiv.org/abs/2206.12780},
	publisher = {{arXiv}},
	author = {Gidney, Craig},
	date = {2022},
	doi = {10.48550/ARXIV.2206.12780},
}

@article{hastings_dynamically_2021,
	title = {Dynamically Generated Logical Qubits},
	volume = {5},
	url = {https://doi.org/10.22331\%2Fq-2021-10-19-564},
	doi = {10.22331/q-2021-10-19-564},
	pages = {564},
	journaltitle = {Quantum},
	author = {Hastings, Matthew B. and Haah, Jeongwan},
	date = {2021-10},
	note = {Publisher: Verein zur Forderung des Open Access Publizierens in den Quantenwissenschaften},
}

@article{battistel_hardware-efficient_2021,
	title = {Hardware-Efficient Leakage-Reduction Scheme for Quantum Error Correction with Superconducting Transmon Qubits},
	volume = {2},
	issn = {2691-3399},
	url = {https://link.aps.org/doi/10.1103/PRXQuantum.2.030314},
	doi = {10.1103/PRXQuantum.2.030314},
	pages = {030314},
	number = {3},
	journaltitle = {{PRX} Quantum},
	shortjournal = {{PRX} Quantum},
	author = {Battistel, F. and Varbanov, B.M. and Terhal, B.M.},
	urldate = {2022-10-10},
	date = {2021-07-26},
	langid = {english},
}

@article{gottesman_heisenberg_1998,
	title = {The Heisenberg Representation of Quantum Computers},
	url = {https://arxiv.org/abs/quant-ph/9807006},
	doi = {10.48550/ARXIV.QUANT-PH/9807006},
	author = {Gottesman, Daniel},
	urldate = {2022-10-13},
	date = {1998},
	note = {Publisher: {arXiv}
Version Number: 1},
}

@article{shor_scheme_1995,
	title = {Scheme for reducing decoherence in quantum computer memory},
	volume = {52},
	issn = {1050-2947, 1094-1622},
	url = {https://link.aps.org/doi/10.1103/PhysRevA.52.R2493},
	doi = {10.1103/PhysRevA.52.R2493},
	pages = {R2493--R2496},
	number = {4},
	journaltitle = {Physical Review A},
	shortjournal = {Phys. Rev. A},
	author = {Shor, Peter W.},
	urldate = {2022-10-18},
	date = {1995-10-01},
	langid = {english},
}

@article{calderbank_good_1996,
	title = {Good quantum error-correcting codes exist},
	volume = {54},
	issn = {1050-2947, 1094-1622},
	url = {https://link.aps.org/doi/10.1103/PhysRevA.54.1098},
	doi = {10.1103/PhysRevA.54.1098},
	pages = {1098--1105},
	number = {2},
	journaltitle = {Physical Review A},
	shortjournal = {Phys. Rev. A},
	author = {Calderbank, A. R. and Shor, Peter W.},
	urldate = {2022-10-18},
	date = {1996-08-01},
	langid = {english},
}

@article{steane_multiple-particle_1996,
	title = {Multiple-particle interference and quantum error correction},
	volume = {452},
	issn = {1364-5021, 1471-2946},
	doi = {10.1098/rspa.1996.0136},
	pages = {2551--2577},
	number = {1954},
	journaltitle = {Proceedings of the Royal Society of London. Series A: Mathematical, Physical and Engineering Sciences},
	shortjournal = {Proc. R. Soc. Lond. A},
	author = {Steane, Andrew},
	urldate = {2022-10-18},
	date = {1996-11-08},
	langid = {english},
}

@thesis{gottesman_stabilizer_1997,
	title = {Stabilizer Codes and Quantum Error Correction},
	rights = {Assumed {arXiv}.org perpetual, non-exclusive license to distribute this article for submissions made before January 2004},
	url = {https://arxiv.org/abs/quant-ph/9705052},
	type = {phdthesis},
	author = {Gottesman, Daniel},
	urldate = {2022-10-18},
	date = {1997},
	note = {Publisher: {arXiv}
Version Number: 1},
}

@article{breuckmann_balanced_2020,
	title = {Balanced Product Quantum Codes},
	rights = {{arXiv}.org perpetual, non-exclusive license},
	url = {https://arxiv.org/abs/2012.09271},
	doi = {10.48550/ARXIV.2012.09271},
	author = {Breuckmann, Nikolas P. and Eberhardt, Jens N.},
	urldate = {2022-10-18},
	date = {2020},
	note = {Publisher: {arXiv}
Version Number: 3},
}

@article{bravyi_quantum_1998,
	title = {Quantum codes on a lattice with boundary},
	rights = {Assumed {arXiv}.org perpetual, non-exclusive license to distribute this article for submissions made before January 2004},
	url = {https://arxiv.org/abs/quant-ph/9811052},
	doi = {10.48550/ARXIV.QUANT-PH/9811052},
	author = {Bravyi, S. B. and Kitaev, A. Yu.},
	urldate = {2022-10-19},
	date = {1998},
	note = {Publisher: {arXiv}
Version Number: 1},
}

@article{kitaev_fault-tolerant_1997,
	title = {Fault-tolerant quantum computation by anyons},
	volume = {303},
	issn = {00034916},
	url = {http://arxiv.org/abs/quant-ph/9707021},
	doi = {10.1016/S0003-4916(02)00018-0},
	pages = {2--30},
	number = {1},
	journaltitle = {Annals of Physics},
	shortjournal = {Annals of Physics},
	author = {Kitaev, A. Yu},
	urldate = {2022-10-19},
	date = {1997},
	eprinttype = {arxiv},
	eprint = {quant-ph/9707021},
}

@article{horsman_surface_2012,
	title = {Surface code quantum computing by lattice surgery},
	volume = {14},
	issn = {1367-2630},
	url = {https://iopscience.iop.org/article/10.1088/1367-2630/14/12/123011},
	doi = {10.1088/1367-2630/14/12/123011},
	pages = {123011},
	number = {12},
	journaltitle = {New Journal of Physics},
	shortjournal = {New J. Phys.},
	author = {Horsman, Clare and Fowler, Austin G and Devitt, Simon and Meter, Rodney Van},
	urldate = {2022-10-20},
	date = {2012-12-07},
}

@article{bravyi_universal_2005,
	title = {Universal quantum computation with ideal Clifford gates and noisy ancillas},
	volume = {71},
	issn = {1050-2947, 1094-1622},
	url = {https://link.aps.org/doi/10.1103/PhysRevA.71.022316},
	doi = {10.1103/PhysRevA.71.022316},
	pages = {022316},
	number = {2},
	journaltitle = {Physical Review A},
	shortjournal = {Phys. Rev. A},
	author = {Bravyi, Sergey and Kitaev, Alexei},
	urldate = {2022-10-20},
	date = {2005-02-22},
	langid = {english},
}

@article{dennis_topological_2002,
	title = {Topological quantum memory},
	volume = {43},
	issn = {0022-2488, 1089-7658},
	url = {http://aip.scitation.org/doi/10.1063/1.1499754},
	doi = {10.1063/1.1499754},
	pages = {4452--4505},
	number = {9},
	journaltitle = {Journal of Mathematical Physics},
	shortjournal = {Journal of Mathematical Physics},
	author = {Dennis, Eric and Kitaev, Alexei and Landahl, Andrew and Preskill, John},
	urldate = {2022-10-20},
	date = {2002-09},
	langid = {english},
}

@article{bravyi_subsystem_2012,
	title = {Subsystem surface codes with three-qubit check operators},
	rights = {{arXiv}.org perpetual, non-exclusive license},
	url = {https://arxiv.org/abs/1207.1443},
	doi = {10.48550/ARXIV.1207.1443},
	author = {Bravyi, Sergey and Duclos-Cianci, Guillaume and Poulin, David and Suchara, Martin},
	urldate = {2022-10-24},
	date = {2012},
	note = {Publisher: {arXiv}
Version Number: 2},
}

@article{fowler_optimal_2013,
	title = {Optimal complexity correction of correlated errors in the surface code},
	rights = {{arXiv}.org perpetual, non-exclusive license},
	url = {https://arxiv.org/abs/1310.0863},
	doi = {10.48550/ARXIV.1310.0863},
	author = {Fowler, Austin G.},
	urldate = {2022-10-25},
	date = {2013},
	note = {Publisher: {arXiv}
Version Number: 1},
}

@article{rigetti_fully_2010,
	title = {Fully microwave-tunable universal gates in superconducting qubits with linear couplings and fixed transition frequencies},
	volume = {81},
	issn = {1098-0121, 1550-235X},
	url = {https://link.aps.org/doi/10.1103/PhysRevB.81.134507},
	doi = {10.1103/PhysRevB.81.134507},
	pages = {134507},
	number = {13},
	journaltitle = {Physical Review B},
	shortjournal = {Phys. Rev. B},
	author = {Rigetti, Chad and Devoret, Michel},
	urldate = {2022-10-25},
	date = {2010-04-05},
	langid = {english},
}

@article{ghosh_understanding_2013,
	title = {Understanding the effects of leakage in superconducting quantum-error-detection circuits},
	volume = {88},
	issn = {1050-2947, 1094-1622},
	url = {https://link.aps.org/doi/10.1103/PhysRevA.88.062329},
	doi = {10.1103/PhysRevA.88.062329},
	pages = {062329},
	number = {6},
	journaltitle = {Physical Review A},
	shortjournal = {Phys. Rev. A},
	author = {Ghosh, Joydip and Fowler, Austin G. and Martinis, John M. and Geller, Michael R.},
	urldate = {2022-11-16},
	date = {2013-12-23},
	langid = {english},
}

@article{ghosh_leakage-resilient_2015,
	title = {Leakage-resilient approach to fault-tolerant quantum computing with superconducting elements},
	volume = {91},
	issn = {1050-2947, 1094-1622},
	url = {https://link.aps.org/doi/10.1103/PhysRevA.91.020302},
	doi = {10.1103/PhysRevA.91.020302},
	pages = {020302},
	number = {2},
	journaltitle = {Physical Review A},
	shortjournal = {Phys. Rev. A},
	author = {Ghosh, Joydip and Fowler, Austin G.},
	urldate = {2022-11-16},
	date = {2015-02-20},
	langid = {english},
}

@article{aaronson_introduction_2022,
	title = {Introduction to Quantum Information Science {II} Lecture Notes},
	url = {https://www.scottaaronson.com/qisii.pdf},
	author = {Aaronson, Scott},
	date = {2022},
}

@article{tucci_introduction_2005,
	title = {An Introduction to Cartan's {KAK} Decomposition for {QC} Programmers},
	rights = {Assumed {arXiv}.org perpetual, non-exclusive license to distribute this article for submissions made before January 2004},
	url = {https://arxiv.org/abs/quant-ph/0507171},
	doi = {10.48550/ARXIV.QUANT-PH/0507171},
	author = {Tucci, Robert R.},
	urldate = {2022-11-17},
	date = {2005},
	note = {Publisher: {arXiv}
Version Number: 1},
}

@article{corcoles_process_2013,
	title = {Process verification of two-qubit quantum gates by randomized benchmarking},
	volume = {87},
	issn = {1050-2947, 1094-1622},
	url = {https://link.aps.org/doi/10.1103/PhysRevA.87.030301},
	doi = {10.1103/PhysRevA.87.030301},
	pages = {030301},
	number = {3},
	journaltitle = {Physical Review A},
	shortjournal = {Phys. Rev. A},
	author = {Córcoles, A. D. and Gambetta, Jay M. and Chow, Jerry M. and Smolin, John A. and Ware, Matthew and Strand, Joel and Plourde, B. L. T. and Steffen, M.},
	urldate = {2022-11-17},
	date = {2013-03-19},
	langid = {english},
}

@article{gottesman_opportunities_2022,
	title = {Opportunities and Challenges in Fault-Tolerant Quantum Computation},
	rights = {{arXiv}.org perpetual, non-exclusive license},
	url = {https://arxiv.org/abs/2210.15844},
	doi = {10.48550/ARXIV.2210.15844},
	author = {Gottesman, Daniel},
	urldate = {2022-11-17},
	date = {2022},
	note = {Publisher: {arXiv}
Version Number: 1},
}

@article{haah_boundaries_2022,
	title = {Boundaries for the Honeycomb Code},
	volume = {6},
	issn = {2521-327X},
	url = {https://quantum-journal.org/papers/q-2022-04-21-693/},
	doi = {10.22331/q-2022-04-21-693},
	pages = {693},
	journaltitle = {Quantum},
	shortjournal = {Quantum},
	author = {Haah, Jeongwan and Hastings, Matthew B.},
	urldate = {2022-11-17},
	date = {2022-04-21},
	langid = {english},
}

@article{aasen_adiabatic_2022,
	title = {Adiabatic paths of Hamiltonians, symmetries of topological order, and automorphism codes},
	rights = {{arXiv}.org perpetual, non-exclusive license},
	url = {https://arxiv.org/abs/2203.11137},
	doi = {10.48550/ARXIV.2203.11137},
	author = {Aasen, David and Wang, Zhenghan and Hastings, Matthew B.},
	urldate = {2022-11-17},
	date = {2022},
	note = {Publisher: {arXiv}
Version Number: 2},
}

@article{divincenzo_multi-qubit_2012,
	title = {Multi-qubit parity measurement in circuit quantum electrodynamics},
	rights = {{arXiv}.org perpetual, non-exclusive license},
	url = {https://arxiv.org/abs/1205.1910},
	doi = {10.48550/ARXIV.1205.1910},
	author = {{DiVincenzo}, David P. and Solgun, Firat},
	urldate = {2022-11-17},
	date = {2012},
	note = {Publisher: {arXiv}
Version Number: 2},
}

@article{livingston_experimental_2022,
	title = {Experimental demonstration of continuous quantum error correction},
	volume = {13},
	issn = {2041-1723},
	url = {https://www.nature.com/articles/s41467-022-29906-0},
	doi = {10.1038/s41467-022-29906-0},
	pages = {2307},
	number = {1},
	journaltitle = {Nature Communications},
	shortjournal = {Nat Commun},
	author = {Livingston, William P. and Blok, Machiel S. and Flurin, Emmanuel and Dressel, Justin and Jordan, Andrew N. and Siddiqi, Irfan},
	urldate = {2022-11-17},
	date = {2022-12},
	langid = {english},
}

@article{reagor_hardware_2022,
	title = {Hardware optimized parity check gates for superconducting surface codes},
	rights = {{arXiv}.org perpetual, non-exclusive license},
	url = {https://arxiv.org/abs/2211.06382},
	doi = {10.48550/ARXIV.2211.06382},
	author = {Reagor, Matthew J. and Bohdanowicz, Thomas C. and Perez, David Rodriguez and Sete, Eyob A. and Zeng, William J.},
	urldate = {2022-11-17},
	date = {2022},
	note = {Publisher: {arXiv}
Version Number: 1},
}

@article{bonilla_ataides_xzzx_2021,
	title = {The {XZZX} surface code},
	volume = {12},
	issn = {2041-1723},
	url = {http://www.nature.com/articles/s41467-021-22274-1},
	doi = {10.1038/s41467-021-22274-1},
	pages = {2172},
	number = {1},
	journaltitle = {Nature Communications},
	shortjournal = {Nat Commun},
	author = {Bonilla Ataides, J. Pablo and Tuckett, David K. and Bartlett, Stephen D. and Flammia, Steven T. and Brown, Benjamin J.},
	urldate = {2022-11-17},
	date = {2021-12},
	langid = {english},
}

@article{tuckett_tailoring_2019,
	title = {Tailoring Surface Codes for Highly Biased Noise},
	volume = {9},
	issn = {2160-3308},
	url = {https://link.aps.org/doi/10.1103/PhysRevX.9.041031},
	doi = {10.1103/PhysRevX.9.041031},
	pages = {041031},
	number = {4},
	journaltitle = {Physical Review X},
	shortjournal = {Phys. Rev. X},
	author = {Tuckett, David K. and Darmawan, Andrew S. and Chubb, Christopher T. and Bravyi, Sergey and Bartlett, Stephen D. and Flammia, Steven T.},
	urldate = {2022-11-17},
	date = {2019-11-12},
	langid = {english},
}

@article{paetznick_performance_2022,
	title = {Performance of planar Floquet codes with Majorana-based qubits},
	rights = {{arXiv}.org perpetual, non-exclusive license},
	url = {https://arxiv.org/abs/2202.11829},
	doi = {10.48550/ARXIV.2202.11829},
	author = {Paetznick, Adam and Knapp, Christina and Delfosse, Nicolas and Bauer, Bela and Haah, Jeongwan and Hastings, Matthew B. and da Silva, Marcus P.},
	urldate = {2022-11-17},
	date = {2022},
	note = {Publisher: {arXiv}
Version Number: 2},
}

@article{gidney_stim_2021,
	title = {Stim: a fast stabilizer circuit simulator},
	volume = {5},
	issn = {2521-327X},
	url = {https://quantum-journal.org/papers/q-2021-07-06-497/},
	doi = {10.22331/q-2021-07-06-497},
	shorttitle = {Stim},
	pages = {497},
	journaltitle = {Quantum},
	shortjournal = {Quantum},
	author = {Gidney, Craig},
	urldate = {2022-11-17},
	date = {2021-07-06},
	langid = {english},
}

@article{royer_qubit_2018,
	title = {Qubit parity measurement by parametric driving in circuit {QED}},
	volume = {4},
	issn = {2375-2548},
	url = {https://www.science.org/doi/10.1126/sciadv.aau1695},
	doi = {10.1126/sciadv.aau1695},
	pages = {eaau1695},
	number = {11},
	journaltitle = {Science Advances},
	shortjournal = {Sci. Adv.},
	author = {Royer, Baptiste and Puri, Shruti and Blais, Alexandre},
	urldate = {2022-11-17},
	date = {2018-11-02},
	langid = {english},
}

@article{bacon_operator_2005,
	title = {Operator Quantum Error Correcting Subsystems for Self-Correcting Quantum Memories},
	rights = {Assumed {arXiv}.org perpetual, non-exclusive license to distribute this article for submissions made before January 2004},
	url = {https://arxiv.org/abs/quant-ph/0506023},
	doi = {10.48550/ARXIV.QUANT-PH/0506023},
	author = {Bacon, Dave},
	urldate = {2022-11-18},
	date = {2005},
	note = {Publisher: {arXiv}
Version Number: 4},
}

@article{roffe_bias-tailored_2022,
	title = {Bias-tailored quantum {LDPC} codes},
	rights = {{arXiv}.org perpetual, non-exclusive license},
	url = {https://arxiv.org/abs/2202.01702},
	doi = {10.48550/ARXIV.2202.01702},
	author = {Roffe, Joschka and Cohen, Lawrence Z. and Quintavalle, Armanda O. and Chandra, Daryus and Campbell, Earl T.},
	urldate = {2022-11-18},
	date = {2022},
	note = {Publisher: {arXiv}
Version Number: 2},
}

@article{chamberland_topological_2020,
	title = {Topological and Subsystem Codes on Low-Degree Graphs with Flag Qubits},
	volume = {10},
	issn = {2160-3308},
	url = {https://link.aps.org/doi/10.1103/PhysRevX.10.011022},
	doi = {10.1103/PhysRevX.10.011022},
	pages = {011022},
	number = {1},
	journaltitle = {Physical Review X},
	shortjournal = {Phys. Rev. X},
	author = {Chamberland, Christopher and Zhu, Guanyu and Yoder, Theodore J. and Hertzberg, Jared B. and Cross, Andrew W.},
	urldate = {2022-11-18},
	date = {2020-01-31},
	langid = {english},
}

@article{miao_overcoming_2022,
	title = {Overcoming leakage in scalable quantum error correction},
	rights = {Creative Commons Attribution 4.0 International},
	url = {https://arxiv.org/abs/2211.04728},
	doi = {10.48550/ARXIV.2211.04728},
	author = {Miao, Kevin C. and {McEwen}, Matt and Atalaya, Juan and Kafri, Dvir and Pryadko, Leonid P. and Bengtsson, Andreas and Opremcak, Alex and Satzinger, Kevin J. and Chen, Zijun and Klimov, Paul V. and Quintana, Chris and Acharya, Rajeev and Anderson, Kyle and Ansmann, Markus and Arute, Frank and Arya, Kunal and Asfaw, Abraham and Bardin, Joseph C. and Bourassa, Alexandre and Bovaird, Jenna and Brill, Leon and Buckley, Bob B. and Buell, David A. and Burger, Tim and Burkett, Brian and Bushnell, Nicholas and Campero, Juan and Chiaro, Ben and Collins, Roberto and Conner, Paul and Crook, Alexander L. and Curtin, Ben and Debroy, Dripto M. and Demura, Sean and Dunsworth, Andrew and Erickson, Catherine and Fatemi, Reza and Ferreira, Vinicius S. and Burgos, Leslie Flores and Forati, Ebrahim and Fowler, Austin G. and Foxen, Brooks and Garcia, Gonzalo and Giang, William and Gidney, Craig and Giustina, Marissa and Gosula, Raja and Dau, Alejandro Grajales and Gross, Jonathan A. and Hamilton, Michael C. and Harrington, Sean D. and Heu, Paula and Hilton, Jeremy and Hoffmann, Markus R. and Hong, Sabrina and Huang, Trent and Huff, Ashley and Iveland, Justin and Jeffrey, Evan and Jiang, Zhang and Jones, Cody and Kelly, Julian and Kim, Seon and Kostritsa, Fedor and Kreikebaum, John Mark and Landhuis, David and Laptev, Pavel and Laws, Lily and Lee, Kenny and Lester, Brian J. and Lill, Alexander T. and Liu, Wayne and Locharla, Aditya and Lucero, Erik and Martin, Steven and Megrant, Anthony and Mi, Xiao and Montazeri, Shirin and Morvan, Alexis and Naaman, Ofer and Neeley, Matthew and Neill, Charles and Nersisyan, Ani and Newman, Michael and Ng, Jiun How and Nguyen, Anthony and Nguyen, Murray and Potter, Rebecca and Rocque, Charles and Roushan, Pedram and Sankaragomathi, Kannan and Schuster, Christopher and Shearn, Michael J. and Shorter, Aaron and Shutty, Noah and Shvarts, Vladimir and Skruzny, Jindra and Smith, W. Clarke and Sterling, George and Szalay, Marco and Thor, Douglas and Torres, Alfredo and White, Theodore and Woo, Bryan W. K. and Yao, Z. Jamie and Yeh, Ping and Yoo, Juhwan and Young, Grayson and Zalcman, Adam and Zhu, Ningfeng and Zobrist, Nicholas and Neven, Hartmut and Smelyanskiy, Vadim and Petukhov, Andre and Korotkov, Alexander N. and Sank, Daniel and Chen, Yu},
	urldate = {2022-11-18},
	date = {2022},
	note = {Publisher: {arXiv}
Version Number: 1},
}

@article{higgott_pymatching_2021,
	title = {{PyMatching}: A Python package for decoding quantum codes with minimum-weight perfect matching},
	rights = {{arXiv}.org perpetual, non-exclusive license},
	url = {https://arxiv.org/abs/2105.13082},
	doi = {10.48550/ARXIV.2105.13082},
	shorttitle = {{PyMatching}},
	author = {Higgott, Oscar},
	urldate = {2022-11-18},
	date = {2021},
	note = {Publisher: {arXiv}
Version Number: 2},
}

@article{aaronson_improved_2004,
	title = {Improved simulation of stabilizer circuits},
	volume = {70},
	issn = {1050-2947, 1094-1622},
	url = {https://link.aps.org/doi/10.1103/PhysRevA.70.052328},
	doi = {10.1103/PhysRevA.70.052328},
	pages = {052328},
	number = {5},
	journaltitle = {Physical Review A},
	shortjournal = {Phys. Rev. A},
	author = {Aaronson, Scott and Gottesman, Daniel},
	urldate = {2022-11-18},
	date = {2004-11-30},
	langid = {english},
}

@article{molmer_multiparticle_1999,
	title = {Multiparticle Entanglement of Hot Trapped Ions},
	volume = {82},
	issn = {0031-9007, 1079-7114},
	url = {https://link.aps.org/doi/10.1103/PhysRevLett.82.1835},
	doi = {10.1103/PhysRevLett.82.1835},
	pages = {1835--1838},
	number = {9},
	journaltitle = {Physical Review Letters},
	shortjournal = {Phys. Rev. Lett.},
	author = {Mølmer, Klaus and Sørensen, Anders},
	urldate = {2022-11-18},
	date = {1999-03-01},
	langid = {english},
}

@article{bombin_optimal_2007,
	title = {Optimal resources for topological two-dimensional stabilizer codes: Comparative study},
	volume = {76},
	issn = {1050-2947, 1094-1622},
	url = {https://link.aps.org/doi/10.1103/PhysRevA.76.012305},
	doi = {10.1103/PhysRevA.76.012305},
	shorttitle = {Optimal resources for topological two-dimensional stabilizer codes},
	pages = {012305},
	number = {1},
	journaltitle = {Physical Review A},
	shortjournal = {Phys. Rev. A},
	author = {Bombin, H. and Martin-Delgado, M. A.},
	urldate = {2022-12-14},
	date = {2007-07-06},
	langid = {english},
}

@article{wen_quantum_2003,
	title = {Quantum Orders in an Exact Soluble Model},
	volume = {90},
	issn = {0031-9007, 1079-7114},
	url = {https://link.aps.org/doi/10.1103/PhysRevLett.90.016803},
	doi = {10.1103/PhysRevLett.90.016803},
	pages = {016803},
	number = {1},
	journaltitle = {Physical Review Letters},
	shortjournal = {Phys. Rev. Lett.},
	author = {Wen, Xiao-Gang},
	urldate = {2022-12-26},
	date = {2003-01-10},
	langid = {english},
}

@article{paraoanu_microwave-induced_2006,
	title = {Microwave-induced coupling of superconducting qubits},
	volume = {74},
	issn = {1098-0121, 1550-235X},
	url = {https://link.aps.org/doi/10.1103/PhysRevB.74.140504},
	doi = {10.1103/PhysRevB.74.140504},
	pages = {140504},
	number = {14},
	journaltitle = {Physical Review B},
	shortjournal = {Phys. Rev. B},
	author = {Paraoanu, G. S.},
	urldate = {2023-01-10},
	date = {2006-10-31},
	langid = {english},
}

@article{lalumiere_tunable_2010,
	title = {Tunable joint measurements in the dispersive regime of cavity {QED}},
	volume = {81},
	issn = {1050-2947, 1094-1622},
	url = {https://link.aps.org/doi/10.1103/PhysRevA.81.040301},
	doi = {10.1103/PhysRevA.81.040301},
	pages = {040301},
	number = {4},
	journaltitle = {Physical Review A},
	shortjournal = {Phys. Rev. A},
	author = {Lalumière, Kevin and Gambetta, J. M. and Blais, Alexandre},
	urldate = {2023-01-18},
	date = {2010-04-01},
	langid = {english},
}

@article{bombin_single-shot_2015,
	title = {Single-Shot Fault-Tolerant Quantum Error Correction},
	volume = {5},
	issn = {2160-3308},
	url = {https://link.aps.org/doi/10.1103/PhysRevX.5.031043},
	doi = {10.1103/PhysRevX.5.031043},
	pages = {031043},
	number = {3},
	journaltitle = {Physical Review X},
	shortjournal = {Phys. Rev. X},
	author = {Bombín, Héctor},
	urldate = {2023-01-24},
	date = {2015-09-28},
	langid = {english},
}

@article{fujiwara_ability_2014,
	title = {Ability of stabilizer quantum error correction to protect itself from its own imperfection},
	volume = {90},
	issn = {1050-2947, 1094-1622},
	url = {https://link.aps.org/doi/10.1103/PhysRevA.90.062304},
	doi = {10.1103/PhysRevA.90.062304},
	pages = {062304},
	number = {6},
	journaltitle = {Physical Review A},
	shortjournal = {Phys. Rev. A},
	author = {Fujiwara, Yuichiro},
	urldate = {2023-01-24},
	date = {2014-12-01},
	langid = {english},
}

@article{raussendorf_fault-tolerant_2006,
	title = {A fault-tolerant one-way quantum computer},
	volume = {321},
	issn = {00034916},
	url = {https://linkinghub.elsevier.com/retrieve/pii/S0003491606000236},
	doi = {10.1016/j.aop.2006.01.012},
	pages = {2242--2270},
	number = {9},
	journaltitle = {Annals of Physics},
	shortjournal = {Annals of Physics},
	author = {Raussendorf, R. and Harrington, J. and Goyal, K.},
	urldate = {2023-01-25},
	date = {2006-09},
	langid = {english},
}

@article{khaneja_cartan_2000,
	title = {Cartan Decomposition of {SU}(2{\textasciicircum}n), Constructive Controllability of Spin systems and Universal Quantum Computing},
	rights = {Assumed {arXiv}.org perpetual, non-exclusive license to distribute this article for submissions made before January 2004},
	url = {https://arxiv.org/abs/quant-ph/0010100},
	doi = {10.48550/ARXIV.QUANT-PH/0010100},
	author = {Khaneja, Navin and Glaser, Steffen},
	urldate = {2023-01-27},
	date = {2000},
	note = {Publisher: {arXiv}
Version Number: 1},
}

@article{bombin_logical_2021,
	title = {Logical blocks for fault-tolerant topological quantum computation},
	rights = {{arXiv}.org perpetual, non-exclusive license},
	url = {https://arxiv.org/abs/2112.12160},
	doi = {10.48550/ARXIV.2112.12160},
	author = {Bombin, Hector and Dawson, Chris and Mishmash, Ryan V. and Nickerson, Naomi and Pastawski, Fernando and Roberts, Sam},
	urldate = {2023-02-03},
	date = {2021},
	note = {Publisher: {arXiv}
Version Number: 1},
}

@article{chao_quantum_2018,
	title = {Quantum Error Correction with Only Two Extra Qubits},
	volume = {121},
	issn = {0031-9007, 1079-7114},
	url = {https://link.aps.org/doi/10.1103/PhysRevLett.121.050502},
	doi = {10.1103/PhysRevLett.121.050502},
	pages = {050502},
	number = {5},
	journaltitle = {Physical Review Letters},
	shortjournal = {Phys. Rev. Lett.},
	author = {Chao, Rui and Reichardt, Ben W.},
	urldate = {2023-02-03},
	date = {2018-08-01},
	langid = {english},
}

@article{chao_flag_2020,
	title = {Flag Fault-Tolerant Error Correction for any Stabilizer Code},
	volume = {1},
	issn = {2691-3399},
	url = {https://link.aps.org/doi/10.1103/PRXQuantum.1.010302},
	doi = {10.1103/PRXQuantum.1.010302},
	pages = {010302},
	number = {1},
	journaltitle = {{PRX} Quantum},
	shortjournal = {{PRX} Quantum},
	author = {Chao, Rui and Reichardt, Ben W.},
	urldate = {2023-02-03},
	date = {2020-09-03},
	langid = {english},
}

@article{chamberland_fault-tolerant_2019,
	title = {Fault-tolerant magic state preparation with flag qubits},
	volume = {3},
	issn = {2521-327X},
	url = {https://quantum-journal.org/papers/q-2019-05-20-143/},
	doi = {10.22331/q-2019-05-20-143},
	pages = {143},
	journaltitle = {Quantum},
	shortjournal = {Quantum},
	author = {Chamberland, Christopher and Cross, Andrew W.},
	urldate = {2023-02-03},
	date = {2019-05-20},
	langid = {english},
}

@article{panteleev_asymptotically_2021,
	title = {Asymptotically Good Quantum and Locally Testable Classical {LDPC} Codes},
	rights = {Creative Commons Attribution 4.0 International},
	url = {https://arxiv.org/abs/2111.03654},
	doi = {10.48550/ARXIV.2111.03654},
	author = {Panteleev, Pavel and Kalachev, Gleb},
	urldate = {2023-02-03},
	date = {2021},
	note = {Publisher: {arXiv}
Version Number: 2},
}

@article{baspin_connectivity_2022,
	title = {Connectivity constrains quantum codes},
	volume = {6},
	issn = {2521-327X},
	url = {https://quantum-journal.org/papers/q-2022-05-13-711/},
	doi = {10.22331/q-2022-05-13-711},
	pages = {711},
	journaltitle = {Quantum},
	shortjournal = {Quantum},
	author = {Baspin, Nouédyn and Krishna, Anirudh},
	urldate = {2023-02-03},
	date = {2022-05-13},
	langid = {english},
}

@article{breuckmann_quantum_2021,
	title = {Quantum Low-Density Parity-Check Codes},
	volume = {2},
	issn = {2691-3399},
	url = {https://link.aps.org/doi/10.1103/PRXQuantum.2.040101},
	doi = {10.1103/PRXQuantum.2.040101},
	pages = {040101},
	number = {4},
	journaltitle = {{PRX} Quantum},
	shortjournal = {{PRX} Quantum},
	author = {Breuckmann, Nikolas P. and Eberhardt, Jens Niklas},
	urldate = {2023-02-03},
	date = {2021-10-11},
	langid = {english},
}

@article{delfosse_spacetime_2023,
	title = {Spacetime codes of Clifford circuits},
	rights = {Creative Commons Attribution 4.0 International},
	url = {https://arxiv.org/abs/2304.05943},
	doi = {10.48550/ARXIV.2304.05943},
	author = {Delfosse, Nicolas and Paetznick, Adam},
	urldate = {2023-05-18},
	date = {2023},
	note = {Publisher: {arXiv}
Version Number: 1},
}

@article{bombin_unifying_2023,
	title = {Unifying flavors of fault tolerance with the {ZX} calculus},
	rights = {{arXiv}.org perpetual, non-exclusive license},
	url = {https://arxiv.org/abs/2303.08829},
	doi = {10.48550/ARXIV.2303.08829},
	author = {Bombin, Hector and Litinski, Daniel and Nickerson, Naomi and Pastawski, Fernando and Roberts, Sam},
	urldate = {2023-05-18},
	date = {2023},
	note = {Publisher: {arXiv}
Version Number: 1},
}

@article{mcewen_data_2023,
    title = {Data for ``Relaxing Hardware Requirements for Surface Code Circuits using Time-dynamics''},
    author = {Matt McEwen and Dave Bacon and Craig Gidney},
    date = {2023-01-31},
    doi = {10.5281/zenodo.7587578},
    url = {https://zenodo.org/record/7587578},
}

\appendix
\renewcommand\thefigure{\thesection.\arabic{figure}} 
\renewcommand\thetable{\thesection.\arabic{table}} 

\setcounter{figure}{0}  
\setcounter{table}{0}  
\section{Stabilizer Formalism and Detecting Regions}\label{app:stabilizers}

The stabilizer formalism provides helpful tools for analysing Clifford circuits, like the ones used in error correction \cite{gottesman_stabilizer_1997}.
We generically use `stabilizer' to mean a signed Pauli term used to describe a quantum state; a state is stabilized by a Pauli term if applying that term as an operation does not affect the state, or equivalently the state is an eigenstate of the given Pauli operator with an eigenvalue $\pm1$ matching the sign of the stabilizer.

In the main text we have described the notion of detectors and detecting regions.  Recall that detectors are the set of  measurement outcomes that display deterministic behavior under noiseless execution.  The choice of detectors is not unique, since products of detectors themselves are also valid detectors.  Given a choice of measurements with deterministic outcomes for noiseless execution, it is useful to define a formalism which allows for identifying detecting regions.

\subsection{Stabilizer Flows}\label{sec:stabilizer_flows}

Operations in stabilizer circuits, including preparation and measurement in a Pauli basis, parity measurements, and Clifford preparations, can be understood entirely by the way they transform stabilizers \cite{gottesman_heisenberg_1998, aaronson_improved_2004, gidney_stim_2021}.
Here, we introduce notation and definitions we use implicitly to describe these transformations in terms of `stabilizer flows'. 
Flows are a generalization of the notion of a stabilizer, which can also be applied to dissipative operations such as resets and measurements.

Let ${\mathcal P}_n$ denote the set of Pauli operators on $n$ qubits: all operators that are a tensor product of $\{I, X, Y, Z\}$ along with a phase $\pm 1$.
Let $g_s$ be an operation applied to a system $s$.
Let $\text{mix}_s$ be an operation that allocates a system $S$ and prepares it into the maximally mixed state.
Let $\text{init}_q$ be an operation that allocates a new qubit $q$ and prepares it into the state $|0\rangle$.
Let $\text{control}_q(x)$ be the operation $x$ controlled by qubit $q$.
Let $\text{expectation}_q(x)$ be the expected probability of measuring $|1\rangle$, if $q$ was measured in the computational basis after the operation $x$.

We define a stabilizer flow $A \xrightarrow{g} B$ in terms of \emph{cancelling phase kickback}.
The expression $A \xrightarrow{g} B$ is an assertion that phase kickback from $A$ before $g$ would be exactly cancelled by phase kickback from $B^{-1}$ after $g$.
Formally:

\begin{definition}Stabilizer flow arrow notation:
\label{definition:stabilizer_flow_arrow}
$$[A \xrightarrow{g} B] \equiv \left[\text{expectation}_q(H_q \cdot \text{control}_q(B_s^{-1}) \cdot g_s \cdot \text{control}_q(A_s) \cdot H_q \cdot \text{mix}_s \cdot \text{init}_q) = 0\right]$$
\end{definition}

This definition is shown as a circuit in \fig[a]{stabilizer_flows}.
Of note are the following properties of stabilizer flows:
\begin{itemize}
    \item The definition avoids using $g^{-1}$, or controlling $g$ with $q$, so that the definition can be used on irreversible operations such as measurement.
    \item The definition can be executed on a quantum computer, by running the prescribed series of operations to estimate the probability of measuring $|1\rangle$.
    When $A$ and $B$ are members of the Pauli group and $g$ is a stabilizer operation, such an experiment can efficiently determine whether a given stabilizer flow is correct because all expectations are either 0\%, 50\%, or 100\%.
    \item For a gate $g$ and a Pauli $A$, there may not exist any Pauli $B$ such that $A \xrightarrow{g} B$.  An example of this is when $g$ is Z-basis measurement and $A$ is the $X$ Pauli.  The measurement anticommutes with $X$, destroying any information that could be used after the measurement to cancel the phase kickback from X before the measurement. This type of missing stabilizer flow also arises in situations involving elided interactions with the classical world, such as dissipative reset operations.
    \item For a gate $g$ and a Pauli $A$, there may be multiple stabilizer flows, $A \xrightarrow{g} B$ and also $A \xrightarrow{g}C$.  An example of this is when $g$ is the gate that prepares the system into the $|0\rangle$ state from no previously existing qubit.  In this case no previously existing qubit is a $1$ dimensional Hilbert space, and we denote identity on this as $1$.  Then $1 \xrightarrow{R} Z$ as well as $1 \xrightarrow{R} I$.  This follows because after preparing $|0\rangle$, measuring $Z$ results in the $+1$ eigenstate of $Z$, while measuring $I$ always results in a $+1$ eigenstate.  This type of gate is at the origin of different choices one can make for detector regions.
    \item For a gate $g$ which results in a measurement record, the Pauli phase may depend on the measurement result.  For example measuring $Z$ with result $r$ of $0$ or $1$ results in the flow $I \xrightarrow{g} (-1)^r Z$
\end{itemize}

\begin{figure}[t!]
    \centering
    \resizebox{\linewidth}{!}{
        \includegraphics{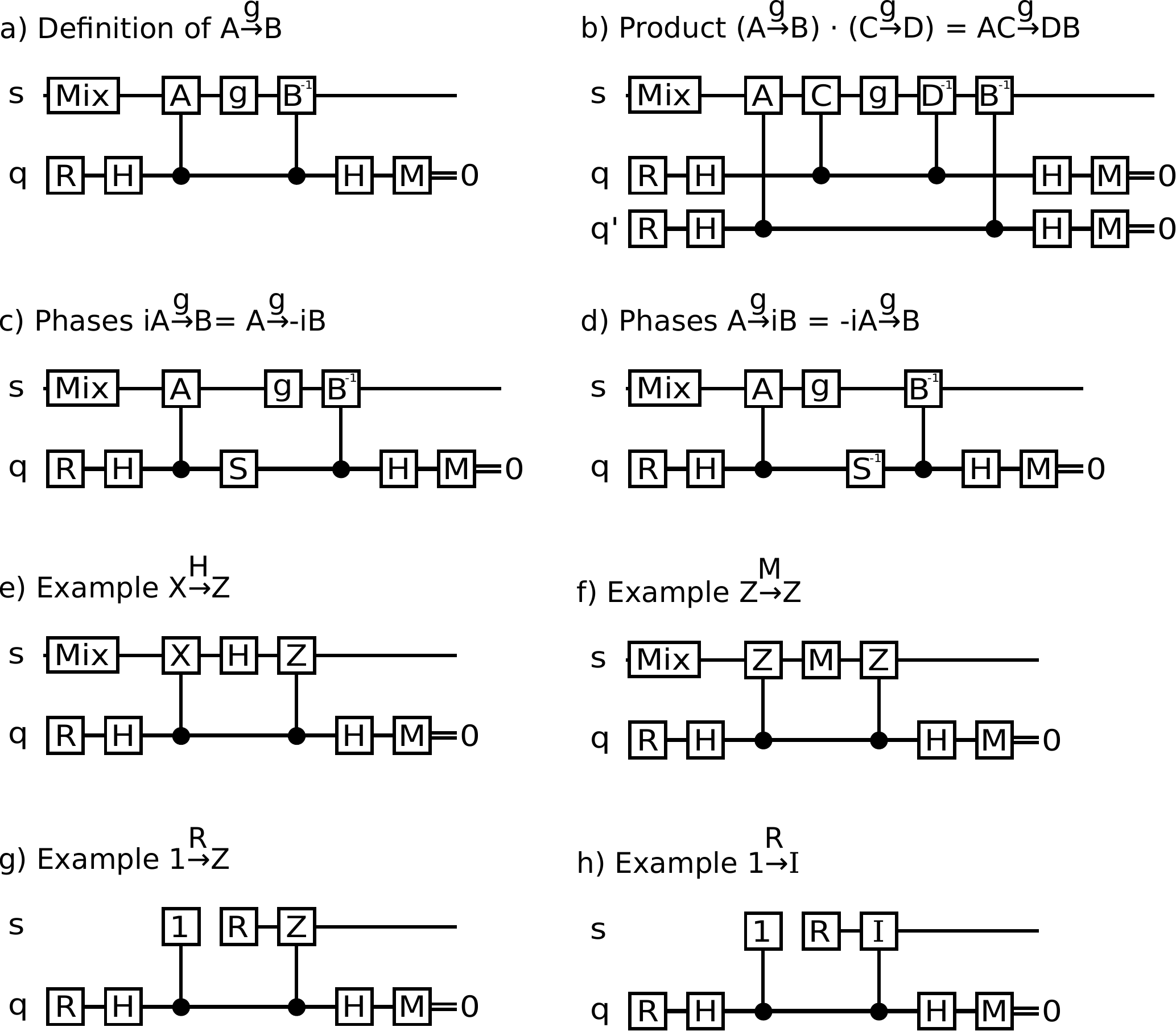}
    }
    \caption{
    \textbf{Definition of stabilizer flow arrow notation, and examples}
    We use the following non-standard gate indicators: \emph{Mix} prepares the maximally mixed state. $R$ here indicates the initialization of a new qubit, rather than the reset of an existing qubit which preserves the dimension of the Hilbert space. In all cases, the measurement outcomes are required to be always $0$.
    a) The definition of $A \xrightarrow{g} B$ in terms of a circuit. Here we show the case where $A$ and $B$ are single qubit Pauli operators, but similar circuits apply when $A$ and $B$ are (possibly differing) multi-qubit Pauli operators.  
    b) An illustration of the circuit defining the product of stabilizer flows, illuminating the reversed order of elements after the flow.
    c,d) Illustration of including a phase in the stabilizer flow circuit definition.
    e) An example of the circuit confirming a simple stabilizer flow $X \xrightarrow{H} Z$.
    f) An example of a circuit confirming a stabilizer flow $Z \xrightarrow{M} Z$ for a non-reversible operation, measurement in the $Z$ basis ($M$).
    g,h) Examples where the dimension of the spaces of the Pauli operators on the left hand and right hand side of a stabilizer flow arrow are different, for $1 \xrightarrow{R} Z$ and $1 \xrightarrow{R} I$. In this case $1$ is the operator which acts on the a $1$ dimensional space representing there being no qubit. 
    }
    \label{fig:stabilizer_flows}
\end{figure}

In \tab{stabilizer_flows} we present a list of stabilizer flows for a large set of Clifford gates, measurements, and preparations.  Note that for unitary gates, the stabilizer flow $A \xrightarrow{g} B$ can be directly calculated using $B = gAg^{-1}$.  The table presents only a set of {\em generators}.  This is because stabilizers flows form a group (for a proof of this see in \sec{stabilizer_flow_group}).  In particular this means that if one has two stabilizer flows $A \xrightarrow{g} B$ and $C \xrightarrow{g} D$, then the stabilizer flow $AC \xrightarrow{g} DB$ is a valid stabilizer flow.

\begin{lemma} Scalar factors can be moved across the flow by conjugation:
$$\forall c \in \mathbb{C} : (cA \xrightarrow{g} B) = (A \xrightarrow{g} B c^\ast)$$
\end{lemma}
Proof. A scalar factor on $A$ is equivalent to a Z rotation on the ancilla qubit $q$ during the controlled application of $A$.
This rotation commutes with $g$, so it can be slid to the other side of the circuit and fused into the controlled application of $B^{-1}$, without changing the function of the circuit.
The scalar ends up conjugated because $B$ is inverted in the circuit.\qed

\begin{definition}Stabilizer flow product:
\begin{align*}
    (A \xrightarrow{g} B) \cdot (C \xrightarrow{g} D) \equiv (AC \xrightarrow{g} DB)
\end{align*}
Informally, this product says that to combine stabilizer flows around a gate you should nest the stabilizer flows inside of each other.
Note the reversed order on the right hand side of the arrow.
Also note that computing the product can introduce scalars, for example $(X \xrightarrow{H} Z) \cdot (Z \xrightarrow{H} X) = (-iY \xrightarrow{H} -iY)= (Y \xrightarrow{H} -Y)$.
\end{definition}

Because stabilizer flows form a group, we need only specify a few generating stabilizer flows, from which the others can be obtained by the group product.

Finally we note that there are a few other ways of generating stabilizer flows:

\begin{definition}Stabilizer flow chaining:
$$\left[A \xrightarrow{g} B \xrightarrow{h} C\right] \equiv \left[(A \xrightarrow{g} B) \land (B \xrightarrow{h} C)\right]$$
\end{definition}

\begin{lemma}Gates can be composed in series by folding chains:
$$(A \xrightarrow{g} B \xrightarrow{h} C) \implies (A \xrightarrow{hg} C)$$
\end{lemma}

\begin{lemma}Gates can be composed in parallel by the tensor product
$$\left[ (A_1 \otimes C_2) \xrightarrow{g_1 \otimes h_2} (B_1 \otimes D_2) \right] = \left[ (A \xrightarrow{g} B) \land (C \xrightarrow{h} D) \right]$$
\end{lemma}

All stabilizer flow generators for Clifford operations can be found by looking in the Clifford operation's stabilizer tableau.
The tableau describes how single-qubit $X$ and $Z$ terms are transformed~\cite{aaronson_improved_2004, gidney_stim_2021}, and these transformations directly correspond to stabilizer flow generators.
Stabilizer flows are useful beyond tableaus because stabilizer flows can express the function of dissipative gates such as initialization and measurement.

\begin{table}
    \centering
    \resizebox{\linewidth}{!}{
    \begin{tabular}{|r|l|}
    \hline
    Gate & Generators \\
    \hline
    I & $\begin{aligned}
        X &\rightarrow X\\
        Z &\rightarrow Z
    \end{aligned}$ \\
    \hline
    X & $\begin{aligned}
        X &\rightarrow X\\
        Z &\rightarrow -Z
    \end{aligned}$ \\
    \hline
    Y & $\begin{aligned}
        X &\rightarrow -X\\
        Z &\rightarrow -Z
    \end{aligned}$ \\
    \hline
    Z & $\begin{aligned}
        X &\rightarrow -X\\
        Z &\rightarrow Z
    \end{aligned}$ \\
    \hline
    H & $\begin{aligned}
        X &\rightarrow Z\\
        Z &\rightarrow X
    \end{aligned}$ \\
    \hline
    S & $\begin{aligned}
        X &\rightarrow Y\\
        Z &\rightarrow Z
    \end{aligned}$ \\
    \hline
    $\text{H}_{YZ}$ & $\begin{aligned}
        X &\rightarrow -X\\
        Z &\rightarrow Y
    \end{aligned}$ \\
    \hline
    $\text{C}_{XYZ}$ & $\begin{aligned}
        X &\rightarrow Y\\
        Z &\rightarrow X
    \end{aligned}$ \\
    \hline
    $\text{Init}_Z $ & $\begin{aligned}
        1 &\rightarrow Z\\
    \end{aligned}$ \\
    \hline
    $\text{Init}_X $ & $\begin{aligned}
        1 &\rightarrow X\\
    \end{aligned}$ \\
    \hline
    $\text{Reset}_Z $ & $\begin{aligned}
        I &\rightarrow Z \\
    \end{aligned}$ \\
    \hline
    $\text{Reset}_X $ & $\begin{aligned}
        I &\rightarrow X \\
    \end{aligned}$ \\
    \hline
    $\text{Measure}_Z$  & $\begin{aligned}
        Z &\rightarrow Z\\
        I &\rightarrow (-1)^{\text{result}} \cdot Z\\
    \end{aligned}$ \\
    \hline
    $\text{Measure}_X$ & $\begin{aligned}
        X &\rightarrow X\\
        I &\rightarrow (-1)^{\text{result}} \cdot X\\
    \end{aligned}$ \\
    \hline
    $\text{DestructiveMeasure}_Z$ & $\begin{aligned}
        Z &\rightarrow (-1)^{\text{result}}\\
    \end{aligned}$ \\
    \hline
    $\text{DestructiveMeasure}_X$ & $\begin{aligned}
        X &\rightarrow (-1)^{\text{result}}\\
    \end{aligned}$ \\
    \hline
\end{tabular}~\begin{tabular}{|r|l|}
    \hline
    Gate & Generators \\
    \hline
    CNOT & $\begin{aligned}
        X_1 &\rightarrow X_1 X_2\\
        X_2 &\rightarrow X_1\\
        Z_1 &\rightarrow Z_1\\
        Z_2 &\rightarrow Z_1 Z_2
    \end{aligned}$ \\
    \hline
    ISWAP & $\begin{aligned}
        Z_1 &\rightarrow Z_2\\
        Z_2 &\rightarrow Z_1\\
        X_1 &\rightarrow Z_1 Y_2\\
        X_2 &\rightarrow Z_2 Y_1\\
    \end{aligned}$ \\
    \hline
    $\text{M}_{ZZ}$ & $\begin{aligned}
        Z_1 &\rightarrow Z_1\\
        Z_2 &\rightarrow Z_2\\
        X_1 X_2 &\rightarrow X_1 X_2 \\
        X_1 X_2 &\rightarrow (-1)^\text{result} \\
    \end{aligned}$ \\
    \hline
    $\text{Spider}_{Z3}$ & $\begin{aligned}
        Z_1 Z_2 &\rightarrow 1\\
        Z_2 Z_3 &\rightarrow 1\\
        X_1 X_2 X_3 &\rightarrow 1\\
    \end{aligned}$ \\
    \hline
    $\text{Spider}_{X4}$ & $\begin{aligned}
        X_1 X_2 &\rightarrow 1\\
        X_2 X_3 &\rightarrow 1\\
        X_3 X_4 &\rightarrow 1\\
        Z_1 Z_2 Z_3 Z_4 &\rightarrow 1\\
    \end{aligned}$ \\
    \hline
    BellMeasure & $\begin{aligned}
        X_1 X_2 &\rightarrow X_1 X_2\\
        Z_1 Z_2 &\rightarrow Z_1 Z_2\\
        X_1 X_2 &\rightarrow (-1)^{\text{result}_{XX}}\\
        Z_1 Z_2 &\rightarrow (-1)^{\text{result}_{ZZ}}\\
    \end{aligned}$ \\
    \hline
\end{tabular}~    \begin{tabular}{|r|l|}
    \hline
    Gate & Unsigned Generators \\
    \hline
    I, X, Y, Z & $\begin{aligned}
        X &\Rightarrow X\\
        Z &\Rightarrow Z
    \end{aligned}$ \\
    \hline
    $H, \sqrt{Y}, \sqrt{Y}^\dagger$ & $\begin{aligned}
        X &\Rightarrow Z\\
        Z &\Rightarrow X
    \end{aligned}$ \\
    \hline
    $S, S^\dagger, H_{XY}$ & $\begin{aligned}
        X &\Rightarrow Y\\
        Z &\Rightarrow Z
    \end{aligned}$ \\
    \hline
    $\sqrt{X}, \sqrt{X}^\dagger, \text{H}_{YZ}$ & $\begin{aligned}
        X &\Rightarrow X\\
        Z &\Rightarrow Y
    \end{aligned}$ \\
    \hline
    $\text{C}_{XYZ}$ & $\begin{aligned}
        X &\Rightarrow Y\\
        Z &\Rightarrow X
    \end{aligned}$ \\
    \hline
    $\text{C}_{ZYX}$ & $\begin{aligned}
        X &\Rightarrow Z\\
        Z &\Rightarrow Y
    \end{aligned}$ \\
    \hline
    $\text{Init}_X$ & $\begin{aligned}
        1 &\Rightarrow X\\
    \end{aligned}$ \\
    \hline
    $\text{Init}_Z$ & $\begin{aligned}
        1 &\Rightarrow Z\\
    \end{aligned}$ \\
    \hline
    $\text{Measure}_Z$ & $\begin{aligned}
        Z &\Rightarrow Z\\
        I &\Rightarrow Z\\
    \end{aligned}$ \\
    \hline
    $\begin{aligned}
        \text{DestructiveMeasure}_Z,\\
        \text{PostSelect}_Z\\
    \end{aligned}$ & $\begin{aligned}
        Z &\Rightarrow 1\\
    \end{aligned}$ \\
    \hline
    $\text{BellMeasure}$ & $\begin{aligned}
        X_1 X_2 &\Rightarrow 1\\
        Z_1 Z_2 &\Rightarrow 1\\
        1 &\Rightarrow X_1 X_2\\
        1 &\Rightarrow Z_1 Z_2
    \end{aligned}$ \\
    \hline
\end{tabular}
    }
    \caption{
        Stabilizer flow generators and unsigned stabilizer generators for various operations expressible in the stabilizer formalism.
        The $\text{H}_{YZ}$ gate is a 180 degree rotation around $Y+Z$.
        The $\text{C}_{XYZ}$ gate is a 120 degree rotation around $X+Y+Z$.
        The Spider gates correspond to phase-free ZX calculus nodes, and assume each edge is considered to be an incoming qubit.
    }
    \label{tab:stabilizer_flows}
\end{table}

\subsection{Unsigned Stabilizers}\label{sec:unsinged_stabilizers}

When working with stabilizers, the exact sign of the stabilizer is often not relevant.
Algorithmic decisions often depend only on whether or not two stabilizers commute, and commutation is independent of sign.
This is because, as Aaronson notes in the context of the computational complexity of stabilizer circuits~\cite{aaronson_introduction_2022}: ``The key observation is that the phases never feed back into the rest of the computation. Stabilizer operations can cause the phases to change, but a phase update never causes any other part of the tableau to change.".

In addition to often being irrelevant, the signs of stabilizers are the most costly values to compute.
For example, suppose you are given a stabilizer tableau $T$ corresponding to an $n$-qubit Clifford operation $C$, and told to use $T$ to compute the preimage of $X_0$ (the Pauli product before $C$ that is equivalent to $X_0$ after $C$).
All terms of this Pauli product can be computed in $\Theta(n)$ time, except the sign, which takes $\Theta(n^2)$ time.

Because of the low utility and high cost of tracking signs, it's often useful to take a view that completely ignores the signs.
Many things become simpler in this unsigned view.
For example, although the $X$ gate has stabilizer generators $X \rightarrow X$ and $Z \rightarrow -Z$ while the $I$ gate has different stabilizer generators $X \rightarrow X$ and $Z \rightarrow Z$, the differences are only in the signs.
From the unsigned perspective, Pauli gates can be ignored as if they were identity gates.

\begin{definition}To indicate that a stabilizer flow is unsigned, we will use a double arrow:
\label{definition:unsigned_gate_stabilizer_arrow}
$$\left[ A \xRightarrow{g} B \right] \equiv \left[\exists f : A \xrightarrow{g} B \cdot (-1)^{f(\text{measurements}_g)} \right]$$
\end{definition}

Here $f$ is a function that must predict which signed stabilizer is in effect based on the measurements performed by $g$.
Note that this definition means measurement results can be omitted when specifying unsigned stabilizers.
For example, $M_Z$ has the signed stabilizer $1 \xrightarrow{M_Z} (-1)^\text{result} \cdot Z$ and therefore, by setting $f$ to the identity function, has the unsigned stabilizer $1 \xRightarrow{M_Z} Z$.
This means that, in the unsigned view, measurement is generated by $1 \xRightarrow{M_Z} Z$ and $Z \xRightarrow{M_Z} 1$.
It looks like, and can be analyzed as if it were, a deterministic process.

Working with unsigned stabilizers removes time dependencies.
The direction of time normally matters because dissipative operations like resets and measurements have no time reversed equivalent.
In the unsigned view, these operations do have time reversed equivalents.
For example, the time reversed variant of an initialization is a measurement that discards the qubit.
One has the generator $1 \Rightarrow Z$ while the other has the generator $Z \Rightarrow 1$.
This insensitivity to the direction of time makes tasks simpler, and is particularly well suited for diagrammatic systems such as the ZX calculus where there isn't a prescribed time direction.

\subsection{Stabilizer flows form a group}\label{sec:stabilizer_flow_group}

In this subsection we go through the proof that stabilizer flows form a group.  This is presented for completeness, but can be easily skipped.

\begin{lemma}The stabilizer flows of a gate form a group.
\end{lemma}
Proof.  More formally the set is the set of stabilizer flows and the group operation is the group product between stabilizer flows defined above. 
Identity element: the identity element is $I \xrightarrow{g} I$.  Note that this is always defined, even when the $I$ on the left hand side of the arrow is a different dimension than the $I$ on the right had side of the arrow.
Inverse elements: the inverse element of $A \xrightarrow{g} B$ is $A^{-1} \xrightarrow{g}
B^{-1}$, and every Pauli operator has an inverse.
Associative product: if $x, y, z$ are stabilizer flows of $g$ then $(xy)z = x(yz)$ because the product definition reduces these expressions into expressions with associative matrix products.
Closed product: if $x, y$ are stabilizer flows of $g$ then $xy$ is also a stabilizer flow of $g$ because the right hand side of the required equality can be rewritten first using $x$ then using $y$ to transform it into the left hand side.
\qed

\subsection{Detecting Regions}

As we discuss at length in the text, detecting regions are a useful primitive concept for understanding quantum error correction circuits. Here, we elaborate on the formal definitions we use for circuits and for detecting regions.

Similar to Gottesman\cite{gottesman_opportunities_2022}, we regard the circuit as a set of \emph{locations} related by gates stabilizers.
A detecting region is then a list of circuit locations decorated by Pauli terms, or equivalently a set of pairs of a location and a Pauli term. A detecting region must satisfy the stabilizer flow for any locations related by a stabilizer flow. At a measurement, the detecting region may satisfy either stabilizer flow for a measurement, either \emph{transiting} through that measurement ($Z \rightarrow Z$), or \emph{terminating} on that measurement ($Z \rightarrow I$ or $I \rightarrow Z$).
A detector is the set of measurements that a detecting region terminates on.
We also say that a detecting region terminates on any reset, initialization, or destructive measurement it touches.




\subsection{KAK Decomposition and equivalence}\label{app:kak}

The KAK decomposition~\cite{khaneja_cartan_2000, tucci_introduction_2005} decomposes any two qubit unitary $U$ into single qubit gates and three commuting interactions parameterized by real numbers.
When $U$ is known to be a Clifford operation, the three parameterized interactions can be reduced to two interactions that are either included or excluded~\cite{corcoles_process_2013}:
$$U = (A \otimes B) \cdot \text{CZ}^{a} \cdot \text{SWAP}^{b} \cdot (C \otimes D)$$
where $A, B, C, D$ are single qubit Cliffords and $a, b$ are bits.
Operations with the same $a, b$ bits are KAK-equivalent ($\eqkak$).
On a machine with good single qubit operations, all KAK-equivalent interactions are effectively interchangeable.

There are four classes of KAK-equivalent Clifford gates.
The identity-like gates ($a=0, b=0$), the CNOT-like gates ($a=1, b=0$), the SWAP-like gates ($a=0, b=1$), and the ISWAP-like gates ($a=1, b=1$).
For example, the Mølmer–Sørensen gate~\cite{molmer_multiparticle_1999} is CNOT-like:
$$
\begin{aligned}
\sqrt{X \otimes X}
&=
(H \otimes H) \cdot \text{CZ} \cdot ((S \cdot H) \otimes (S \cdot H))
\\&\eqkak
(I \otimes H) \cdot \text{CZ} \cdot (I \otimes H)
\\&=
\text{CNOT}
\end{aligned}
$$
while a CNOT paired with a SWAP is ISWAP-like:
$$
\begin{aligned}
\text{CNOT} \cdot \text{SWAP}
&=
(I \otimes H) \cdot \text{CZ} \cdot \text{SWAP} \cdot (H \otimes I)
\\&\eqkak
(S \otimes S) \cdot \text{CZ} \cdot \text{SWAP}
\\&=
\text{ISWAP}
\end{aligned}
$$

Identity-like or swap-like gates are inappropriate for compiling QEC circuits because these gates aren't sufficient to create entanglement.
There are many papers that use CNOT-like gates to build the surface code circuits.
In \sec{iswap}, we present a circuit for the surface code using ISWAP-like gates.

\setcounter{figure}{0} 
\setcounter{table}{0}  
\section{Constructs for the Repetition Code}\label{app:step_code}

In the main text, we discussed the repetition code only briefly when introducing detecting regions. 
In practice, we developed the walking and ISWAP constructions on the simpler case of the repetition code before we progressed to the surface code. 
We include these constructions here for their pedagogical value. As previously noted, the overlapping structure of detecting regions is easier to visualize in the 2D-space-time of the repetition code.

\subsection{The Stepping Repetition Code}
As we have already noted in \fig{overlapping_det_regions}, the repetition code also features a half-cycle state where all qubits are involved in code stabilizers. As with the surface code state, there are many ways to map the half-cycle state to a state appropriate for measuring the stabilizers. 

We can approach the circuit for the repetition code as going from half-cycle to half-cycle rather than end-cycle to end-cycle. 
First, the circuit maps half of the half-cycle stabilizers to one-body stabilizers that can be measured, and then reconstructs the half-cycle state. 
This makes clear that any strategy that measures half the stabilizers in the half-cycle state and then reconstructs that state is a reasonable cycle circuit, regardless of whether the state is reconstructed on the same physical qubits.
The simplicity of the repetition code allows us to consider all possible cycles that achieve this, as shown in \fig[a]{step_code}.
We see that in addition to the two equivalent standard cycles for the bit flip repetition code, there are two other non-trivial cycles which involve constructing and deconstructing the half cycle stabilizers onto different qubits.

Allowing the freedom to use any of these four cycles makes what we call \emph{step code} circuits. These are logically equivalent to the repetition code, and are decoded in the same way. Similar to the walking circuits for the surface code, the step code circuits permit the freedom to move the code a distance up to 1 physical qubit in either direction in each cycle. \fig[b]{step_code} shows an example of a step code circuit. 

\begin{figure}[p]
    \centering
    \resizebox{\linewidth}{!}{
        \includegraphics{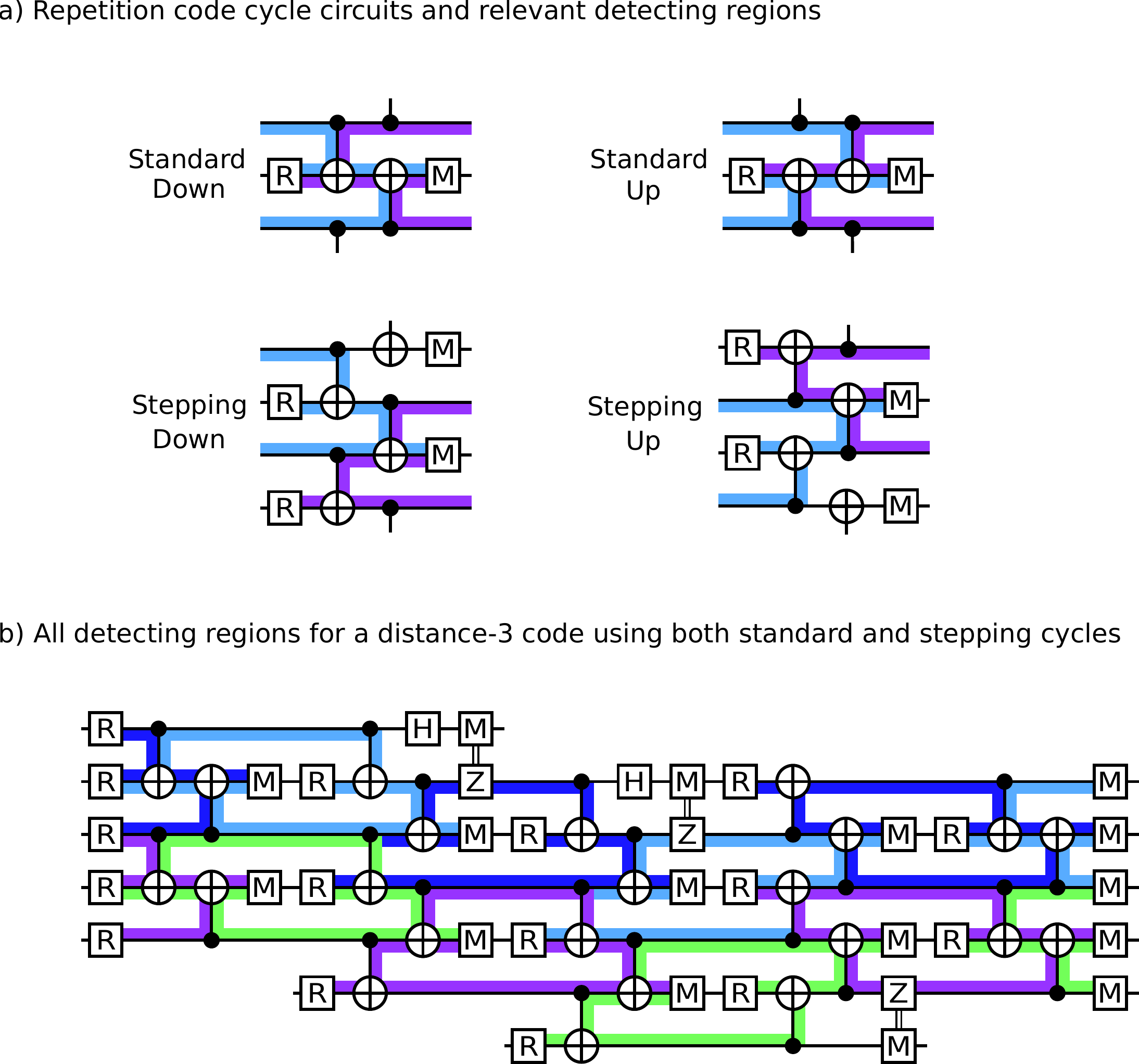}
    }
    \caption{
    \textbf{Repetition code circuits with stepping.}
    The bit-flip repetition code can also be implemented using a wider variety of cycles than are typically used.
    a) Four possible cycle circuits. The Standard cycles are equivalent typical circuits for the bit-flip code. The Stepping cycles measure the same stabilizers as the standard cycle, but the qubits that are measured are not the same qubits that were reset at the beginning of the cycle. This effectively exchanges the roles of data and measure qubits.
    b) All detecting regions for a distance-3 step code. In order, the cycles are [Standard Down, Step Down, Step Down, Step Up, Standard Down]. The combination of these cycles produces a variety of shapes of detecting regions, but their size and overlapping structure is the same as for the standard repetition code. 
    Note the inclusion of effective X-basis measurement and Z-basis classical feedback at boundaries being stepped away from; these correct the distance-1 X observable, and are unnecessary when considering the repetition code as a classical code. By combining various cycles, the repetition code state can step in either direction by one physical qubit per cycle.
    }
    \label{fig:step_code}
\end{figure}

\subsection{CXSWAP Repetition Code}

\begin{figure}[p]
    \centering
    \resizebox{\linewidth}{!}{
        \includegraphics{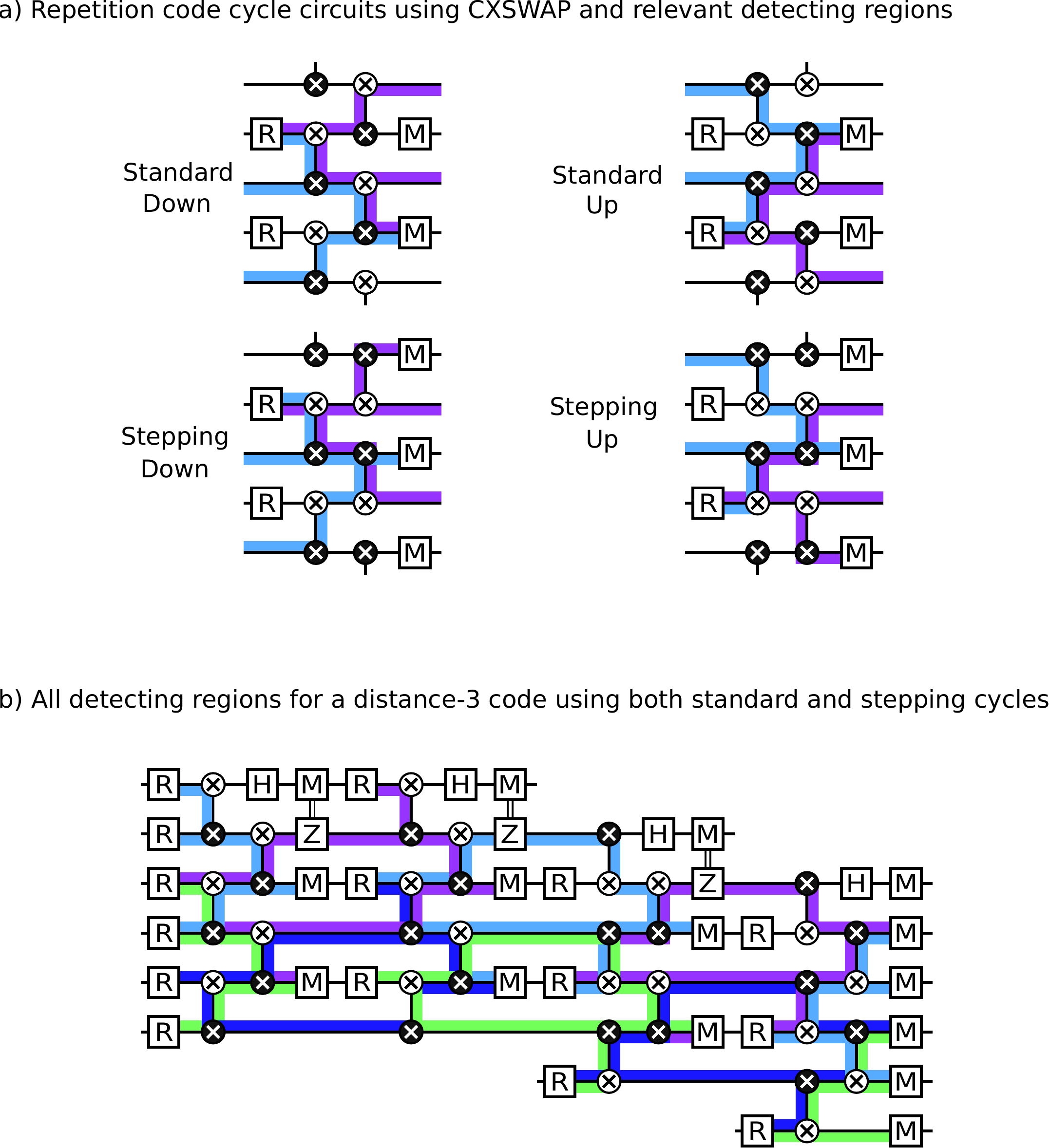}
    }
    \caption{
    \textbf{Repetition code circuits with CXSWAP gates.}
    The bit-flip repetition code can also be implemented using CXSWAP gates.
    a) Four possible cycle circuits. The Standard cycles preserve qubit roles, while the stepping cycles exchange qubit roles.
    b) All detecting regions for a distance-3 step code. In order, the cycles are [Standard Down, Standard Down, Step up, Standard Up]. The combination of these cycles produces a variety of shapes of detecting regions, but their size and overlapping structure is the same as for the standard repetition code. 
    Note the inclusion of effective X-basis measurement and Z-basis classical feedback at boundaries being stepped away from; these correct the distance-1 X observable, and are unnecessary when considering the repetition code as a classical code.
    }
    \label{fig:step_swap_code}
\end{figure}

We can similarly use freedom in the shape of detecting regions to provide cycle circuits that use CXSWAP gates instead of CX gates. This substitution also allows both standard circuits that preserve measure and data qubit roles, and stepping cycles where they are exchanged. This is illustrated in \fig{step_swap_code}.

\setcounter{figure}{0} 
\setcounter{table}{0}  
\section{Sliding Surface Codes for Logical Compilation}\label{app:sliding}

In \sec{walking}, we focused on explaining the walking circuit and describing its benefits in simplifying hardware. 
The capability of walking logical patches also provides new possible freedoms at a higher level, presenting a new primitive operation at the level of lattice surgery which can permit some logical operations to be cheaper. 
Here, we detail one such application, using the sliding behaviour to perform a lateral shift on a densely packed register of logical qubit patches.

In algorithms compiled down to lattice surgery primitives, shifting a dense linear register of logical qubit patches is a common and desirable operation~\cite{gidney_how_2021}. A typical use case is when processing register bits iteratively, treating the register as a queue to be consumed; here, being able to shift the register enables the compilation to operate on the least-significant bit and then shift the register, rather than needing to operate on each bit in its original location.

However, register shifting is not a very natural operation in lattice surgery, as the shifting of each logical patch must be done serially, as shown in \fig[a]{register_sliding}.
For a linear register of $N$ logical patches with cost distance $d$ attempting to shift $S$ patch distances along its length, the naive total cost of this strategy in space-time volume is $N(N+S) = N^2 + NS$ in units of $d^3$. This operation has a cost quadratic in the size of the register, and is particularly expensive for shifting a large register over a short distance.

Comparatively, we can use the sliding circuit for $2d$ cycles to shift all the logical patches in parallel by a distance $d$, as shown in \fig[b]{register_sliding}. This requires the existence of an additional qubit row parallel to the direction of shifting for the walking circuit to execute, but does not effect the overall volume consumed by the register as it walks, especially if surrounding patches are wiggling in-phase with the sliding. This can greatly reduces the cost of some logical compilations, such as using a large register as a FIFO queue.

\begin{figure}[h]
    \centering
    \resizebox{\linewidth}{!}{
        \includegraphics{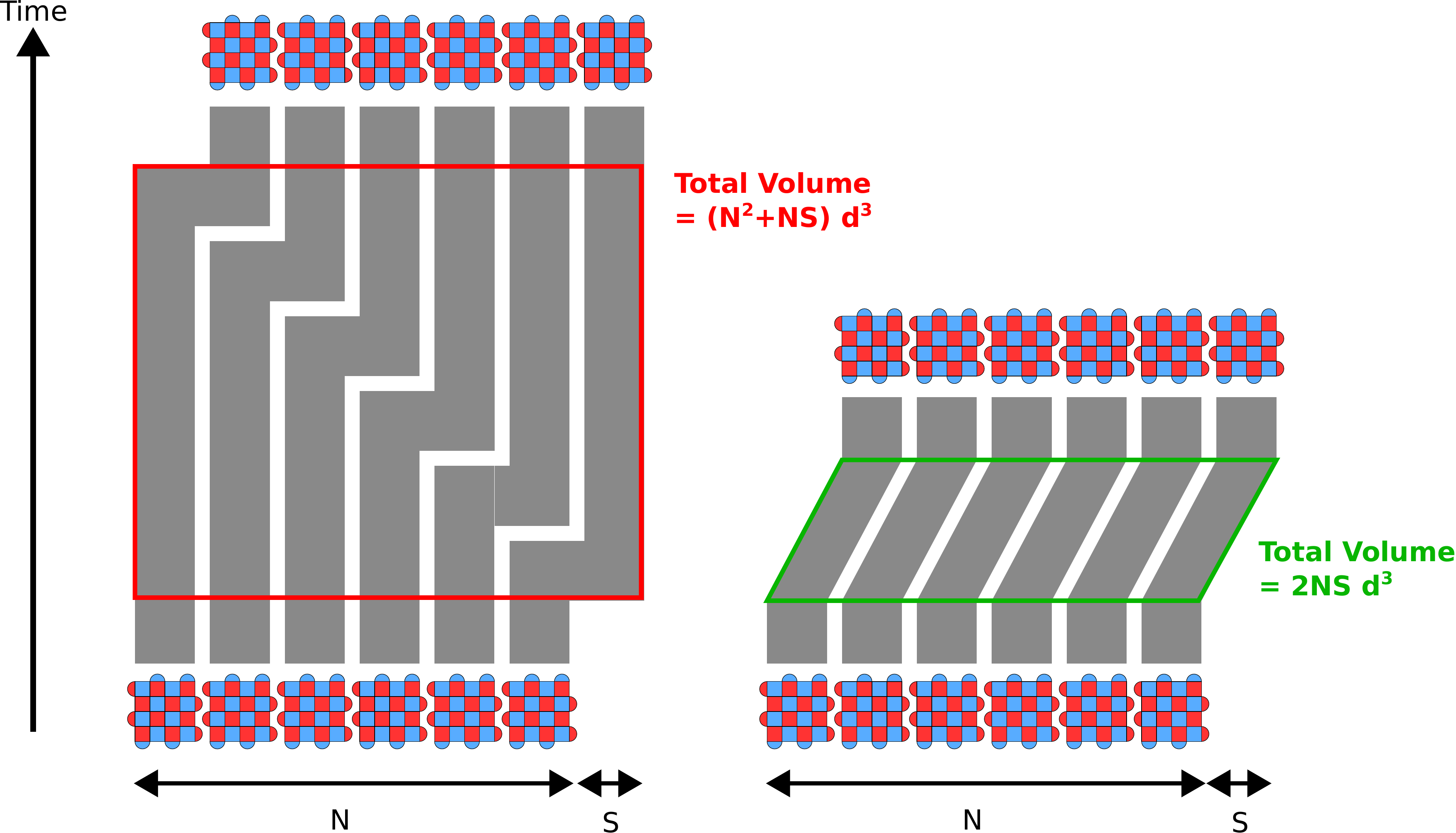}
    }
    \caption{
    \textbf{Cost comparison for register shifting using lattice surgery vs using sliding.}
    Left: A 2D projection of the naive strategy using lattice surgery expansions and contractions to shift a logical register of $N=6$ qubits by a distance $S=1$, consuming $(N^2+NS)$ space-time blocks each $d \times d \times d$ in volume.
    Right: the same operation achieved using the sliding behaviour on all patches making up the register, consuming only $2NS$ space-time blocks. For shifts $S<N$, this strategy lowers the overall space-time cost. 
    }
    \label{fig:register_sliding}
\end{figure}

\setcounter{figure}{0} 
\setcounter{table}{0}  
\section{Numerical Benchmarking and Noise Model}\label{app:noise_model}

For numerical benchmarking of our constructions, we use Stim~\cite{gidney_stim_2021}. In the main text, we compile the relevant circuit for superconducting hardware using CZ gates, and use the \texttt{SI1000} noise model introduced in~\cite{gidney_benchmarking_2022}. This noise model is inspired by the hardware errors experienced by superconducting transmon qubit arrays.  Here, we augment this error model with an ISWAP gate with the same leading fidelity. This elides any different in fidelity between performing a CZ or ISWAP gate, whereas we expect the error models in hardware to be noticeably different. 

Other constructions using CNOT or entangling measurements are benchmarked using the simpler \texttt{UniformDepolarizing} noise model, which does not have ratios of errors adjusted to minic real devices. 
The definitions of the noise operations in all three models are given in \tab{noise_gates}, and the models themselves are detailed in \tab{noise_model}.

The code used to run the numerical simulations is included in our data and code repository~\cite{mcewen_data_2023}. We used the Sinter interface and took up to $1\times10^8$ shots for each circuit implementation, completing early if we collected $1\times10^3$ logical errors for a specific circuit case. For decoding, we used an internal decoder that performs correlated-error-aware minimum weight perfect matching~\cite{google_quantum_ai_suppressing_2022}. We combined the decoded results for memory experiments in the X and Z basis individually into statistics for a combined XZ basis by projecting the likelihood that neither X nor Z bases experience a logical error. The benchmarking for all circuits collectively took $65 345 126$ core-seconds in total, which is just over $2$ core-years. It executed in a little under a 8 days on a 96 core machine.

To produce the teraquop footprint, we fit a line to the XZ logical error rate versus code distance to extrapolate to a logical error rate of $1\times10^{-12}$ to find the code distance and therefore footprint required, which we plot as the teraquop footprint in the main text.
Our plots generally include maximum likelihood highlighting to indicate confidence in our numerical estimates. In the plots of logical error rate and lambda, these correspond to the region of hypotheses up to 1000 times less likely than the maximum likelihood hypothesis. In the plots of teraquop footprint, these correspond to how far each point can be moved by using line fits whose least-squares error terms are at
most 1.0 higher than the error term for the optimal least-squares fit when fitting against the natural log of the logical error rates.
The procedures we used for producing these envelopes are also included in our code repository. 

\begin{table}[ht]
    \centering
    \resizebox{\linewidth}{!}{
    \begin{tabular}{|c|l|}
         \hline
         \textbf{Noisy Gate} & \textbf{Definition} \\
         \hline
         $\text{AnyClifford}_2(p)$ & \text{Any two-qubit Clifford gate, followed by a two-qubit depolarizing channel of strength $p$.} \\
         \hline
         $\text{AnyClifford}_1(p)$ & Any one-qubit Clifford gate, followed by a one-qubit depolarizing channel of strength $p$. \\
         \hline
         $\text{R}_{Z}(p)$ & Initialize the qubit as $\ket{0}$, followed by a bitflip channel of strength $p$. \\
         \hline
         $\text{R}_{X}(p)$ & Initialize the qubit as $\ket{+}$, followed by a phaseflip channel of strength $p$. \\
         \hline
         $M_Z(p, q)$ & Measure the qubit in the $Z$-basis, followed by a one-qubit depolarizing channel of strength $p$, \\
         & and flip the value of the classical measurement result with probability $q$.\\
         \hline
         $M_X(p, q)$ & Measure the qubit in the $X$-basis, followed by a one-qubit depolarizing channel of strength $p$, \\
         & and flip the value of the classical measurement result with probability $q$. \\
         \hline
         $M_{PP}(p, q)$ & Measure a Pauli product $PP$ on a pair of qubits, \\
         & followed by a two-qubit depolarizing channel of strength $p$, \\
         & and flip the classically reported measurement value with probability $q$. \\
         \hline
         $\text{Idle}(p)$ & If the qubit is not used in this time step, apply a one-qubit depolarizing channel of strength $p$. \\
         \hline
         $\text{ResonatorIdle}(p)$ & If the qubit is not measured or reset in a time step during which other qubits are \\ &  being measured or reset, apply a one-qubit depolarizing channel of strength $p$. \\
         \hline
    \end{tabular}
    }
    \caption{
        Modified from \cite{gidney_benchmarking_2022}. Noise channels and the rules used to apply them.
        Noisy rules stack with each other - for example, Idle($p$) and ResonatorIdle($p$) can both apply depolarizing channels in the same time step.
    }
    \label{tab:noise_gates}
\end{table}

\begin{table}[ht]
    \centering
    \begin{tabular}{|r|l|l|}
        \hline
         \textbf{Name}
             & \begin{tabular}{@{}l@{}}Uniform\\Depolarizing\end{tabular}
             & \begin{tabular}{@{}l@{}}Superconducting\\Inspired (SI1000)\end{tabular}
        \\\hline
        \textbf{Noisy Gateset}
            &\noindent\begin{tabular}{@{}l@{}}
                $\text{CX}(p)$\\
                $\text{CXSWAP}(p)$\\
                $\text{AnyClifford}_1(p)$\\
                $\text{R}_{Z/X}(p)$\\
                $M_{Z/X}(p, p)$\\
                $M_{PP}(p, p)$\\
                $\text{Idle}(p)$\\
            \end{tabular}
            &\begin{tabular}{@{}l@{}}
                \vspace{-0.25cm}
                {} \\
                $\text{CZ}(p)$\\
                $\text{ISWAP}(p)$\\
                $\text{AnyClifford}_1(p/10)$\\
                $\text{Init}_Z(2p)$\\
                $M_Z(p, 5p)$\\
                $M_{ZZ}(p, 5p)$\\
                $\text{Idle}(p/10)$\\
                $\text{ResonatorIdle}(2p)$\\
            \end{tabular}
        \\\hline
    \end{tabular}
    \caption{
        Modified from \cite{gidney_benchmarking_2022}. Details of the error models used in this paper.
        Circuits compiled for CX or CXSWAP gates are benchmarked using \texttt{UniformDepolarizing}. 
        Circuits compiled for CZ or ISWAP gates are benchmarked using \texttt{SI1000}.
        The superconducting-inspired acronym refers to an expected cycle time of about $1000$ nanoseconds for the standard surface code circuit cycle~\cite{google_quantum_ai_exponential_2021, google_quantum_ai_suppressing_2022}.
        See \tab{noise_gates} for definitions of the noisy gates.
    }
    \label{tab:noise_model}
\end{table}

\setcounter{figure}{0} 
\setcounter{table}{0}  
\section{Further Benchmarking}\label{app:benchmarking}

In this section, we present benchmarking for each of the constructions listed in \tab{big_table}.

\begin{figure}[p]
    \centering
    \resizebox{\linewidth}{!}{
        \includegraphics{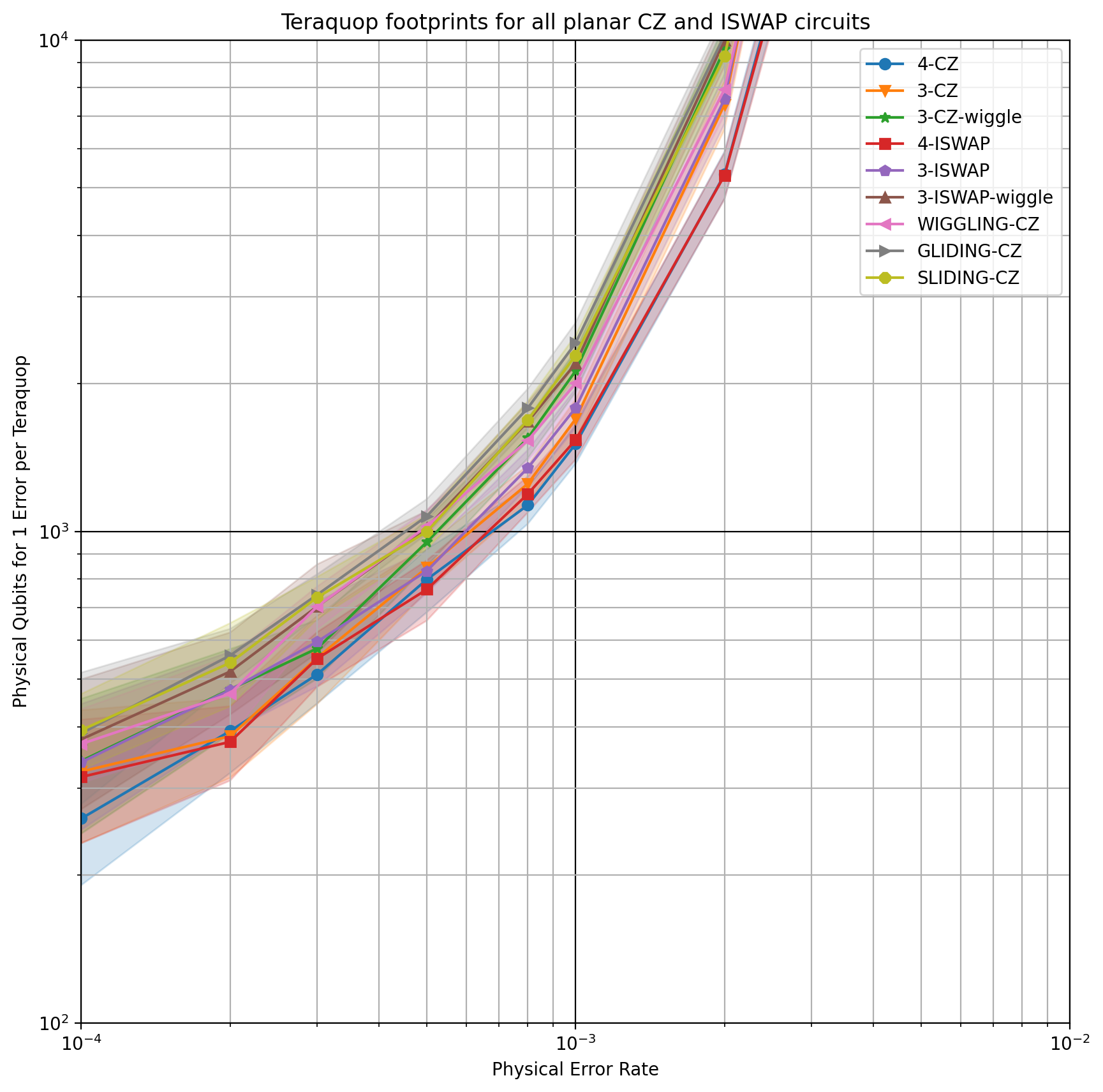}
    }
    \caption{
    \textbf{Teraquop footprints for planar circuits using the \texttt{SI1000} error model. }
    The number of physical qubits required for a single code patch to achieve a logical error rate of $1\times10^{-12}$ over a $d \times d \times d$ space-time block.
    Each curve averages memory experiments in both Z and X basis, and uses the \texttt{SI1000} noise model as described in \app{noise_model}. 
    Circuit details for each curve are included in \tab{big_table}. 
    }
    \label{fig:si1000_benchmarking}
\end{figure}

\begin{figure}[p]
    \centering
    \resizebox{\linewidth}{!}{
        \includegraphics{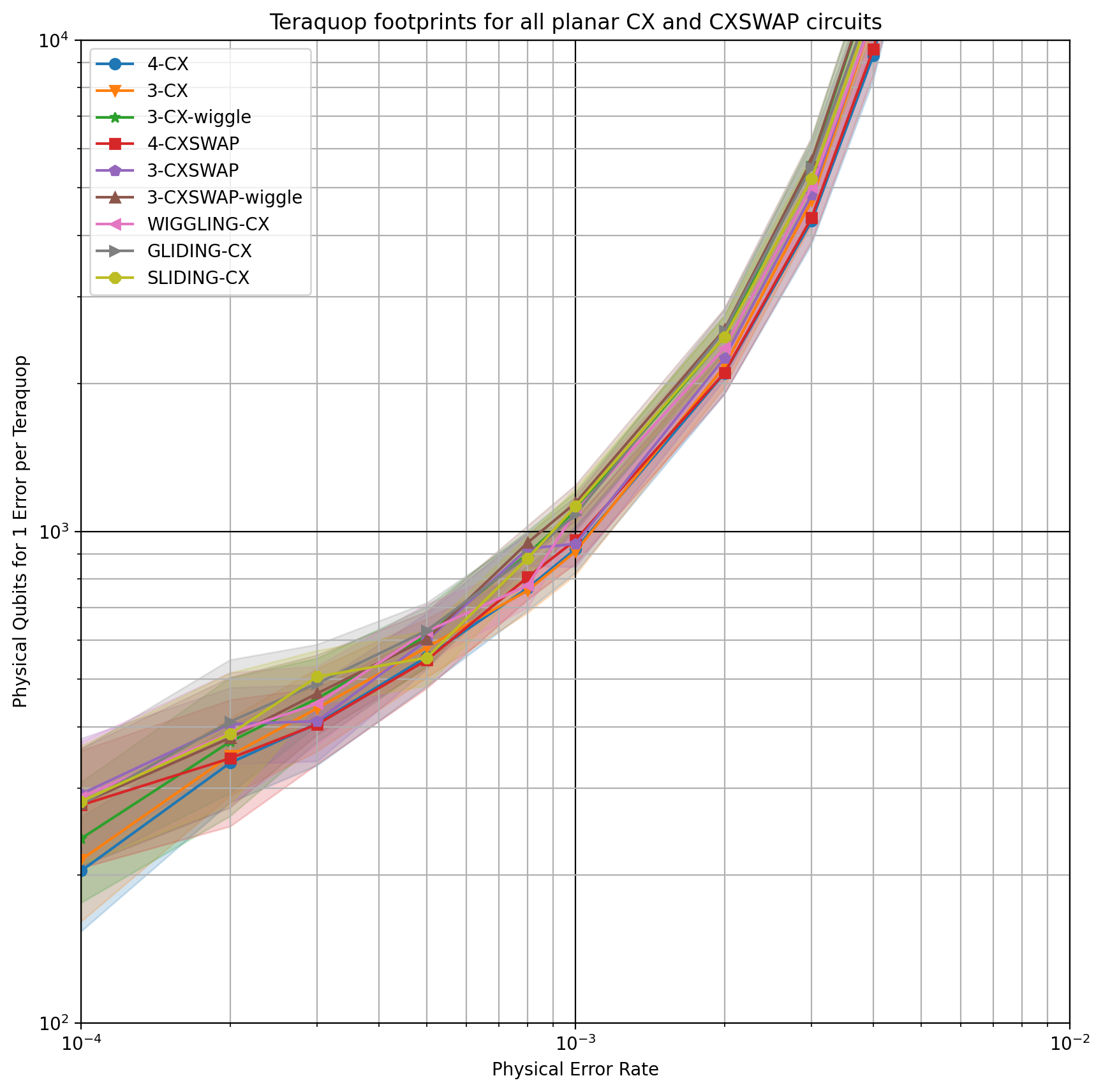}
    }
    \caption{
    \textbf{Teraquop footprints for planar circuits using the \texttt{UniformDepolarizing} error model. }
    The number of physical qubits required for a single code patch to achieve a logical error rate of $1\times10^{-12}$ over a $d \times d \times d$ space-time block.
    Each curve averages memory experiments in both Z and X basis, and uses the \texttt{UniformDepolarizing} noise model as described in \app{noise_model}. 
    Circuit details for each curve are included in \tab{big_table}.
    }
    \label{fig:ud_benchmarking}
\end{figure}

\begin{figure}[p]
    \centering
    \resizebox{\linewidth}{!}{
        \includegraphics{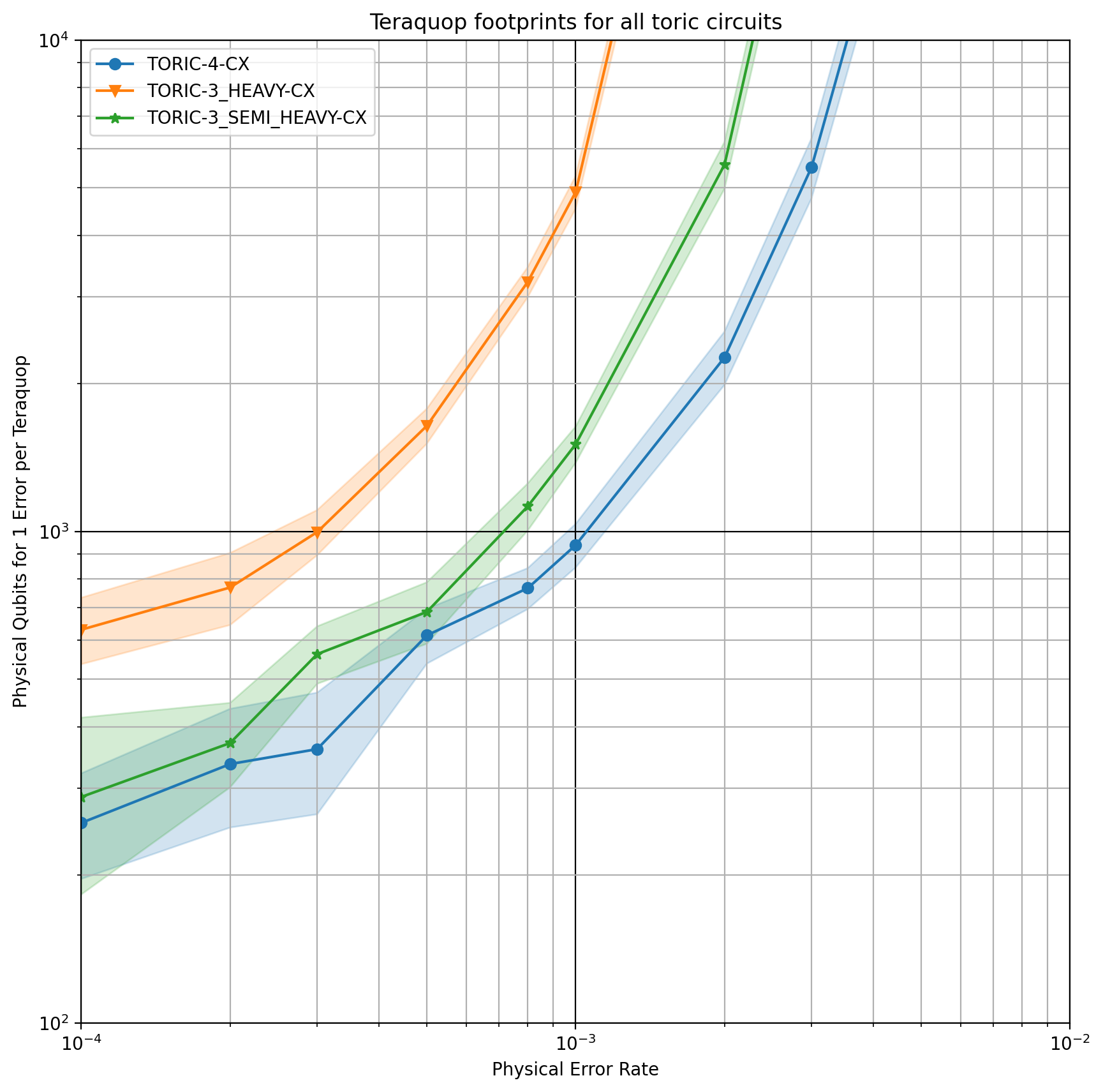}
    }
    \caption{
    \textbf{Teraquop footprints for circuits benchmarked with toric boundary conditions. }
    The number of physical qubits required for a single code patch to achieve a logical error rate of $1\times10^{-12}$ over a $d \times d \times d$ space-time block.
    Each curve averages memory experiments in both Z and X basis, and uses the \texttt{UniformDepolarizing} noise model as described in \app{noise_model}. 
    Circuit details for each curve are included in \tab{big_table}.
    }
    \label{fig:toric_benchmarking}
\end{figure}

\begin{figure}[p]
    \centering
    \resizebox{\linewidth}{!}{
        \includegraphics{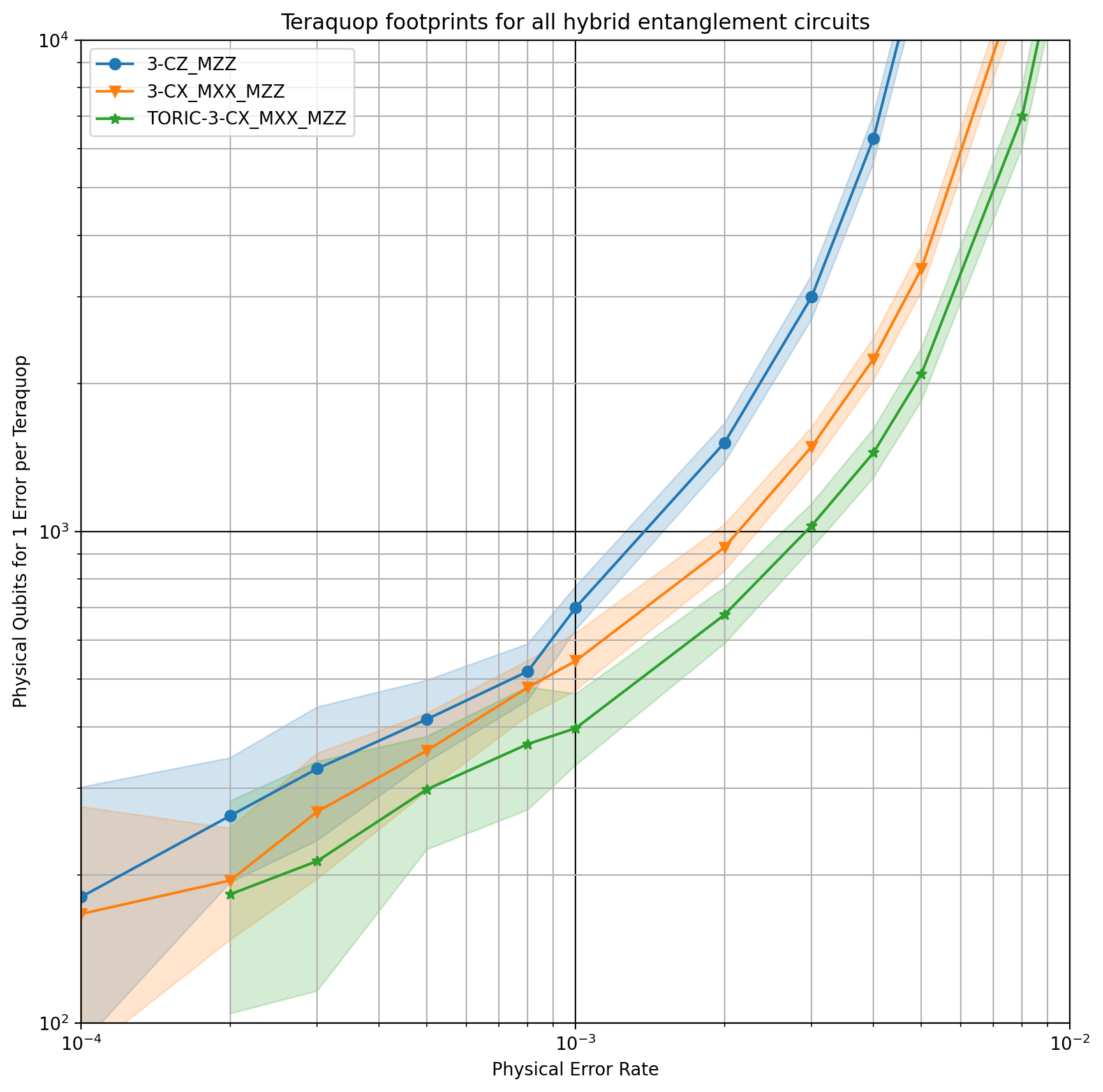}
    }
    \caption{
    \textbf{Teraquop footprints for circuits benchmarked with hybrid entangling operations. }
    The number of physical qubits required for a single code patch to achieve a logical error rate of $1\times10^{-12}$ over a $d \times d \times d$ space-time block.
    Each curve averages memory experiments in both Z and X basis.
    \texttt{3-CZ\_MZZ} uses the \texttt{SI1000} noise model,
    \texttt{3-CX\_MZZ\_MXX} and \texttt{TORIC-3-CX\_MZZ\_MXX} use the \texttt{UniformDepolarizing} noise model, 
    as described in \app{noise_model}. 
    Circuit details for each curve are included in \tab{big_table}.
    Note that \texttt{TORIC-3-CX\_MZZ\_MXX} has no data point for $p=1\times 10^{-4}$; No errors were sampled for distances above $d=4$, so extrapolation to the teraquop regime was not possible.
    }
    \label{fig:hybrid_benchmarking}
\end{figure}

We provide teraquop footprints for each circuit in groups sharing an error model. 
\fig{si1000_benchmarking} includes all planar circuits using CZ or ISWAP gates and benchmarked using the \texttt{SI1000} error model.
\fig{ud_benchmarking} includes all planar circuits using CX or CXSWAP gates and benchmarked using the \texttt{UniformDepolarizing} error model. 
\fig{toric_benchmarking} includes all toric circuits using CX gates, benchmarked using the \texttt{UniformDepolarizing} error model. 
Finally, \fig{hybrid_benchmarking} includes all circuits using entangling measurements. As detailed in \app{noise_model}, our noise models both treat entangling measurements as having the same depolarizing error as two qubit gates, so circuits using entangling measurements perform significantly better than circuits using only two qubits gates. 

 In the ancillary files available with the paper, and in the data repository~\cite{mcewen_data_2023}, we additionally provide a set of supplementary figures. For each circuit, we provide a visualization of the circuit schedule, a plot of the logical error versus physical error rate and the logical error rate versus code distance. The circuit schedules were automatically generated using Stim's diagram functionality, and can be reproduced from the circuits also included in the data repository. The raw data for the benchmarking plots, and the code that produces them, is also available in our code repository.

\setcounter{figure}{0} 
\setcounter{table}{0}  
\input{content/F_crumble_links}

\end{document}